\pdfoutput=1
\RequirePackage{fix-cm}
\documentclass[twocolumn]{svjour3}          

\smartqed  
\usepackage{graphicx}
\usepackage{mathptmx}      
\usepackage{latexsym}
\usepackage[numbers,comma]{natbib}  
\usepackage{verbatim}  
\usepackage[section]{placeins}
\usepackage{amsmath}
\usepackage{amssymb}
\usepackage{physics}
\usepackage[bottom]{footmisc}
\usepackage{float}
\usepackage{listings}		
\usepackage{xr}
\usepackage{xcolor}
\usepackage{mathtools,cuted}
\usepackage{float}

\usepackage{tcolorbox}

\renewcommand{\texttt}[1]{%
  \begingroup
  \ttfamily
  \begingroup\lccode`~=`/\lowercase{\endgroup\def~}{/\discretionary{}{}{}}%
  \begingroup\lccode`~=`[\lowercase{\endgroup\def~}{[\discretionary{}{}{}}%
  \begingroup\lccode`~=`.\lowercase{\endgroup\def~}{.\discretionary{}{}{}}%
  \catcode`/=\active\catcode`[=\active\catcode`.=\active
  \scantokens{#1\noexpand}%
  \endgroup
}
\usepackage{lipsum}

\newcommand{\url}[1]{%
  \begingroup
  \ttfamily
  \begingroup\lccode`~=`/\lowercase{\endgroup\def~}{/\discretionary{}{}{}}%
  \begingroup\lccode`~=`.\lowercase{\endgroup\def~}{.\discretionary{}{}{}}%
  \catcode`/=\active\catcode`.=\active
  \scantokens{#1\noexpand}%
  \endgroup
}

\definecolor{codegreen}{rgb}{0,0.4,0}
\definecolor{codegray}{rgb}{0.5,0.5,0.5}
\definecolor{codepurple}{rgb}{0.58,0,0.82}
\definecolor{backcolour}{rgb}{1,0.98,0.85}
\definecolor{codeblue}{rgb}{0,0,1}

\lstdefinestyle{matlabstyle}{
    backgroundcolor=\color{backcolour},   
    commentstyle=\color{codegreen},
    numberstyle=\tiny\color{codegray},
    stringstyle=\color{codepurple},
    basicstyle=\ttfamily\small,
    breakatwhitespace=false,         
    breaklines=true,                 
    captionpos=b,                    
    keepspaces=true,                 
    numbers=left,                    
    numbersep=4pt,                  
    showspaces=false,                
    showstringspaces=false,
    showtabs=false,                  
    tabsize=2,
    aboveskip=1.4em,
    belowskip=1.4em,
    frame=single, 
    framerule=0pt
}
\newcommand{\Evac}{$E_\mathrm{vac}$}
\newcommand{\ECB}{$E_\mathrm{CB}$}
\newcommand{\EVB}{$E_\mathrm{VB}$}
\newcommand{\Eg}{$E_\mathrm{g}$}

\newcommand{\EFn}{$E_\mathrm{Fn}$}
\newcommand{\EFp}{$E_\mathrm{Fp}$}
\newcommand{\Voc}{$V_\mathrm{OC}$}

\newcommand{\Jsc}{$J_\mathrm{SC}$}
\newcommand{\JV}{$J$-$V$}

\newcommand{\Vapp}{$V_{\mathrm{app}}$}

\newcommand{\Spiro}{Spiro-OMeTAD}

\newcommand{\mpTi}{mp-TiO$_2$}

\newcommand{\Ti}{TiO$_2$}
\newcommand{\kTPV}{$k_{\mathrm{TPV}}$}

\newcommand{\mobunit}{cm$^2$ V$^{-1}$ s$^{-1}$}
\newcommand{\rateunit}{cm$^{-3}$ s$^{-1}$}
\newcommand{\densunit}{cm$^{-3}$}

\newcommand{\Junit}{mA cm$^{-2}$}
\newcommand{\Jun}{A cm$^{-2}$}

\newcommand{\df}{\textsf{Driftfusion}}
\newcommand{\im}{IonMonger}

\newcommand{\SI}{Supplemental Information}
\journalname{}
\begin{document}

\title{Driftfusion \thanks{UK Engineering and Physical Sciences Research Council grant No. EP/J002305/1, EP/M025020/1, EP/M014797/1, EP/R020574/1, EP/R023581/1, EP/L016702/1, EP/T028513/1 (ATIP) and European Union's Horizon 2020 research and innovation program grant agreement No. 742708.}}
\subtitle{An open source code for simulating ordered semiconductor devices with mixed ionic-electronic conducting materials in one dimension}

\titlerunning{Driftfusion: An open source code for simulating ordered semiconductor devices}

\author{Philip Calado$^1$ \and Ilario Gelmetti$^2$ \and Benjamin Hilton$^1$ \and Mohammed Azzouzi$^1$ 
\and Jenny Nelson$^1$ \and Piers R. F. Barnes$^1$}


\institute{P. Calado	\\
              \email{p.calado13@imperial.ac.uk}             \\
$^1$ Department of Physics, Imperial College London, London SW7 2AZ, UK.	\\
$^2$ Institute of Chemical Research of Catalonia (ICIQ), Barcelona Institute of Science and Technology (BIST), Avda. Paisos Catalans 16, 43007 Tarragona, Spain.}

\date{}
\maketitle

\begin{abstract}
The recent emergence of lead-halide perovskites as active layer materials for thin film semiconductor devices including solar cells, light emitting diodes, and memristors has motivated the development of several new drift-diffusion models that include the effects of both electronic and mobile ionic charge carriers. In this work we introduce \df , a versatile simulation tool built for modelling one-dimensional ordered semiconductor devices with mixed ionic-electronic conducting layers. \df\ enables users to model devices with multiple, distinct, material layers using up to four charge carrier species: electrons and holes plus up to two ionic species. The time-dependent carrier continuity equations are fully-coupled to Poisson's equation enabling transient optoelectronic device measurement protocols to be simulated. In addition to material and device-wide properties, users have direct access to adapt the physical models for carrier transport, generation and recombination. Furthermore, a discrete interlayer interface approach circumvents the requirement for boundary conditions at material interfaces and enables interface-specific properties to be introduced.

\keywords{semiconductor device simulation \and numerical modelling \and drift-diffusion \and solar cells \and perovskites \and ionic-electronic conductors \and device physics}
\end{abstract}

\section{Introduction}
\label{intro}
Accurate models of semiconductor devices are essential to further our understanding of the key physical processes governing these systems and hence rationally optimise them. One approach to modelling devices on the mesoscopic scale is to use continuum mechanics, whereby charge carriers are treated as continuous media as opposed to discrete particles. Typically, electronic carriers are modelled at discrete energy levels with a transport model describing the dynamics of carriers in response to an electric field (drift) and carrier density gradients (diffusion). This drift-diffusion (Poisson-Nernst-Planck) treatment leads to a system of coupled partial differential equations (the van Roosbroeck system\cite{VanRoosbroeck1950}): a set of continuity equations, defining how the density of each charge carrier changes with time at each spatial location, are coupled with Poisson's equation (Gauss' Law), which relates the space-charge density to the electrostatic potential. For many architectures of thin-film semiconductor device (with the notable exception of transistors), provided that the materials are homogeneous and isotropic, it is sufficient to model devices with properties that vary in a single spatial dimension. In all but the most elementary of cases the resulting system of equations must be solved numerically.

\subsection{The emergence of lead-halide perovskites and recent progress in mixed electronic-ionic conductor device models}
\label{ssec:intro_lit_review}
The recent emergence of lead-halide perovskites (referred to herein as perovskites) as active layer materials for thin film semiconductor devices including solar cells, light emitting diodes (LEDs), and memristors has motivated the development of several new drift-diffusion models that include mobile ionic species in addition to electronic carriers.\cite{Foster2014, VanReenen2015a, Richardson2016, Neukom2017, Sherkar2017a, Garcia-Rosell2018b, Courtier2019, Bertoluzzi2019, Huang2020} Ab initio calculations and experimental evidence has shown that the charge density distribution, and consequently the electric field, in perovskite materials is dominated by high densities of relatively slow-moving mobile ionic defects.\cite{Xiao2014a,Eames2015,Yang2015,Haruyama2015} This has a profound impact on the optoelectronic response of devices with perovskite active layers, leading to strong hysteresis effects in experimental measurements on timescales from microseconds to hundreds-of-seconds.\cite{Moia2019, Calado2019_ideality}

To date, both experimental and theoretical research into perovskites has primarily focussed on their application as a photovoltaic absorber material for solar cells and we now review the recent advances in device-level modelling in this field of application.
Van Reenen, Kemerink \& Snaith were the first to publish perovskite solar cell (PSC) simulations using a coupled model that included continuity equations for three charge carriers: electrons, holes and a single mobile ionic species.\cite{VanReenen2015a} They found that current-voltage (\JV ) hysteresis in PSCs could only be reproduced by including a density of trap states close to one of the interfaces acting as a recombination centre.\cite{VanReenen2015a} Later calculations of a $1.5$ nm Debye length\footnote{Based on an ion density of $1.6\times10^{19}$ cm$^{-3}$.}\cite{Richardson2016} suggested, however, that the choice of a $4$ nm mesh spacing in the simulations was too coarse to properly resolve the ionic charge profiles at the perovskite active layer-transport layer interfaces (described herein simply as interfaces).
Richardson and co-workers overcame the numerical challenge of high ionic carrier and potential gradients at the interfaces by using an asymptotic analytical model to calculate the potential drop in the Debye layers of a single mixed electronic-ionic conducting material layer.\cite{Foster2014, Richardson2016} While this approach enabled the reproduction of hysteresis effects using high rates of bulk recombination, the inability to accurately model interfacial recombination limited the degree to which the simulation could represent real-world devices.\cite{Richardson2016} In a later publication by the same group modelling dark current transients, interfacial recombination was implemented, but only at the inner boundary of the Debye layer.\cite{OKane2017} Furthermore, since these models were limited to a single layer, unrealistically large ionic charge densities were calculated at the interfaces as compared to three-layer models with discrete electron and hole transport layers (ETL and HTL respectively).

Our own work simulating PSCs began with a three-layer p-i-n dual homojunction model in which the p- and n-type regions simulated the HTL and ETL and where interfacial recombination was approximated by including high rates of recombination throughout these layers. Our results supported van Reenen et al.'s conclusion that both mobile ions and high rates of interfacial recombination are required to reproduce \JV\ hysteresis effects and other comparatively slow transient optoelectronic phenomena in p-i-n solar cells.\cite{Calado2016} Shortly after Neukom et al. published a modelling study with similar conclusions.\cite{Neukom2017} They used the commercial package SETFOS\cite{fluxim2019semiconducting} to solve for electronic carriers in combination with a separate \texttt{MATLAB} code that solved for the ionic carrier distributions. More recently, Courtier et al. published results from \im , a freely-available, fully-coupled, three-layer device model that included a single ionic charge carrying species and boundary conditions at the interfaces such that surface recombination of electronic carriers at interfaces between the different layers could be explicitly included.\cite{Courtier2019, Courtier2019a} There remained some limitations with the model however; only the majority carriers were calculated in the ETL and HTL, excluding the possibility of simulating single carrier devices, and intrinsic or low-doped transport layers such as organic semiconductors; ions were confined to the perovskite layer and; users only have the possibility to simulate three-layer devices. Jacobs et al. also published results from a three-layer coupled electronic-ionic carrier simulation implemented using COMSOL Multiphyics\textregistered \cite{COMSOL2019} and \texttt{MATLAB} Livelink\texttrademark .\cite{MATLAB2017, Jacobs2018} Most recently, Tessler \& Vaynzof published impressive results from a similar three-layer PSC device model that included the option to use either Boltzmann or Fermi-Dirac statistics.\cite{Tessler2020} Notwithstanding, the methodological details from both Jacobs et al.\cite{Jacobs2018} and Tessler \& Vaynzof\cite{Tessler2020} are sparse and at the time of writing neither code is publicly available.

\subsection{Driftfusion: An open source code for simulating ordered semiconductor devices with mixed ionic-electronic conducting materials in one-dimension}
\label{ssec:intro_Driftfusion}
Here we present a comprehensive guide to \df , our open source simulation tool designed for simulating semiconductor devices with mixed ionic-electronic conducting layers in one dimension. The software (based in \texttt{MATLAB}) enables users to simulate devices with any number of distinct material layers and up to four charge carrying species: electrons and holes by default plus up to two ionic species. The time-dependent continuity equations are fully-coupled to Poisson's equation enabling transient optoelectronic measurements to be accurately simulated. In addition to common material parameters, users have direct access to adapt the carrier transport, recombination and generation models as well as the system's initial and boundary conditions.\cite{Singh2021} \df\ uses a discrete interlayer interface approach for junctions between material layers (heterojunctions) such that energetic and carrier density properties are graded between adjacent layers using a range of grading options. This method has the added benefits that it both circumvents the requirement for boundary conditions at heterojunctions, and enables interface-specific properties to be defined within the interface regions. While the example architectures and outputs given in this work use PSCs as a model system, \df\ can, in principle, be used to model any ordered one-dimensional mixed ionic-electronic semiconductor or redox system for which the drift-diffusion approach is valid.

This work is divided into four main sections; we begin with a general overview of the simulation tool in Section \ref{sec:general_overview}; in Section \ref{sec:theory} the default physical models for charge carrier transport, generation and recombination are outlined; Section \ref{sec:system_architecture} provides a detailed description of the system architecture and a step-by-step guide of the important commands and functions that will enable readers to get started with using \df ; we conclude in Section \ref{sec:comparisons} by comparing calculations from \df\ to two analytical and two numerical models to validate the simulation solutions and the discrete interface approach.

\section{General overview of \df }
\label{sec:general_overview}

\subsection{Workflow}
\begin{figure*}
\centering
\includegraphics[width=0.8\textwidth]{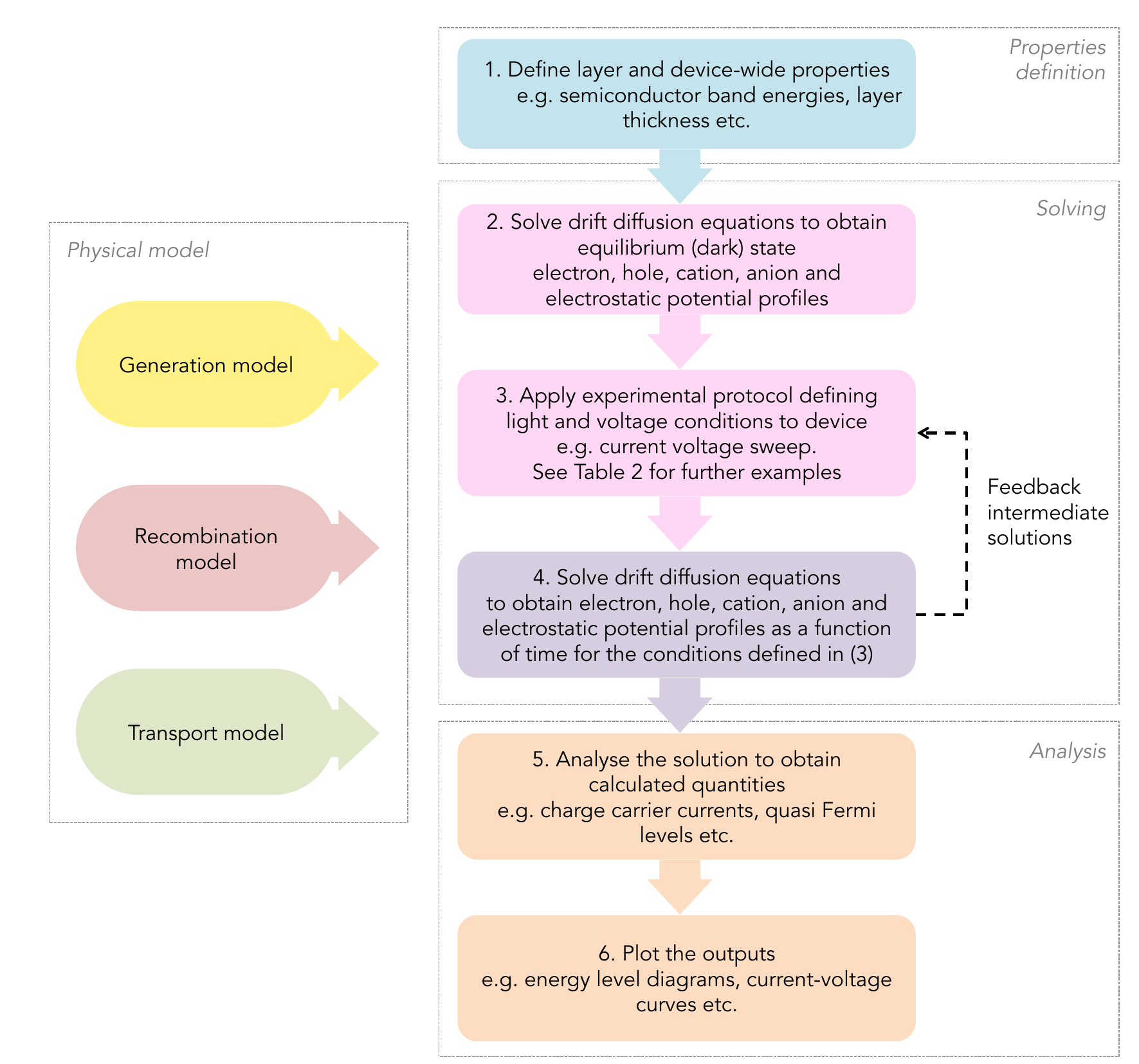}
\caption[The general workflow of \df .]{\textbf{The general workflow of \df .} 1. The user defines a device in the properties definition step; 2. The device equilibrium state is solved for using the given recombination and transport models; 3. An experimental protocol, which defines time-dependent voltage and optical generation conditions, is applied to the equilibrium solution; 4. A solution is obtained that may be fed back into the protocol until the desired solution is reached; 5. The solution is analysed to obtain calculated outputs; 6. The outputs are visualised using plotting tools.}	
\label{fig:overview_flow_diagram}
\end{figure*}

A flow diagram summarising \df 's general workflow is given in Figure \ref{fig:overview_flow_diagram}. The system is designed such that the user performs a linear series of steps to obtain a solution; (1) the process begins with the creation of a semiconductor device object for which both device-wide properties, such as the carrier extraction coefficients, and layer-specific properties, such carrier mobilities are defined. A user-definable physical model comprised of one-dimensional generation, recombination and transport models determines the continuity equation for each charge carrier (see Section \ref{sec:theory} for the default expressions); (2) the continuity equations are solved simultaneously with Poisson's equation (Equation \ref{eq:Poisson}) to obtain a solution for the electron density, hole density, cation density (optional), anion density (optional), and electrostatic potential distributions at equilibrium; (3) an experimental protocol such as a current-voltage scan is defined using the appropriate input parameters e.g. scan rate and voltage limits. The protocol generates time-dependent voltage and light conditions that are subsequently applied to the device, typically using the equilibrium solution as the initial conditions. In more sophisticated protocols the solution is broken into a series of steps whereby intermediate solutions are fed back into the solver. Likewise, protocols can be cascaded such that the solution from one protocol supplies the initial conditions for the next; (4) the desired solution is output as a \texttt{MATLAB} data structure (see \textit{Solution structures} highlighted box); (5) the solution structure can be analysed to obtain calculated outputs such as the charge carrier currents, quasi-Fermi levels etc; (6) a multitude of plotting tools are available to visualise the simulation outputs. Instructions on how to run each step programmatically and further details of the system architecture, protocols, solutions, and analysis functions are given in Section \ref{sec:system_architecture}.

\subsection{Licensing information}
The front end code of \df\ has been made open-source under the GNU Affero General Public License v3.0 in order to accelerate the rate of development and expand our collective knowledge of mixed ionic-electronic conducting devices.\cite{Calado2017} It is important to note, however, that \df\ currently uses \texttt{MATLAB}'s Partial Differential Equation solver for Parabolic and Elliptic equations (\texttt{pdepe}), licensed under the MathWorks, Inc. Software License Agreement, which strictly prohibits modification and distribution. If you use \df\ please consider giving back to the project by providing feedback and/or contributing to its continued development and dissemination.

We now proceed to describe the default physical models underlying this release of \df .\cite{Calado2017} Herein relevant \df\ functions and commands are highlighted using boxes and referred to using \texttt{command line typeface}.\\

\begin{tcolorbox}
\label{box:sol_structures}
\paragraph{Solution structures}
Following successful completion of the steps given in Figure \ref{fig:overview_flow_diagram}, \df\ outputs a \texttt{MATLAB} structure \texttt{sol} (known herein as a \textit{solution structure}) containing the following elements:
\begin{itemize}
\item The solution matrix \texttt{u}: a three-dimensional matrix for which the dimensions are \texttt{[time, space, variable]}. The order of the variables are as follows: 
\begin{enumerate}
\item Electron density
\item Hole density
\item Cation density (where 1 or 2 mobile ionic carriers are stipulated)
\item Anion density (where 2 mobile ionic carriers are stipulated)
\end{enumerate}
\item The spatial mesh \texttt{x}.
\item The time mesh \texttt{t}.
\item The parameters object \texttt{par}.
\end{itemize} 

As illustrated in Section \ref{box:sol_structures}, \texttt{sol} can be used as the input argument for analysis functions contained within \texttt{dfana} or plotting functions within \texttt{dfplot}. See Section \ref{sec:system_architecture} for further details.
\end{tcolorbox}

\section{Implementation of established semiconductor theoretical principles in \df }
\label{sec:theory}
The device physics implemented in \df\ is principally based on established semi-classical semiconductor transport and continuity principles, which are well described in Sze \& Kwok\cite{Sze1981} and Nelson\cite{Nelson2003}. Elements of this section are adapted from Ref. \cite{calado2017transient} and are provided here as a direct reference for the reader. The equations described herein are written in terms of a single spatial dimension and can only be applied to devices with one-dimensional architectures and for which the material layers are homogeneous.

\df\ evolved from a diffusion-only code written to simulate transient processes in dye sensitised solar cells\cite{Barnes2011} and uses \texttt{MATLAB}'s\cite{MATLAB2017} built-in \texttt{pdepe} solver.\cite{pdepe2013} The code solves the continuity equations and Poisson's equation for electron density $n$, hole density $p$, cation density $c$ (optional), anion density $a$ (optional), and the electrostatic potential $V$ as a function of position $x$, and time $t$.

The full details of the numerical methods employed by the \texttt{pdepe} solver for discretising the equations are given in Skeel \& Berlizns 1990.\cite{Skeel1990}

\subsection{Semiconductor energy levels}		\label{sec:Band_diagrams}
\label{ssec:energy_levels}
\begin{figure*}[h]
\centering
\includegraphics[width=0.8\textwidth]{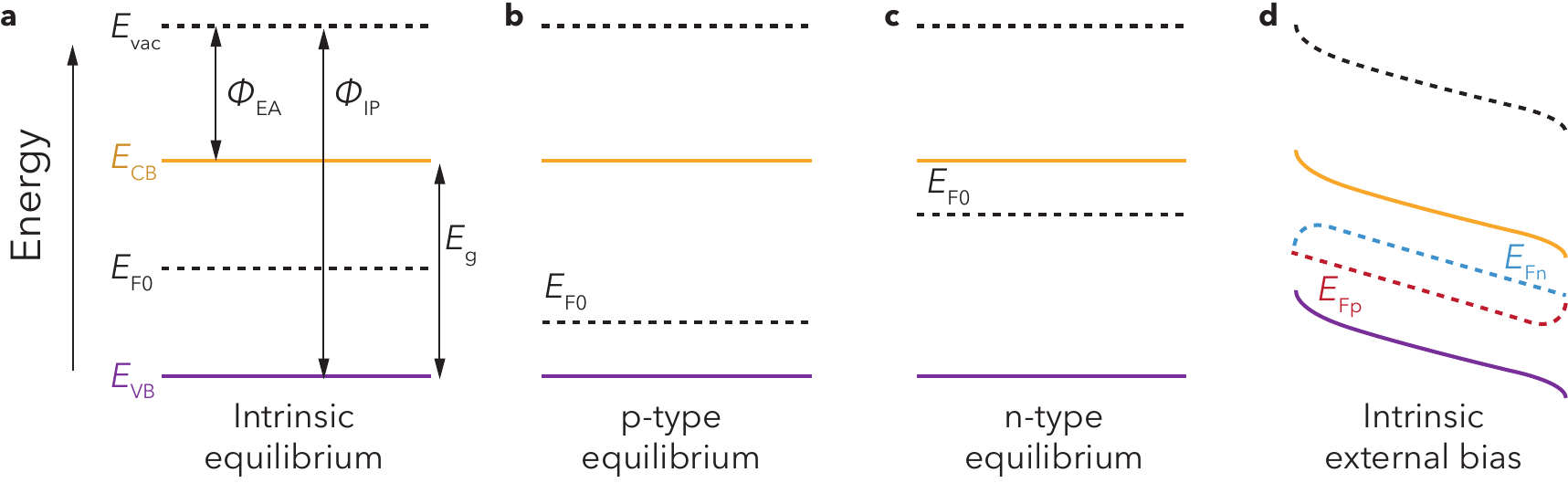}
\caption[Semiconductor energy levels]{\textbf{Semiconductor energy levels.} (\textbf{a}) An intrinsic semiconductor material showing the  vacuum level \Evac , electron affinity $\Phi_\mathrm{EA}$, ionisation potential $\Phi_\mathrm{IP}$, conduction and valence band energies \ECB\ and \EVB , band gap \Eg , equilibrium Fermi energy $E_\mathrm{F0}$ and electron and hole quasi-Fermi levels \EFn\ and \EFp . (\textbf{b}) A p-type material: $E_\mathrm{F0}$ lies closer to the VB due to acceptor impurities adding holes to the VB. (\textbf{c}) An n-type material: $E_\mathrm{F0}$ lies closer to the CB as donor impurities increase the CB electron density.}
\label{fig:Energies}
\end{figure*}

Figure \ref{fig:Energies}a shows the energy levels associated with an idealised intrinsic semiconductor. The electron affinity $\Phi_\mathrm{EA}$ and ionisation potential $\Phi_\mathrm{IP}$ are the energies required to add an electron to the conduction band (CB) from the vacuum level \Evac\ and to remove an electron from the valence band (VB) to \Evac\ respectively. Note that in contrast to the established convention and the description given here, in \df\ $\Phi_\mathrm{EA}$ and $\Phi_\mathrm{IP}$ are input as negative values for consistency with other energetic properties referenced using the electron energy scale.
 
The electronic band gap \Eg\ of the material can be defined as:

\begin{equation} \label{eq:Bandgap}
E_\mathrm{g} =  \Phi_\mathrm{IP} - \Phi_\mathrm{EA}
\end{equation}

\subsubsection{The vacuum energy}
\label{sssec:vacuum_energy}
The vacuum energy, \Evac\ is defined as the energy at which an electron is free of all forces from a solid including atomic and external potentials.\cite{Nelson2003} Spatial changes in the electrostatic potential, $V$ are therefore reflected in \Evac\ such that, at any point in space,

\begin{equation} \label{eq:E_vac}
E_{\mathrm{vac}}(x,t) = -qV(x,t) ,
\end{equation}

\noindent where $q$ is the elementary charge. 

\subsubsection{Conduction and valence band energies}
\label{sssec:conduction_valence_band_energy}
The conduction and valence band energies $E_{\mathrm{CB}}$ and $E_{\mathrm{VB}}$ are defined as the difference between the vacuum energy and $\Phi_\mathrm{EA}$ and $\Phi_\mathrm{IP}$ respectively. 

\begin{equation} \label{eq:E_CB}
E_{\mathrm{CB}}(x,t) = E_{\mathrm{vac}}(x,t)-\Phi_\mathrm{EA}(x),
\end{equation}

\begin{equation} \label{eq:E_VB}
E_{\mathrm{VB}}(x,t) = E_{\mathrm{vac}}(x,t)-\Phi_\mathrm{IP}(x).
\end{equation}

The band energies then, include both the potential energy associated with the specific molecular orbitals of the solid and the electrostatic potential arising from the existence of charge both within and external to the material.

\subsection{Electronic carrier densities and quasi-Fermi levels}
\subsubsection{The occupation probability distribution function and electronic equilibrium carrier densities}
\label{sssec:prob_dist_fun}
At equilibrium the net exchange of mass and energy into and out of, as well as between different locations within a system is zero. Under these conditions the average probability, $f$ that an electron will occupy a particular state of energy, $E$ at equilibrium in a semiconductor at temperature $T$ is given by the Fermi-Dirac (F-D) distribution function:

\begin{equation} \label{eq:Fermi_Dirac}
f(E, E_\mathrm{F0}, T) = \left( e^{\frac{E - E_\mathrm{F0}}{k_{\mathrm{B}} T}} + 1 \right)^{-1},
\end{equation}

\noindent where $k_\mathrm{B}$ is Boltzmann's constant. The equilibrium Fermi energy $E_\mathrm{F0}$ defines the energy at which a hypothetical electronic state has a $50$\% probability of occupation. At equilibrium, and for $V=0$, the Fermi energy is identical to the chemical potential of the material. For an intrinsic semiconductor, $E_\mathrm{F0}$ lies close to the middle of the gap. Where the semiconductor is p-type, dopant atoms accept electrons from the bands, shifting $E_\mathrm{F0}$ towards the valence band (Figure \ref{fig:Energies}b). Similarly, where the semiconductor is n-type, dopants donate electrons to the bands, shifting $E_\mathrm{F0}$ towards the conduction band (Figure \ref{fig:Energies}c). Note that herein the subscript `$0$' denotes the value or expression of properties at equilibrium e.g. $E_{CB,0}$ is the conduction band energy at equilibrium.\footnote{It follows that $E_{CB,0}$ is only dependent on position, while $E_{CB}$ is also time dependent due to the time-dependent nature of $V$.}

To obtain the density of free electrons in the conduction band, $n$, the product of the probability distribution function, $f$ and the conduction band density of states (DOS) function, $g_{CB}$ is integrated across energies above the conduction band edge, $E_{CB}$:

\begin{equation} \label{eq:Fermi_integral_n}
n_0 = \int_{E_\mathrm{CB,0}}^{\infty} g_\mathrm{CB}(E) f(E, E_\mathrm{F0}, T) dE.
\end{equation}

Similarly, to obtain the density of free holes in the valence band $p$, the product of the average probability that an electron is \textit{not} at energy $E$ (i.e. $(1 - f)$) with the valence band DOS function, $g_\mathrm{VB}$ is integrated across all energies up to the valence band edge $E_\mathrm{VB}$:

\begin{equation} \label{eq:Fermi_integral_p}
p_0 = \int_{-\infty}^{E_\mathrm{VB,0}} g_\mathrm{VB}(E) (1 - f(E, E_\mathrm{F0}, T)) dE.
\end{equation}

For semiconductor materials, $g_\mathrm{CB}$ and $g_\mathrm{VB}$ are typically modelled as parabolic functions with respect to electron energy at energies close to the band edges. Specifcally, $g_\mathrm{CB} = 4 \pi (2m_e^*/h^2)^{3/2}(E - E_\mathrm{CB})$ and $g_\mathrm{VB} = 4 \pi (2m_h^*/h^2)^{3/2}(E_\mathrm{VB} - E)$, where $m_e^*$ and $m_h^*$ are the effective electron and hole masses and $h$ is Planck's constant. Unfortunately, closed-form solutions cannot be found to Equations \ref{eq:Fermi_integral_n} and \ref{eq:Fermi_integral_p} using the parabolic band approximation and the F-D distribution function (Equation \ref{eq:Fermi_Dirac}). Hence, we use Blakemore's approximation\cite{Blakemore1982} to the above integrals to obtain closed-form expressions for the equilibrium carrier densities:

\begin{equation} \label{eq:n0}
n_0(x) = N_\mathrm{CB}(x) \left( e^{\frac{E_\mathrm{CB,0}(x) - E_\mathrm{F0}(x)}{k_{\mathrm{B}} T}} + \gamma \right)^{-1},
\end{equation}

\begin{equation} \label{eq:p0}
p_0(x)= N_\mathrm{VB}(x) \left( e^{\frac{E_\mathrm{F0}(x) - E_\mathrm{VB,0}(x)}{k_{\mathrm{B}} T}} + \gamma \right)^{-1},
\end{equation}

\noindent where $N_\mathrm{CB}$ and $N_\mathrm{VB}$ are the temperature-dependent\footnote{For simplicity, the temperature-dependence of $N_\mathrm{CB}$ and $N_\mathrm{VB}$ has been omitted from the equations herein and it should further be noted that this temperature dependence is not explicitly dealt with in this release of \df .} effective density of states (eDOS)\footnote{$N_\mathrm{CB} = 2(2\pi m_e^* k_{\mathrm{B}} T/h^2)^{3/2}$ and $N_\mathrm{VB} = 2(2\pi m_h^* k_{\mathrm{B}} T/h^2)^{3/2}$} of the conduction and valence bands respectively. $\gamma$ is a constant defining how close the approximation is to the Boltzmann regime (note that Equations \ref{eq:n0} and \ref{eq:p0} reduce to the Boltzmann approximation for $\gamma  = 0$). Following Farrell et al., we set $\gamma = 0.27$ by default for ordered materials.\cite{Farrell2017a} This results in a close agreement to F-D statistics for $E_\mathrm{VB} - 1.3 k_\mathrm{B} T  < E_\mathrm{F0} < E_\mathrm{CB} + 1.3 k_\mathrm{B} T$, such that degenerate semiconductor states are permissible within the scope of the model.

\subsubsection{Equilibrium Fermi levels in doped materials}
In charge-neutral n-type materials the equilibrium electron density is approximately equal to the density of donor dopant atoms such that $n_0 \approx N_\mathrm{D}$. Similarly in p-type materials, $p_0 \approx N_\mathrm{A}$, where $N_\mathrm{A}$ is the density of acceptor dopants. In \df\ users input values for $E_{\mathrm{F0}}$ for each material layer and the corresponding equilibrium carrier and doping densities, $n_0$, $p_0$, $N_\mathrm{D}$, and $N_\mathrm{A}$ are calculated during creation of the device parameters object (see Section \ref{ssec:parameters}) according to Equations \ref{eq:n0} and \ref{eq:p0}.

\begin{tcolorbox}
The equilibrium carrier densities \texttt{n0} and \texttt{p0} and Fermi levels \texttt{EF0} \textit{for individual material layers} are calculated and stored as a function of position in the device structures \texttt{dev} and \texttt{dev\_sub} of the device parameters object \texttt{par}. See Subsection \ref{sssec:device_structures} for further details. Note that when the material layers with different equilibrium Fermi levels are brought into contact, $n_0$, $p_0$, and $E_{\mathrm{F0}}$ become position-dependent owing to the creation of a space charge regions and associated electric fields.
\end{tcolorbox}

\subsubsection{Quasi-Fermi levels}
\label{sssec:qfls}
A key approximation in semiconductor physics is the assumption that, under external optical or electrical bias, the electron and hole populations at any particular location can be treated separately, with individual distribution functions and associated quasi-Fermi levels (QFLs), \EFn\ and \EFp\ (Figure \ref{fig:Energies}d). This is permitted because thermal relaxation of carriers to the band edges is typically significantly faster than interband relaxation, resulting in quasi-equilibrium states for each population.\cite{Nelson2003} Under these circumstances a similar approach to that taken for the true equilibrium state can be used to derive expressions for the QFLs:

\begin{equation} \label{eq:Fermi_e}
E_{\mathrm{Fn}}(x,t) = E_{\mathrm{CB}}(x,t)+k_\mathrm{B}T \ln\left( \dfrac{n(x,t)}{N_{\mathrm{CB}}(x)}\ - \gamma \right).
\end{equation}

\begin{equation} \label{eq:Fermi_h}
E_{\mathrm{Fp}}(x,t) = E_{\mathrm{VB}}(x,t)-k_\mathrm{B}T \ln\left( \dfrac{p(x,t)}{N_{\mathrm{VB}}(x)}\ - \gamma \right).
\end{equation}

It can be helpful to conceptualise the QFLs as the sum of the electrostatic ($V$) and average chemical potential energy ($k_\mathrm{B}T\ln (n/N_\mathrm{CB} - \gamma) - \Phi_\mathrm{EA}$ for electrons, $-k_\mathrm{B}T\ln (p/N_\mathrm{VB} - \gamma) - \Phi_\mathrm{IP}$ for holes) components of the carriers at each location. It follows that the gradient of the QFLs provides a convenient way to determine the direction of the current since, from the perspective of the electron energy scale, electrons move `downhill', and holes move `uphill' in response to electrochemical potential gradients. Moreover, the electron and hole currents, $J_n$ and $J_p$, can be expressed in terms of the product of the electron and hole QFL gradients with the corresponding carrier conductivities, $\sigma _n$ and $\sigma _p$:

\begin{equation} \label{eq:Fermi_grade}
J_n(x,t) = \dfrac{\sigma _n}{q} \dv{E_{\mathrm{Fn}}(x,t)}{x},
\end{equation}

\begin{equation} \label{eq:Fermi_gradh}
J_p(x,t) = \dfrac{\sigma _p}{q} \dv{E_{\mathrm{Fp}}(x,t)}{x}.
\end{equation}

\noindent Here, the conductivities are the product of the electronic carrier mobilities $\mu_{n}$ and $\mu_{p}$ with their corresponding concentrations and the elementary charge:

\begin{equation} \label{eq:Conduct_e}
\sigma _n(x,t) = qn(x,t)\mu_n(x),
\end{equation}

\begin{equation} \label{eq:Conduct_h}
\sigma _p(x,t) = qp(x,t)\mu_p(x).
\end{equation}

\begin{tcolorbox}
The band energies, \texttt{Ecb} and \texttt{Evb} and electron and hole QFLs, \texttt{Efn} and \texttt{Efp} can be calculated from a \df\ solution structure, \texttt{sol} by using the function:

\begin{lstlisting}[numbers=none]
[Ecb,Evb,Efn,Efp] =...
					 dfana.calcEnergies(sol)
\end{lstlisting}

The energies are output as a two dimensional matrices for which the dimensions are \texttt{[time, space]}. Please refer to Table \ref{tbl:symbols} for a complete list of \df\ variable names and their corresponding symbols. 
For further details on the \texttt{dfana.my\_calculation} syntax used in this section see Section \ref{ssec:dfana}. 
\end{tcolorbox}

\subsubsection{Open circuit voltage}	
\label{sec:Open_circuit_voltage}
The open circuit voltage, \Voc\ is the maximum energy per unit charge that can be extracted from an electrochemical cell for a given charge state at open circuit. The \Voc\ can be calculated using the difference in the electron QFL at the location of the cathode ($x_\mathrm{cathode}$) and the hole QFL at the location of the anode ($x_\mathrm{anode}$) with the cell at open circuit. 

\begin{equation} \label{eq:Voc}
qV_{\mathrm{OC}}(t) = E_{\mathrm{Fn}}(x_\mathrm{cathode}, t) - E_{\mathrm{Fp}}(x_\mathrm{anode}, t) ,
\end{equation}

\begin{tcolorbox}
The open circuit voltage can be output using the command:

\begin{lstlisting}[numbers=none]
Voc = dfana.calcDeltaQFL(sol_OC)
\end{lstlisting}

\noindent Here, \texttt{sol\_OC} is an open circuit solution obtained either by applying \Vapp\ $=$ \Voc\ or approximated by setting the external series resistance, $R_S$ to a high value (e.g. $1$ M$\Omega$ cm$^2$) using the \texttt{lightonRs} protocol (see Section \ref{ssec:protocols_general} for further information on protocols).
\end{tcolorbox}

\subsection{Poisson's equation}								
\label{ssec:Poisson_eq}

Poisson's equation (Gauss's Law) relates the electrostatic potential to the space charge density $\rho$ and the material dielectric constant $\varepsilon_r$ via the Divergence Theorem. The space charge density is the sum of the mobile carrier and static charge densities at each spatial location. Doping is simulated via the inclusion of fixed charge density terms for ionising donor $N_{\mathrm{D}}$ and acceptor $N_{\mathrm{A}}$ atoms. In the default version of \df\ mobile ionic carriers are modelled as Schottky defects\cite{Walsh2015} for which every ion has an oppositely charged counterpart, maintaining overall ionic defect charge neutrality within the device.\footnote{For clarity only charge neutrality of the Schottky defect terms is guaranteed, this does not include the contribution from dopant atoms, which may not be compensated by electronic carriers in regions where an electric field is present.} The mobile cation density $c$ is initially balanced by a uniform static counter-ion density $N_\mathrm{cat}$ and the mobile anion density $a$ is similarly balanced by a static density $N_\mathrm{ani}$. For the one-dimensional system described, Poisson's equation can be explicitly stated as:

\begin{small}
\begin{multline} \label{eq:Poisson}
\dfrac{\partial^2 V(x,t)}{\partial x^2}= -\dfrac{\rho(x,t)}{\varepsilon_0\varepsilon_r(x)} \\ = -\dfrac{q}{\varepsilon_0\varepsilon_r(x)}(p(x,t) - n(x,t) + N_{\mathrm{D}}(x) - N_{\mathrm{A}}(x) +...\\ z_c c(x,t) + z_a a(x,t) - z_c N_\mathrm{cat}(x) - z_a N_\mathrm{ani}(x)),
\end{multline}
\end{small}

\noindent where $\varepsilon_0$ is permittivity of free space. We emphasise that $p$, $n$, $c$, and $a$ represent mobile species, while $N_{\mathrm{A}}$, $N_{\mathrm{D}}$, $N_\mathrm{cat}$ and  $N_\mathrm{ani}$ are static ion densities. $z_c$, and $z_a$ are the integer charge states for the ionic species (by default $z_c=1$, and $z_a=-1$). 

\begin{tcolorbox}
Terms can easily be added or removed from Poisson's equation by editing the \texttt{S\_potential} term in the Equation Editor in \texttt{dfpde} subfunction of the core \texttt{df} code. See Subsection \ref{ssec:df_master_function} and Listing \ref{lst:equation_editor} for further details.
\end{tcolorbox}

\begin{tcolorbox}
The space charge density \texttt{rho} can be output from a \df\ solution structure \texttt{sol} using the command:

\begin{lstlisting}[numbers=none]
rho = dfana.calcrho(sol)
\end{lstlisting}

\texttt{rho} is output as a two dimensional matrix for which the dimensions are \texttt{[time, space]}.
\end{tcolorbox}

\subsection{Charge transport: Drift and diffusion}
\label{ssec:transport_drift_diffusion}
As the name suggests, the drift-diffusion (Poisson-Nernst-Planck) model assumes that charge transport within semiconductors is driven by two processes:

\begin{enumerate}
\item \textit{Drift} arising from the Lorentz force on charges due to an electric field $F$, where $F = -dV/dx$.
\item \textit{Diffusion} arising from the entropic drive for carriers to move from regions of high to low concentration.
\end{enumerate}

\subsubsection{Bulk transport}

Within the bulk of material layers the expressions for the flux density of electrons $j_n$, holes $j_p$, anions $j_a$, and cations $j_c$ with mobility $\mu_{y}$ and diffusion coefficient $D_{y}$ (where $y$ denotes a generic charge carrier) are given by:

\begin{equation} \label{eq:flux_e}
j_n(x,t) = -\mu_n(x)n(x,t)F(x,t) - D_n(n,x) \dfrac{\partial n(x,t)}{\partial x},
\end{equation}

\begin{equation} \label{eq:flux_h}
j_p(x,t) = \mu_p(x)p(x,t)F(x,t) - D_p(p,x) \dfrac{\partial p(x,t)}{\partial x},
\end{equation}

\begin{equation} \label{eq:flux_c}
j_c(x,t) = \mu_c(x) z_c c(x,t)F(x,t) - D_c(c,x) \dfrac{\partial c(x,t)}{\partial x},
\end{equation}

\begin{equation} \label{eq:flux_a}
j_a(x,t) = \mu_a(x) z_a a(x,t)F(x,t) - D_a(a,x) \dfrac{\partial a(x,t)}{\partial x}.
\end{equation}

Figure \ref{fig:Fluxes} illustrates how the direction of electron and hole flux densities is determined from gradients in the electric potential and charge carrier densities. An analogous diagram can be drawn for mobile ionic species by substituting cations for holes and anions for electrons. The carrier (particle) currents are calculated as the product of the flux densities with the specific carrier charge $q z_y$ such that $J_y  = q z_y j_y$.

\begin{tcolorbox}

The electric field calculated from the gradient of the potential (\texttt{FV}) and by integrating the space-charge density\footnote{$-dV/dx$ is used to obtain the boundary values} (\texttt{Frho}) can be obtained from a \df\ solution structure \texttt{sol} using the syntax:

\begin{lstlisting}[numbers=none]
[FV, Frho] = dfana.calcF(sol)
\end{lstlisting}

\texttt{FV} and \texttt{Frho} are output as a two dimensional matrices for which the dimensions are \texttt{[time, space]}.
\end{tcolorbox}

\subsubsection{Diffusion enhancement}
\label{sssec:diffusion_enhancement}
The implementation of electronic carrier statistics beyond the Boltzmann approximation (Section \ref{sssec:prob_dist_fun}) necessitates the inclusion of a generalised Einstein relation to define the relationship between the carrier mobilities and diffusion coefficients as a function of band state occupancy.\cite{Farrell2017a} The result is a non-linear diffusion enhancement as the QFLs approach and move into the bands. Under Blakemore's approximation,\cite{Blakemore1982} the diffusion coefficient-mobility relationships for electrons and holes can be expressed using the closed-forms:

\begin{equation} \label{eq:Diff_enhance_n}
D_n(n, x) = \dfrac{k_\mathrm{B}T}{q}\mu_n(x)\left( \dfrac{N_\mathrm{CB}(x)}{N_\mathrm{CB}(x) - \gamma n(x,t)} \right),
\end{equation}

\begin{equation} \label{eq:Diff_enhance_p}
D_p(p, x) = \dfrac{k_\mathrm{B}T}{q}\mu_p(x)\left( \dfrac{N_\mathrm{VB}(x)}{N_\mathrm{VB}(x) - \gamma p(x,t)} \right).
\end{equation}

We use similar expressions to those above for the ionic carriers from a model proposed by Kilic et al.\cite{Kilic2007} to account for steric effects at high ion densities:

\begin{equation} \label{eq:Diff_enhance_c}
D_c(c, x) = \dfrac{k_\mathrm{B}T}{q}\mu_c(x)\left( \dfrac{c_{max}(x)}{c_{max}(x) - c(x,t)} \right),
\end{equation}

\begin{equation} \label{eq:Diff_enhance_a}
D_a(a, x) = \dfrac{k_\mathrm{B}T}{q}\mu_a(x)\left( \dfrac{a_{max}(x)}{a_{max}(x) - a(x,t)} \right).
\end{equation}

Here $a_{max}$ and $c_{max}$ denote the limiting anion and cation densities. In the first instance these are set to the lattice cite density for the corresponding ions.

\subsubsection{Transport across heterojunctions}
At the interface between two different semiconductor materials there is a change in the band energies and electronic density of states. In \df\ we choose to model the mixing of states at the interface using a smooth transition in material properties over a discrete interlayer region, in contrast to the commonly employed abrupt interface model\footnote{We note that while either model may be closer in one or more aspects to the real situation, both lack a comprehensive quantum mechanical treatment.} (see Figure \ref{fig:Interface_schematic} below for a schematic illustrating the difference between the two models). To accommodate this approach, Equations \ref{eq:flux_e} and \ref{eq:flux_h} are modified to include additional gradient terms for spatial changes in $\Phi_\mathrm{EA}$, $\Phi_\mathrm{IP}$, $N_{\mathrm{CB}}$, and $N_{\mathrm{VB}}$. This leads to an adapted set of flux equations for electrons and holes within the interfaces:\cite{Zeman2011} 

\begin{multline} \label{eq:flux_e_het}
j_n(x,t) = \mu_n(x,t) n\left(-F(x,t)-\dfrac{\partial \Phi_\mathrm{EA}(x)}{\partial x} \right) \\- 
D_n(n,x) \left(\dfrac{\partial n(x,t)}{\partial x}  -  \dfrac{n(x,t)}{N_{\mathrm{CB}}(x)}\dfrac{\partial N_{\mathrm{CB}}(x)}{\partial x}\right)
\end{multline}

\begin{multline} \label{eq:flux_h_het}
j_p(x,t) = \mu_p(x,t) p\left(F(x,t) + \dfrac{\partial \Phi_\mathrm{IP}(x)}{\partial x} \right) \\- 
D_p(p,x) \left(\dfrac{\partial p(x,t)}{\partial x}  -  \dfrac{p(x,t)}{N_{\mathrm{VB}}(x)}\dfrac{\partial N_{\mathrm{VB}}(x)}{\partial x}\right)
\end{multline}

Values of between $1 - 2$ nm have been extensively tested for the interfacial region thickness and are used in the example parameter files accompanying \df . By default, $\Phi_\mathrm{EA}$ and $\Phi_\mathrm{IP}$ are graded linearly, while $N_{\mathrm{CB}}$ and $N_{\mathrm{VB}}$ are graded exponentially within the interfacial regions. 

\begin{tcolorbox}
The transport equations of \df\ can be edited using the carrier flux terms \texttt{F\_n}, \texttt{F\_p}, \texttt{F\_c}, and \texttt{F\_a} of the Equation Editor in the \texttt{dfpde} subfunction of the core \texttt{df} code. See Subsection \ref{ssec:df_master_function} for further details.
\end{tcolorbox}

\subsubsection{Displacement current}
\label{ssec:displacement_current}
The displacement current $J_{disp}$, as established in the Maxwell-Ampere law, is the rate of change of the electric displacement field, $\partial D/ \partial t$. In terms of the electric field the displacement current can be expressed as:
\begin{equation} \label{eq:Jdisp}
J_{disp}(x,t) = \varepsilon_0 \varepsilon_r(x) \dfrac{\partial F(x,t)}{\partial t}.
\end{equation}

\subsubsection{Total current}
\label{ssec:total_current}
The total current, $J$ is the sum of the individual current components at each point in space and time:

\begin{equation} \label{eq:total}
J(x,t) = J_n(x,t) + J_p(x,t) + J_a(x,t) + J_c(x,t) + J_{disp}(x,t)
\end{equation}

\begin{tcolorbox}
Fluxes and currents are calculated from the \df\ solution structure \texttt{sol} using the command:

\begin{lstlisting}[numbers=none]
[J, j, xout] = dfana.calcJ(sol, "sub")
\end{lstlisting}

\noindent \texttt{J} is a structure containing the individual carrier particle currents \texttt{J.n}, \texttt{J.p}, \texttt{J.c}, and \texttt{J.a}, the displacement current \texttt{J.disp}, and the total current \texttt{J.tot} at each spatial location and time calculated by integrating the continuity equations. \texttt{j} is a structure containing the corresponding carrier and total fluxes. The second input argument \texttt{"sub"} indicates that the currents and fluxes are requested on the subinterval spatial mesh, which is output as \texttt{xout} (see Subsection \ref{ssec:spatial_mesh}).
\end{tcolorbox}

\subsubsection{Validity criteria for the drift-diffusion equations}
The drift-diffusion approach set out above is valid for semiconductor materials that satisfy the following criteria:\cite{Nelson2003} 

\begin{enumerate}
\item The electron and hole populations are at quasi-thermal equilibrium.
\item The electron and hole population temperatures are the same as that of the lattice.
\item Changes in state occupancy are more likely to be due to scattering collisions within a band than generation and recombination events between bands or trapping events.
\item The electron and hole states can be described by a quantum number, $k$.
\item The mean free path length of carriers, $\bar{L}$ is significantly shorter than the layer thickness, $d$ ($\bar{L} << d$).
\end{enumerate}

\subsection{Charge continuity}							
\label{ssec:continuity_equations}
The continuity equations are a set of `book-keeping' equations, based on the conservation of charge, describing how charge carrier densities change in time at each location. In one-dimension, the continuity equation for a generic carrier density $y$ with flux density $j_y$, and source/sink term $S_y$ can be expressed as:

\begin{equation} \label{eq:continuity_general}
\pdv{y(x,t)}{t} = -\dfrac{\partial j_y(x,t)}{\partial x} + S_y(x,t).
\end{equation}

\begin{figure}[h]
\centering
\includegraphics[width=0.48\textwidth]{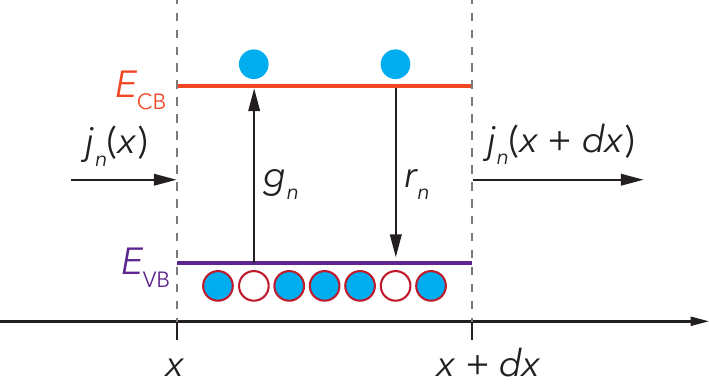}
\caption[Continuity of electrons in a one-dimensional semiconductor.]{\textbf{Continuity of electrons in a one-dimensional semiconductor.} Schematic illustrating the principle of continuity for electrons in a thin slab of material $dx$. A difference in the incoming and outgoing flux density $j_n$, generation $g_n$, and recombination $r_n$ of electrons results in changes in the electron concentration over $dx$ (Equation \ref{eq:cont_n}). The conduction and valence band energies are denoted $E_{\mathrm{CB}}$ and $E_{\mathrm{VB}}$ respectively. Electrons are represented by solid blue circles and holes by open red circles. Figure concept adapted from Ref.\cite{Zeghbroeck2011}.}
\label{fig:Continuity}
\end{figure}

For electronic carriers $S$ is composed of two components; 1. Generation, $g$ of carriers by both thermal and photo excitation and; 2. Recombination, $r$ of carriers through radiative (photon emission) and non-radiative pathways. Figure \ref{fig:Continuity} illustrates the principle of continuity: changes in the electron concentration with time within a thin slab $dx$ are determined by the generation, recombination, and difference between the incoming and outgoing flux density of carriers.

Where chemical reactions take place within devices, additional generation and recombination terms for carriers may also contribute to $S$. In the current version of \df , mobile ionic charge carriers are treated as inert such that $g_{c} = g_{a} = r_{c} = r_{a} = 0$. Users are, however, free to edit the default source terms using the Equation Editor (Section \ref{ssec:df_master_function}). A guide describing how to do this is included in the \SI\ Section \ref{sec:edit_equations}.

In one-dimension the continuity equations for electrons, holes, cations and anions are given by:

\begin{equation} \label{eq:cont_n}
\dfrac{\partial n(x,t)}{\partial t} = -\dfrac{\partial j_n(x,t)}{\partial x} + g_n(x,t) - r_n(x,t),
\end{equation}

\begin{equation} \label{eq:cont_p}
\dfrac{\partial p(x,t)}{\partial t} = -\dfrac{\partial j_p(x,t)}{\partial x} + g_p(x,t) - r_p(x,t),
\end{equation}

\begin{equation} \label{eq:cont_c}
\dfrac{\partial c(x,t)}{\partial t} = -\dfrac{\partial j_c(x,t)}{\partial x(x,t)} + g_c(x,t) - r_c(x,t),
\end{equation}

\begin{equation} \label{eq:cont_a}
\dfrac{\partial a(x,t)}{\partial t} = -\dfrac{\partial j_a(x,t)}{\partial x} + g_a(x,t) - r_a(x,t).
\end{equation}

Equation \ref{eq:Poisson} and Equations \ref{eq:cont_n} - \ref{eq:cont_a} then form the complete set of equations to be solved.

\subsubsection{Steady-state approximation to electronic carrier densities and fluxes within the interfacial regions}
\label{ssec:interface_carriers_MT}
To better understand the discrete interface model employed by \df\ we solve the electron and hole continuity equations (Equations \ref{eq:flux_e_het}, \ref{eq:flux_h_het}, \ref{eq:cont_n} and \ref{eq:cont_p}) to obtain analytical expressions for the electronic carrier densities within the discrete interfacial regions using the following approximations and assumptions:
\begin{enumerate}
\item Carriers within an interface are at steady-state with respect to the surroundings layers ($dn/dt = 0$, $dp/dt = 0$).
\item There is no optical generation within the interface ($g = 0$).
\item The electric field can be treated as approximately constant throughout the interfacial region ($dF/dx = 0$).
\item The recombination rate, $r$ within the interfacial region is constant and distributed uniformly.
\item QFLs remain within the Boltzmann regime ($E_\mathrm{CB} - E_\mathrm{Fn} > 3 k_\mathrm{B} T$ and $E_\mathrm{Fp} - E_\mathrm{VB} > 3k_\mathrm{B} T$)
\end{enumerate}

As detailed in the \SI\ Section \ref{sec:ss_carriers_interfaces}, using the boundary conditions $n(x_n = 0) = n_\mathrm{s}$, $p(x_p = 0) = p_\mathrm{s}$, $j_n(x_n = 0) = j_{n,s}$, and $j_p(x_p = 0) = j_{p,s}$ (see Figure \ref{fig:Interface_schematic}b), the following expressions can be obtained for the carrier densities within the interfacial regions:

\begin{figure}[h]
\centering
\includegraphics[width=0.49\textwidth]{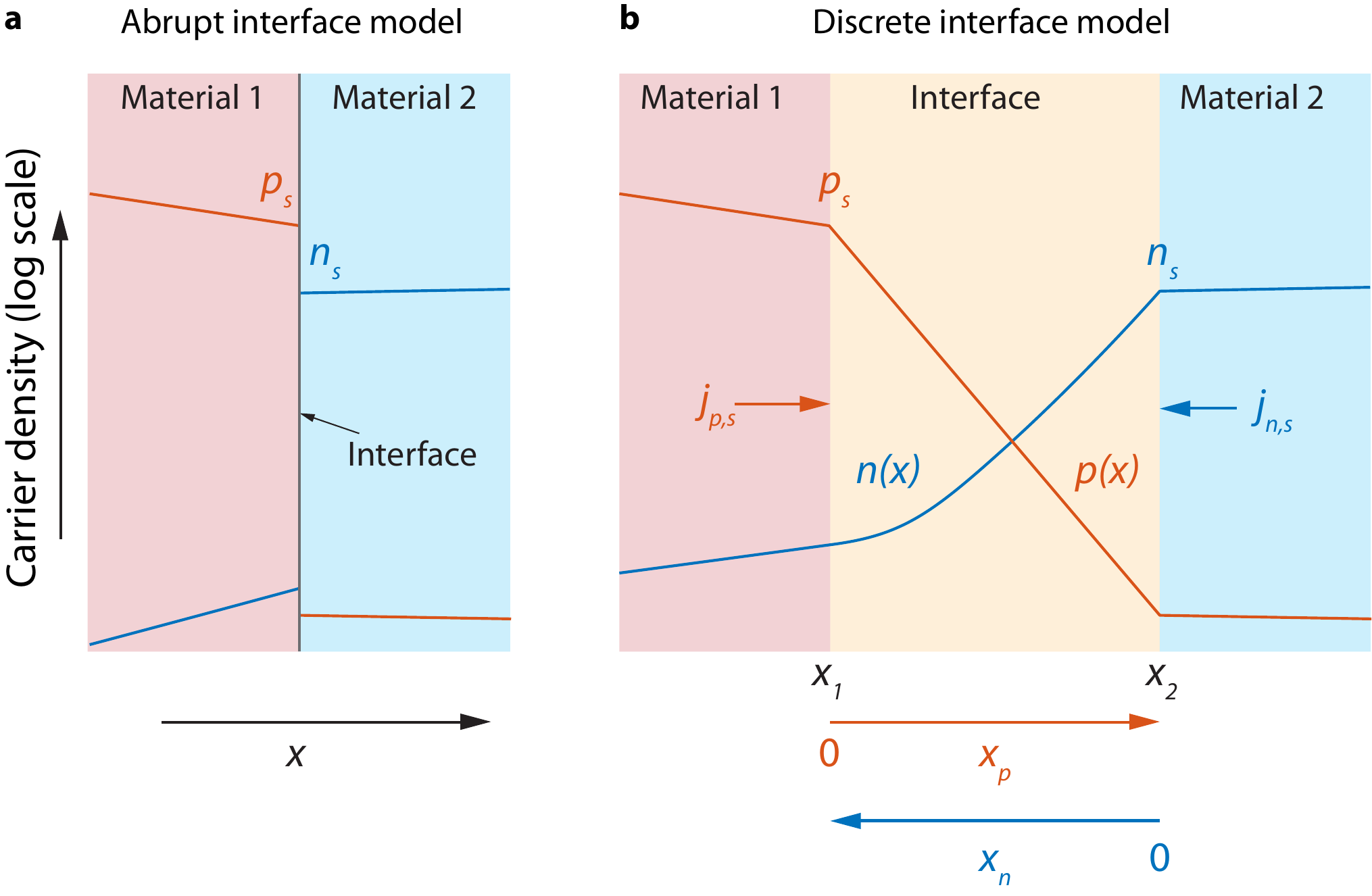}
\caption[Schematic of carrier densities at (a) abrupt and (b) discrete interface models.]{\textbf{Schematic of carrier densities at (a) abrupt and (b) discrete interface models.} $n_s$ and $p_s$ are the boundary electron and hole densities, while $j_{n,\mathrm{s}}$ and $j_{p,\mathrm{s}} $ are the boundary fluxes. The pure exponential change in hole density, $p(x)$ across the interfacial region (\textbf{b}) implies the hole mobility is large such that the $j_{p,\mathrm{s}}$ and $r$ terms in Equation \ref{eq:px_interface} are negligible. By contrast, the curvature in the logarithm of the electron density, $n(x)$ profile indicates that the $j_{n,\mathrm{s}}$ and $r$ terms in Equation \ref{eq:nx_interface} are of a similar order  to the $n_s$ term close to $x_1$. The red, yellow and blue regions indicate Material 1 (p-type), Interface, and Material 2 (n-type) layers respectively. $x_n$ and $x_p$ are the translated position ($x$) co-ordinates.}
\label{fig:Interface_schematic}
\end{figure}

\begin{multline} \label{eq:nx_interface}
n(x_n) = n_\mathrm{s} e^{\alpha x_n}  + \dfrac{j_{n,s}}{k_B T \alpha \mu_n}(1-e^{\alpha x_n})\\
 - \dfrac{r}{k_B T \alpha^2 \mu_n}(1 -e^{\alpha x_n} + \alpha x_n),
\end{multline}

\begin{multline} \label{eq:px_interface}
p(x_p) = p_\mathrm{s} e^{\beta x_p}  + \dfrac{j_{p,s}}{k_B T \beta \mu_p}(1- e^{\beta x_p})\\
 - \dfrac{r}{k_B T \beta^2 \mu_p}(1 -e^{\beta x_p} + \beta x_p),
\end{multline}
\\
\noindent where,

\begin{equation} \label{eq:alpha_interface}
\alpha = -\frac{1}{k_\mathrm{B}T} \left( \dfrac{\partial \Phi_\mathrm{EA}(x_n)}{\partial x_n} - q\dfrac{\partial V}{\partial x_n} \right) + \dfrac{1}{N_{\mathrm{CB}}(x_n)}\dfrac{\partial N_{\mathrm{CB}}(x_n)}{\partial x_n},
\end{equation}

\begin{equation} \label{eq:beta_interface}
\beta = \frac{1}{k_\mathrm{B}T} \left( \dfrac{\partial \Phi_\mathrm{IP}(x_p)}{\partial x_p} - q\dfrac{\partial V}{\partial x_p} \right) + \dfrac{1}{N_{\mathrm{VB}}(x_p)}\dfrac{\partial N_{\mathrm{VB}}(x_p)}{\partial x_p}.
\end{equation}

The corresponding fluxes are given by:

\begin{equation} \label{eq:n_general_sol_jns}
j_{n}(x_n) = j_{n,\mathrm{s}} - rx_n,
\end{equation}
\begin{equation} \label{eq:p_general_sol_jps}
j_{p}(x_p) = j_{p,\mathrm{s}} - rx_p.
\end{equation}

As illustrated in Figure \ref{fig:Interface_schematic}b, the translated co-ordinates $x_n$ and $x_p$ are taken to be in the direction for which $\alpha$ and $\beta$ are negative and typically the direction for which carrier densities decay.

Example solutions comparing the analytical approximations to numerical solutions calculated using \df\ under different transport and recombination regimes are given in the \SI , Section \ref{sec:interface_ana_validation}. Where transport is a limiting factor within the interfaces the solutions become strongly dependent on the boundary flux and recombination rates. It is noteworthy however that in the limiting case of infinitely fast transport ($\mu_{n,p} \to \infty$) Equations \ref{eq:nx_interface} and \ref{eq:px_interface} converge towards purely exponential forms for which the carrier densities change by a Boltzmann factor ($\Delta n = N_\mathrm{CB} e^{\alpha d_\mathrm{int}}$ and $\Delta p = N_\mathrm{VB} e^{\beta d_\mathrm{int}}$) across the width of the interface, $d_\mathrm{int}$. For the special case where $F=0$, the result is a change in carrier densities equivalent to that expected from an abrupt interface model using Boltzmann statistics. The results presented in this section are applied below in Section \ref{ssec:recombination_interfaces} to the interfacial volumetric surface recombination model.
%
%

\subsection{Electronic carrier generation}			
\label{ssec:generation}
Two optical models for electronic carrier generation are currently available for use in \df ; uniform generation for which a uniform volumetric generation rate, $g_0$ is defined for each layer (excluding interfacial regions) and; Beer-Lambert law generation as described below. Irrespective of the choice of optical model the generation rate is zeroed within the interfacial regions to avoid potential stability issues.

\subsubsection{Beer-Lambert law generation}
\label{sssec:beer_lambert}
The Beer-Lambert law models the photon flux density as falling exponentially within a material with a characteristic photon energy-dependent absorption coefficient $\alpha _{\mathrm{abs}}$. The volumetric generation rate $g$, over a range of photon energies $E_{\gamma}$ with incident photon flux density $\varphi _0$, is given by the integral across the spectrum:

\begin{multline} \label{eq:Beer_Lam}
g(x) = \\
(1-\kappa) \int _0 ^\infty \alpha _{\mathrm{abs}}(E_{\gamma},x) \varphi_0 (E_{\gamma}) \exp(-\alpha _{\mathrm{abs}}(E_{\gamma},x) x) \ dE_{\gamma},
\end{multline}

\noindent where $\kappa$ is the reflectance. For simplicity, we assume that a single electron-hole pair is generated by a single photon.

\subsubsection{Arbitrary generation profiles}
\label{sssec:arbitrary_generation}
An arbitrary generation profile can be inserted following creation of the parameters object for users who wish to use profiles calculated from different models using an external software package. Details on how to do this are given in Section \ref{sssec:generation_function}.

\subsection{Recombination}			
\label{ssec:recombination}
\begin{figure}[h]
\centering
\includegraphics[width=0.48\textwidth]{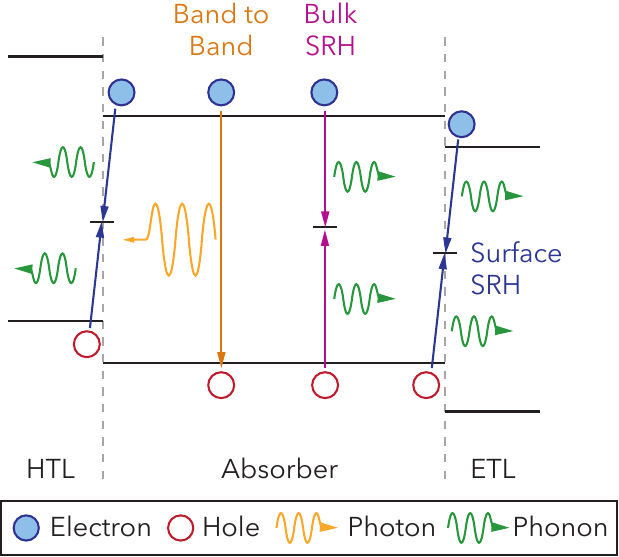}
\caption[Schematic of different recombination mechanisms in a hole transport layer (HTL)-Absorber-electron transport layer (ETL) device.]{\textbf{Schematic of different recombination mechanisms in a hole transport layer (HTL)-Absorber-electron transport layer (ETL) device.} Figure adapted from Ref \cite{Calado2019_ideality}.}
\label{fig:Rec_mech2}
\end{figure}

By default, two established models for recombination are included in \df : band-to-band recombination and trap-mediated Shockley-Read-Hall (SRH) recombination. Figure \ref{fig:Rec_mech2} is a simplified energy level schematic illustrating these mechanisms. The recombination expressions can be modified in the \texttt{source} terms of the Equation Editor (Section \ref{ssec:df_master_function}).

\subsubsection{Band-to-band recombination}
\label{sssec:rec_btb}
The rate of band-to-band recombination $r_{\mathrm{btb}}$ (also commonly termed direct, radiative or bimolecular recombination) is proportional to the product of the electron and hole densities at a given location such that:

\begin{equation} \label{eq:Band2band}
r_{\mathrm{btb}}(x,t) = B(x)(n(x,t)p(x,t) - n_{\mathrm{i}}(x)^2),
\end{equation}
\\
where $B$ is the band-to-band recombination rate coefficient. The $n_{\mathrm{i}}^2$ term is equivalent to including an expression for thermal generation and ensures that $np \geqslant n_i^2$ at steady-state.

\subsubsection{Shockley-Read-Hall (SRH) recombination}
\label{sssec:rec_srh}
Recombination via trap states is modelled using a simplified Shockley-Read-Hall (SRH) recombination\cite{Shockley1952} expression $r_{\mathrm{SRH}}$ for which the capture cross section, mean thermal velocity of carriers, and trap density are collected into SRH time constants, $\tau_{n,\mathrm{SRH}}$ and $\tau_{p,\mathrm{SRH}}$ for electrons and holes respectively:

\begin{multline}
\label{eq:SRH}
r_{\mathrm{SRH}}(x,t) = \\
\dfrac{n(x,t)p(x,t) - n_{\mathrm{i}}(x)^2}{\tau_{n,\mathrm{SRH}}(x)(p(x,t)+p_{\mathrm{t}}(x)) + \tau_{p,\mathrm{SRH}}(x)(n(x,t)+n_{\mathrm{t}}(x))}.
\end{multline}

Here, $n_{\mathrm{t}}$ and $p_{\mathrm{t}}$ are parameters that define the dependence of the recombination rate to the trap level and are given by the electron and hole densities when their respective QFLs are at the position of the trap energy, $E_{\mathrm{t}}$:

\begin{equation} \label{eq:nt}
n_{\mathrm{t}} = n_{\mathrm{i}}	\exp\left({\dfrac{E_{\mathrm{i}}- E_{\mathrm{t}}}{k_\mathrm{B}T}}\right),
\end{equation}
\begin{equation} \label{eq:pt}
p_{\mathrm{t}} = n_{\mathrm{i}}	\exp\left({\dfrac{E_{\mathrm{t}}- E_{\mathrm{i}}}{k_\mathrm{B}T}}\right).
\end{equation}

It should be noted that Equation \ref{eq:SRH} is valid only when trapped carriers are in thermal equilibrium with those in the bands. It follows that the rate of trapping and de-trapping of carriers is assumed to be fast compared to the timescale being simulated, such that the approximation is reasonable. In the current version of \df\ we also assume that the quantity of trapped carriers is negligible compared to that of the free carriers such that trapped carriers can be neglected in Poisson's equation. To more completely simulate the dynamics of capture and emission of carriers, and the associated contribution to the chemical capacitance of devices, one or more additional variables could be included.

\begin{tcolorbox}
The volumetric recombination rate can be obtained from a \df\ solution structure using the command:
\begin{lstlisting}[numbers=none]
r = dfana.calcr(sol);
\end{lstlisting}

\noindent \texttt{r} is a structure containing three matrices \texttt{r.btb}, \texttt{r.srh}, \texttt{r.vsr}, and \texttt{r.tot} in which the band-to-band, SRH, volumetric surface recombination (VSR, see below) and the total recombination rates are stored as two dimensional matrices with dimensions \texttt{[time, space]}.
\end{tcolorbox}

\begin{tcolorbox}
The recombination models used in the simulation can be edited using the carrier source terms \texttt{S\_electron},  \texttt{S\_hole}, \texttt{S\_cation}, and \texttt{S\_anion} in the Equation Editor in \texttt{dfpde} subfunction of the core \texttt{df} code. Note that the models used in \texttt{dfana} must also be updated in accordance with any changes to \texttt{dfpde} as \textit{these functions are not coupled}. See Subsection \ref{ssec:df_master_function} for further details.
\end{tcolorbox}

\subsubsection{Surface recombination at interfaces}			
\label{ssec:recombination_interfaces}

Abrupt interface models typically use a SRH surface recombination model to determine the recombination flux, $R_\mathrm{int}$ between majority carriers $n_\mathrm{s}$ and $p_\mathrm{s}$ at the interface between two materials (see Figure \ref{fig:Interface_schematic}a):\cite{Courtier2019a}

\begin{equation} \label{eq:SRH_2D}
R_\mathrm{int}(t) = \dfrac{n_\mathrm{s}(t)p_\mathrm{s}(t) - n_{\mathrm{i}}^2}{\frac{1}{s_n} (p_\mathrm{s}(t)+p_{\mathrm{t}}) + \frac{1}{s_p} (n_\mathrm{s}(t)+n_{\mathrm{t}})}.
\end{equation}

%

Here, $s_n$ and $s_p$ are the surface recombination velocities for electrons and holes at the interface. This model implies that the electron and hole populations in the two materials have delocalised wave functions that overlap significantly such that recombination events are probable.

Since \df\ uses discrete interfacial regions, in order to obtain an equivalent recombination flux to the abrupt interface model, we convert Equation \ref{eq:SRH_2D} into a volumetric surface recombination rate, $r_{\mathrm{vsr}}$ by distributing the recombination uniformly across a zone of thickness, $d_\mathrm{vsr}$ within the interface (see Figure \ref{fig:rec_zone}), such that $r_{\mathrm{vsr}} = R_\mathrm{int}/d_\mathrm{vsr}$. By default the recombination zone is automatically located next to the interface with the highest minority carrier density at equilibrium. To obtain an expression for $r_{\mathrm{vsr}}$, Equations \ref{eq:nx_interface} and \ref{eq:px_interface} can be rearranged to express $n_\mathrm{s}$ and $p_\mathrm{s}$ in terms of $n(x_n)$ and $p(x_p)$:

\begin{small}
\begin{multline} \label{eq:ns_interface}
n_\mathrm{s} = e^{-\alpha x_n} \left(n(x_n) - \dfrac{j_{n,s}}{k_B T \alpha \mu_n}(1-e^{\alpha x_n}) +... \right. \\
\left. \dfrac{r}{k_B T \alpha^2 \mu_n}(1 -e^{\alpha x_n} + \alpha x_n) \right),
\end{multline}
\end{small}
\begin{small}
\begin{multline} \label{eq:ps_interface}
p_\mathrm{s} = e^{-\beta x_p} \left(p(x_p) - \dfrac{j_{p,s}}{k_B T \beta \mu_p}(1- e^{\beta x_p}) +... \right. \\
\left. \dfrac{r}{k_B T \beta^2 \mu_p}(1 -e^{\beta x_p} + \beta x_p) \right).
\end{multline}
\end{small}

For sufficiently high values of $\mu_n$, and $\mu_p$ the $n(x_n)$ and $p(x_p)$ terms dominate Equations \ref{eq:ns_interface} and \ref{eq:ps_interface} and the carrier density profiles within the interfaces tend towards purely exponential functions, such that $n_\mathrm{s} \approx n(x_n) e^{-\alpha x_n} $ and $p_\mathrm{s} \approx p(x_p) e^{-\beta x_p}$. In many instances it is then sufficient to approximate the volumetric surface recombination rate within the recombination zone as:

\begin{small}
\begin{multline}
\label{eq:SRH_vsr}
r_{\mathrm{vsr}}(x,t) = \\
\dfrac{n(x,t)e^{-\alpha x_n} p(x,t)e^{-\beta x_p} - n_{\mathrm{i}}^2}{\tau_{n,\mathrm{vsr}}(p(x,t)e^{-\beta x_p}+p_{\mathrm{t}}) + \tau_{p,\mathrm{vsr}}(n(x,t)e^{-\alpha x_n}+n_{\mathrm{t}})},
\end{multline}
\end{small}


\noindent where $\alpha$ and $\beta$ are given in Equations \ref{eq:alpha_interface} and \ref{eq:beta_interface}, and the $d_\mathrm{vsr}$ term is subsumed into the volumetric surface recombination time constants, $\tau_{n,\mathrm{vsr}}$ and $\tau_{p,\mathrm{vsr}}$ such that:

\begin{equation} \label{eq:taun_int}
\tau_{n,\mathrm{vsr}} = \frac{d_\mathrm{vsr}}{s_n}
\end{equation}
\begin{equation} \label{eq:taup_int}
\tau_{p,\mathrm{vsr}} = \frac{d_\mathrm{vsr}}{s_p}
\end{equation}

We stress here that this is not a physically motivated model in the sense that we do not anticipate recombination to be distributed uniformly throughout a recombination zone in reality. This approach does however result in a good approximation to the established abrupt interface surface recombination model for a wide variety of devices and conditions (see Section \ref{ssec:comparison_ionmonger}).

\begin{tcolorbox}
The volumetric surface recombination model can be toggled on and off by using the property \texttt{par.vsr\_mode}. Where \texttt{par.vsr\_mode = 1}, $E_\mathrm{t}$ and the $n_\mathrm{i}$, $n_\mathrm{t}$ and $p_\mathrm{t}$ terms are set to constant and calculated from the energy levels defined for the interfacial region. This ensures that $r_{\mathrm{vsr}}$ remains approximately constant throughout the recombination zone. Where \texttt{par.vsr\_mode = 0}, the standard SRH expression in Equation \ref{eq:SRH} is assumed. In this case $E_t$ is graded linearly and $n_\mathrm{i}$, $n_\mathrm{t}$ and $p_\mathrm{t}$ are graded exponentially.
\end{tcolorbox}

\begin{tcolorbox}
The assumptions used in the derivation of Equation \ref{eq:SRH_vsr} breakdown when the transport within the interface is limited or where recombination fluxes are particularly high (see Section \ref{fig:interface_solutions}). Since both the flux and recombination terms in Equations \ref{eq:ns_interface} and \ref{eq:ps_interface} are unknowns without well-defined limits, \df\ performs a check for self-consistency directly following calculation of the solution when VSR mode is switched on: the function \texttt{compare\_rec\_flux} calculates the sum of the interfacial recombination fluxes using the values of $n_s$ and $p_s$ from the solution and the SRH model given in Equation \ref{eq:SRH_2D}. This sum is compared to that of the integrated recombination rate calculated using the VSR model (Equation \ref{eq:SRH_vsr}) across all interfaces. If the fractional difference in the two calculations is greater than \texttt{par.RelTol\_vsr}, for fluxes above \texttt{par.AbsTol\_vsr}, a warning is displayed. In such cases users could consider increasing the electronic carrier mobilities within the interfacial regions or reducing the recombination coefficients.
\end{tcolorbox}

\subsection{Initial conditions}					
\label{ssec:initial_condish}
At present two sets of initial conditions are used in \df , dependent on the number of layers. These conditions are designed to be consistent with the boundary conditions and to minimise the error in the space charge density at junctions which can lead to large electric fields and convergence failure.

\subsubsection{Single layer device}
\label{sssec:ic_single_layer}
A linearly varying electrostatic potential and exponentially varying electronic carrier densities over the layer thickness $d$ are used as the initial conditions (Equations \ref{eq:IC_n_single_layer}, \ref{eq:IC_p_single_layer}, and \ref{eq:IC_V_single_layer}) when simulating a single layer. Uniform ionic carrier density profiles are used throughout the layer to guarantee ionic defect charge neutrality (Equations \ref{eq:IC_a_single_layer} and \ref{eq:IC_c_single_layer}).

\begin{equation} \label{eq:IC_V_single_layer}
V(x) = \dfrac{x}{d}V_{bi},
\end{equation}
\begin{equation} \label{eq:IC_n_single_layer}
n(x) = n_{0,l} \exp \left( \ln \left( \dfrac{n_{0,r}}{n_{0,l}} \right) \dfrac{x}{d} \right),
\end{equation}
\begin{equation} \label{eq:IC_p_single_layer}
p(x) = p_{0,l} \exp \left( \ln \left( \dfrac{p_{0,r}}{p_{0,l}} \right) \dfrac{x}{d} \right),
\end{equation}
\begin{equation} \label{eq:IC_c_single_layer}
c(x) = N_\mathrm{cat}(x),
\end{equation}
\begin{equation} \label{eq:IC_a_single_layer}
a(x) = N_\mathrm{ani}(x).
\end{equation}

Here, the built-in potential $V_{\mathrm{bi}}$ of the device is determined by the difference in boundary electrode workfunctions $\Phi_{l}$ and $\Phi_{r}$:

\begin{equation} \label{eq:Vbi}
qV_{\mathrm{bi}} = \Phi_{r} - \Phi_{l}
\end{equation}

\subsubsection{Multilayer device}
\label{sssec:ic_multi_layer}
For multilayer devices the electrostatic potential is set to fall uniformly throughout the device (Equation \ref{eq:IC_V}), while the electronic carrier densities are chosen to be the equilibrium densities for the individual layers ($n_0$ and $p_0$). As with the single layers, the ionic carriers are given a uniform density (Equations \ref{eq:IC_n} - \ref{eq:IC_c}), thus guaranteeing local electro-neutrality. 

\begin{equation} \label{eq:IC_V}
V(x) = \dfrac{x}{d_\mathrm{dev}}V_{bi},
\end{equation}
\begin{equation} \label{eq:IC_n}
n(x) = n_0(x),
\end{equation}
\begin{equation} \label{eq:IC_p}
p(x) = p_0(x),
\end{equation}
\begin{equation} \label{eq:IC_c}
c(x) = N_\mathrm{cat}(x),
\end{equation}
\begin{equation} \label{eq:IC_a}
a(x) = N_\mathrm{ani}(x).
\end{equation}
\\
Here, the device thickness $d_\mathrm{dev}$ is the sum of the individual layer thicknesses $d_i$ ($d_\mathrm{dev} = \sum_{i} d_i$). \df\ auto-detects the number of layers in the device and uses the appropriate set of initial conditions when running the \texttt{equilibrate} protocol to obtain the equilibrium solutions for the device (Section \ref{ssec:protocols_equilibrate}).

\begin{tcolorbox}
The initial conditions of the simulation can be edited in the \texttt{dfic} subfunction of the core \texttt{df} code. See Subsection \ref{ssec:df_master_function} for further details.
\end{tcolorbox}

\subsection{Boundary conditions}
\label{ssec:boundary_conditions}
Solving Equation \ref{eq:Poisson} and Equations \ref{eq:cont_n} - \ref{eq:cont_a} requires two constants of integration for each variable, which are provided by the system boundary conditions. For the charge carriers, Neumann (defined-flux value) conditions are used to set the flux density into and out of the system. The electrostatic potential uses Dirichlet conditions (defined-variable value) such that the potential is fixed at both boundaries at each point in time as detailed in Section \ref{sec:builtin}. The details of these boundary conditions are discussed in the following subsections.

\subsubsection{Electrostatic potential boundary conditions}	 \label{sec:builtin}
\label{ssec:potential_carrier_BCs}
In \df\ the electrostatic potential at the left-hand boundary is set to zero (Equation \ref{eq:Vx0}) and used as the reference potential. The applied electrical bias, \Vapp\ and an effective potential arising from series resistance $V_{Rs}$ are applied to the right-hand boundary as described in Equation \ref{eq:Vxd}.

\begin{equation} \label{eq:Vx0}
V_l(t)=0
\end{equation}
\begin{equation} \label{eq:Vxd}
V_r(t)=V_{\mathrm{bi}}-V_{\mathrm{app}}(t)+V_{\mathrm{Rs}}(t)
\end{equation}
\\
Here, Ohm's law is used to calculate $V_{Rs}$ from the electron and hole flux densities:

\begin{equation} \label{eq:VRs}
V_{Rs}(t)= q(j_{p,r}(t)	- j_{n,r}(t))R_s,
\end{equation}
\\
where $R_s$ is the area-normalised series resistance, given by the product of the external series resistance and the device active area. Setting $R_s$ to a high value (e.g. $R_s = 10^6$ $\mathrm{\Omega}$ cm$^2$) approximates an open circuit condition for devices with metal electrodes. Technically this can be achieved using the \texttt{lighton$\_$Rs} protocol (see Section \ref{ssec:protocols_general} for a description of protocols).\footnote{Note that at the time of writing this method is not stable for all input parameter sets.}

\begin{tcolorbox}
The boundary conditions of the simulation can be edited in the \texttt{dfbc} subfunction of the core \texttt{df} code. See Subsection \ref{ssec:df_master_function} for further details.
\end{tcolorbox}

\begin{tcolorbox}
The function generator \texttt{fun\_gen} defines the applied potential \texttt{Vapp} as a function of time \texttt{t}, which can be recalculated from a \df\ solution structure \texttt{sol} using the command:

\begin{lstlisting}[numbers=none]
Vapp = dfana.calcVapp(sol)
\end{lstlisting}
\end{tcolorbox}

\subsubsection{Carrier selectivity and surface recombination at the system boundaries}				\label{ssec:electronic_carrier_BCs}
Many architectures of semiconductor device, including solar cells and LEDs employ selective contact layers that block minority carriers from being extracted (or injected) via energetic barriers. These are known variously as transport layers, blocking layers, blocking contacts, or selective contacts. For solar cells semiconductor layers are typically sandwiched between two metallic electrodes constituted of metals or highly-doped semiconductors. Such materials can be numerically challenging to simulate owing to their high charge carrier densities and thin depletion widths. Consequently, a common approach is to use boundary conditions defining charge carrier extraction and recombination flux densities to simulate the properties of either the contact or electrode material. It should be noted, however, that the employment of fixed electrostatic potential boundary conditions (as defined in Section \ref{sec:builtin}) implies that the potential falls only \textit{within the discrete system} and not within the electrodes. This approximation is only realistic for contact materials with vanishingly small depletion widths (infinite interfacial capacitances) i.e. metals and highly doped semiconductors. It follows that to accurately simulate semiconductor contacts layers with finite depletion regions these layers must also be included within the discretised system.

For electronic carriers the surface recombination velocity coefficients $s_n$ and $s_p$ determine the carrier extraction/recombination rate at the boundaries of the system. For majority carriers in solar cells, high values of $s_n$ and $s_p$ (e.g. $>10^7$ cm s$^{-1}$\cite{Tress2011}) are advantageous for carrier extraction, while low values imply poor contact extraction properties. For minority carriers, high values of $s_n$ and $s_p$ are typically undesirable as they imply high rates of surface recombination at the electrode. In \df\ the expressions for electronic boundary carrier flux densities, $j_n$ and $j_p$ are given by the typical first-order expressions:

\begin{equation} \label{eq:jn_l}
j_{n,l} (t) = s_{n,l}(n_l(t) - n_{0, l}),
\end{equation}
\begin{equation} \label{eq:jp_l}
j_{p,l} (t) = s_{p,l}(p_l(t) - p_{0, l}),
\end{equation}
\begin{equation} \label{eq:jn_r}
j_{n,r} (t) = s_{n,r}(n_r(t) - n_{0, r}),
\end{equation}
\begin{equation} \label{eq:jp_r}
j_{p,r} (t) = s_{p,r}(p_r(t) - p_{0, r}),
\end{equation}
\\
\noindent where $n_{0, l}$, $n_{0, r}$, $p_{0, l}$, and $p_{0, r}$ are the equilibrium carrier densities at the left ($x=0$) and right-hand ($x=d$) boundaries, calculated using Equations \ref{eq:n0} and \ref{eq:p0} under the assumption that the semiconductor QFLs are at the same energy as the electrode Fermi energy, which is assumed to remain constant. For the left-hand boundary $n_{0, l}$ and $p_{0, l}$ are given by Equations \ref{eq:n_left_right} and \ref{eq:p_left_right}.

\begin{equation} \label{eq:n_left_right}
n_{0, l} = N_{\mathrm{CB}} \left(e^{\frac{\Phi_l - \Phi_\mathrm{EA}}{k_\mathrm{B}T}} + \gamma \right)^{-1},
\end{equation}

\begin{equation} \label{eq:p_left_right}
p_{0, l} = N_{\mathrm{VB}} \left( e^{\frac{\Phi_\mathrm{IP} - \Phi_l}{k_\mathrm{B}T}} + \gamma \right)^{-1},
\end{equation}

\noindent where $\Phi_{l}$ is the left-hand electrode work function. Analogous expressions are used for $n_{r}$ and $p_{r}$ at the right-hand boundary. Extraction barriers can also be modelled with this approach by including a term for the barrier energy in the exponent of Equations \ref{eq:n_left_right} and \ref{eq:p_left_right}. At present, however, quantum mechanical tunnelling and image charge density models for energetic barriers at the system boundaries are not accounted for in \df .

%
%

\subsubsection{Ionic carrier boundary conditions}
\label{ssec:ionic_carrier_BCs}
In the simplest case, ionic carriers are confined to the device and do not react at the electrode boundaries. This leads to a set of zero flux density boundary conditions for mobile anions and cations:
\begin{equation} \label{eq:jc_BCs_l}
j_{c,l}(t) = 0	,
\end{equation}
\begin{equation} \label{eq:ja_BCs_l}
j_{a,l}(t)  = 0,
\end{equation}
\begin{equation} \label{eq:jc_BCs_r}
j_{c,r}(t)  = 0,
\end{equation}
\begin{equation} \label{eq:ja_BCs_r}
j_{a,r}(t)  = 0.
\end{equation}

Where an infinite reservoir of ions exists at a system boundary (such as an electrolyte), a Dirichlet boundary condition defining a constant ion density could alternatively be imposed.

%
%
This concludes our description of the physical models employed in \df . In the following section the system architecture and key commands are introduced as well as a guide on how to get started with using \df .

\section{System architecture and how to use \df }
\label{sec:system_architecture}
\begin{figure}[h]
\centering
\includegraphics[width=0.48\textwidth]{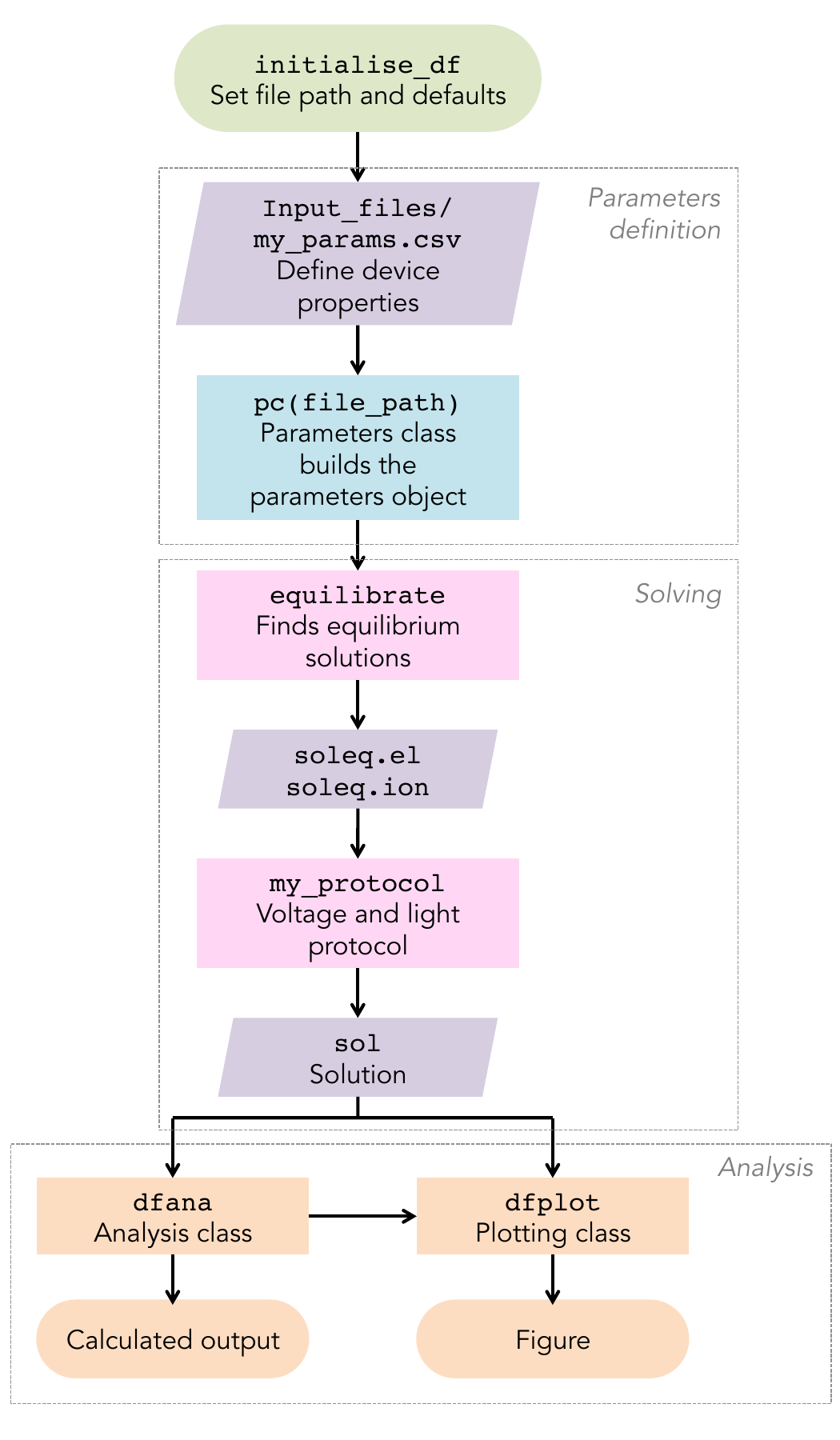}
\caption[Flow diagram showing the key steps to obtain a solution.]{\textbf{Flow diagram showing the key steps to obtain a solution.} A key for the classification of the box shapes is given in Figure \ref{fig:parameters_flow_diagram}.}	
\label{fig:system_architecture}
\end{figure}

\df\ is designed such that the user performs a linear sequence of simple procedures to obtain a solution. The key steps are summarised in Figure \ref{fig:system_architecture}; Following initialisation of the system, the user defines a device by creating a parameters object containing all the individual layer and device-wide properties; The equilibrium solutions (\texttt{soleq.el} and \texttt{soleq.ion}) for the device are then obtained before applying a voltage and light protocol, which may involve intermediate solutions; Once a desired solution (\texttt{sol}) has been obtained, analysis and plotting functions can be called to calculate outputs and visualise the solutions. Below, the principal functions are discussed in further detail.

 
\subsection{Initialising the system: \texttt{initialise$\_$df}}
At the start of each \texttt{MATLAB} session, \texttt{initialise$\_$df} needs to be called from within the \df\ parent folder (\textit{not one of its subfolders}) to add the program folders to the file path and set plotting defaults. This action must be completed before any saved data objects are loaded into the \texttt{MATLAB} workspace to ensure that objects are associated with their corresponding class definitions.

\subsection{Defining device properties and creating a parameters object: \texttt{pc(file$\_$path)}}
\label{ssec:parameters}
The parameters class, \texttt{pc} contains the default device properties and functions required to build a parameters object, which we shall denote herein as \texttt{par}. The parameters object defines both layer-specific and device-wide properties. Layer-specific properties must be a cell or numerical array containing the same number of elements as there are layers, \textit{including} \texttt{interface} layers. For example, a three-layer device with two heterojunctions requires layer-specific property arrays to have five elements. Examples can be found in the \texttt{/Input\_files} folder (see Section \ref{sssec:import_properties}).

\begin{figure*}
\centering
\includegraphics[width=0.98\textwidth]{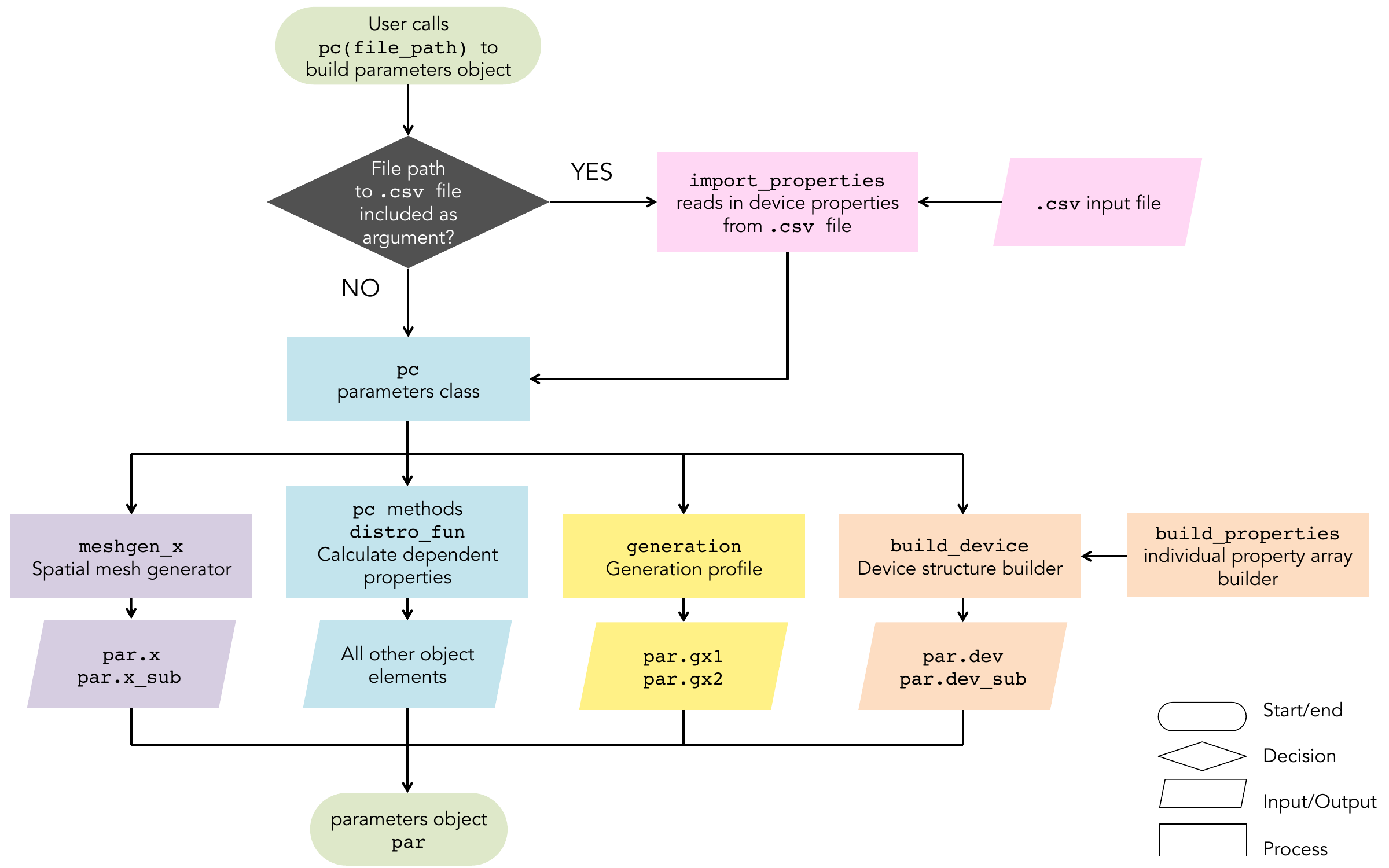}
\caption[Flow diagram showing the key processes involved in building a parameters object.]{\textbf{Flow diagram showing the key processes involved in building a parameters object.}}
\label{fig:parameters_flow_diagram}
\end{figure*}

Figure \ref{fig:parameters_flow_diagram} shows the processes through which the main components of the parameters object are built; The user is required to define a set of material properties for each layer. The comments in \texttt{pc} describe each of the parameters in detail and give their units (also see \SI , Table \ref{tbl:symbols} for quick reference); Several subfunctions (methods) of \texttt{pc} then calculate dependent properties, such as the equilibrium carrier densities for example, from the choice of probability distribution function (\texttt{prob$\_$distro$\_$function}) and other user-defined properties. The treatment of properties in the device interfaces is dealt with, and can be changed, using the device builder \texttt{build$\_$device} (see below).

\subsubsection{Importing properties}
\label{sssec:import_properties}
Typically, the most important user-definable properties are stored in a \texttt{.csv} file, which is easily editable with a spreadsheet editor such as LibreOffice or Microsoft Excel. The file path to the \texttt{.csv} file can then be used as an input argument for \texttt{pc}, for example:

\begin{lstlisting}[numbers=none]
par = pc(`Input_files/spiro_mapi_tio2.csv');
\end{lstlisting}

This allows the user to easily create and store sets of key device parameters without editing \texttt{pc}. Default values for properties set in the parameters class \texttt{pc} will be overwritten by the values in the \texttt{.csv} file during creation of the parameters object \texttt{par}. New properties defined in \texttt{pc} can easily be added to the \texttt{.csv} file provided that they are also included in \texttt{import$\_$properties}, which tests to see which properties are present in the text file and reads them into the parameters object where present. Following the properties read-in step performed by \texttt{import$\_$properties}, the number of rows in the \texttt{layer$\_$type} column (see Section \ref{sssec:layer_types}) is used for error checking all other entries to confirm that properties have been defined for each discrete layer of the system (this does not include the \texttt{electrode} rows, which are pseudo-layers). To avoid potential incompatibility, new user-defined material properties defined in \texttt{pc}, which have distinct values for each layer, should be included in the \texttt{.csv} file and added to the list of importable properties in \texttt{import$\_$properties}, as well as \texttt{build$\_$device}. An example of how to do this is given in the \SI , Section \ref{sec:edit_equations}.

\begin{table*}
\caption{Key to properties contained in external \texttt{.csv} parameters files and their constraints. $^1$Note that contrary to convention, \texttt{Phi$\_$EA} and \texttt{Phi$\_$IP} take negative values for consistency with other energies referenced to the electron energy scale. $^2$These properties are required for \texttt{electrode} pseudo-layers, but all other properties are ignored in these rows. $^3$For \texttt{electrode} layers, entries for \texttt{EF0} are stored in the parameters object (\texttt{par}) as the distinct properties \texttt{Phi$\_$left} and \texttt{Phi$\_$right} rather than as part of the \texttt{EF0} array. \texttt{Phi$\_$left} and \texttt{Phi$\_$right} take negative values for consistency with other energies referenced to the electron energy scale. $^4$ For \texttt{electrode} layers, entries for \texttt{sn} and \texttt{sp} are stored in the parameters object (\texttt{par}) as the distinct properties \texttt{sn$\_$l}, \texttt{sn$\_$r}, \texttt{sp$\_$l}, and \texttt{sp$\_$r} rather than as part of the \texttt{sn} and \texttt{sp} arrays.}
\label{tbl:input_files}
\begin{tabular}{p{2cm} p{6cm} p{5.2cm} p{1.6cm} p{1.6cm}}
\hline\noalign{\smallskip}
Column heading				&	Description							& Options/range					& Units		&	Section ref. \\
\noalign{\smallskip}\hline\noalign{\smallskip}					
\texttt{layer$\_$type}		& 	Layer type 							& \texttt{electrode}, \texttt{layer}, \texttt{active}, \texttt{interface}				&	- 		&	\ref{sssec:layer_types}		\\
\texttt{material}			&	Chemical short-form name			& Materials contained within \texttt{./Libraries/Index$\_$of$\_$Refraction$\_$library.xls}			& 	-		&	\ref{sssec:generation_function} \\
\texttt{thickness}			&	Layer thickness						&		$>0$			&	cm		&	\ref{ssec:spatial_mesh}				\\
\texttt{layer$\_$points}	&	Number of points in the layer			&		$\geqslant 3$			&	-		&	\ref{ssec:spatial_mesh}				\\
\texttt{xmesh$\_$ceoff}	&	A parameter defining how densely the spatial mesh points are concentrated at the boundaries of the layer for \texttt{xmesh$\_$type = `erf-linear'}	 	&	$0.1-0.9$ 	&	norm.	&	\ref{ssec:spatial_mesh}	\\
\texttt{Phi$\_$EA}			&	Electron affinity$^1$								&	-		&	eV		&	\ref{ssec:energy_levels}		\\
\texttt{Phi$\_$IP}			&	Ionisation potential$^1$							&	-		&	eV		&	\ref{ssec:energy_levels}		\\
\texttt{EF0}				&	Equilibrium Fermi energy$^{2,3}$	&	$\geqslant $\texttt{Phi$\_$IP}, $\leqslant$\texttt{Phi$\_$EA}	&	eV		&	\ref{sssec:prob_dist_fun}		\\
\texttt{Et}					&	SRH trap energy level (single trap level model)	&	$\geqslant $\texttt{Phi$\_$IP}, $\leqslant$\texttt{Phi$\_$EA}	&	eV		&	\ref{ssec:recombination}, \ref{ssec:recombination_interfaces}	\\
\texttt{Nc}					&	Effective density of conduction band states		&	-		&	cm$^{-3}$	&	\ref{sssec:prob_dist_fun}				\\
\texttt{Nv}					&	Effective density of valence band states			&	-		&	cm$^{-3}$	&	\ref{sssec:prob_dist_fun}				\\
\texttt{Ncat}				&	Intrinsic Schottky defect density (mobile cations) at equilibrium	&	$<$\texttt{c$\_$max}		&	cm$^{-3}$	&	\ref{ssec:Poisson_eq} \\
\texttt{Nani}				&	Intrinsic Schottky defect density (mobile anions) at equilibrium 	&	$<$\texttt{a$\_$max}		&	cm$^{-3}$	&	\ref{ssec:Poisson_eq}	\\
\texttt{c$\_$max}			&	Limiting mobile cation density							&	$>$\texttt{Ncat}	&	cm$^{-3}$	&	\ref{sssec:diffusion_enhancement} 		\\
\texttt{a$\_$max}			&	Limiting mobile anion density 							&	$>$\texttt{Nani}	&	cm$^{-3}$	&	\ref{sssec:diffusion_enhancement}		\\
\texttt{mu$\_$n}			&	Electron mobility									& 	$\geqslant 0$		&	\mobunit	&	\ref{ssec:transport_drift_diffusion}	\\
\texttt{mu$\_$p}			&	Hole mobility										&	$\geqslant 0$		&	\mobunit	&	\ref{ssec:transport_drift_diffusion}	\\
\texttt{mu$\_$c}			&	Cation mobility										&	$\geqslant 0$		&	\mobunit	&	\ref{ssec:transport_drift_diffusion}	\\
\texttt{mu$\_$a}			&	Anion mobility										&	$\geqslant 0$		&	\mobunit	&	\ref{ssec:transport_drift_diffusion}	\\
\texttt{epp}				&	Relative dielectric constant						&	$>0$	&		-		&	\ref{ssec:Poisson_eq} 					\\
\texttt{g0}					&	Uniform generation rate							&	$\geqslant 0$		&	cm$^{-3}$ s$^{-1}$	&	\ref{ssec:generation}, \ref{sssec:generation_function}	\\
\texttt{B}					&	Band-to-band recombination rate coefficient		& 	$>0$			&	\rateunit		&	\ref{sssec:rec_btb}	\\
\texttt{taun}				&	SRH electron lifetime								& 	$>0$			&	s				&	\ref{sssec:rec_srh}	\\
\texttt{taup}				&	SRH hole lifetime									& 	$>0$			&	s				&	\ref{sssec:rec_srh}	\\
\texttt{sn}					&	Electron surface recombination velocity$^{2,4}$ 	&	$>0$ 		&	cm s$^{-1}$		& \ref{ssec:recombination_interfaces}, \ref{ssec:boundary_conditions}		\\
\texttt{sp}					&	Hole surface recombination velocity$^{2,4}$	 	&	$>0$		&	cm s$^{-1}$		& \ref{ssec:recombination_interfaces}, \ref{ssec:boundary_conditions}		\\
\texttt{vsr$\_$zone$\_$loc}	&	Volumetric interfacial surface recombination zone location	&	\texttt{auto}, \texttt{L}, \texttt{C}, \texttt{R}	&	-	&	\ref{ssec:recombination_interfaces}	\\
\texttt{Red}				&	Layer colour red RGB triplet component			&	$0 - 1$		&	norm.		&	-			\\
\texttt{Green}				&	Layer colour green RGB triplet component			&	$0 - 1$		&	norm.		&	-			\\
\texttt{Blue}				&	Layer colour blue RGB triplet component			&	$0 - 1$		&	norm.		&	-			\\
\texttt{optical$\_$model}	&	Optical model										&	\texttt{uniform}, \texttt{Beer-Lambert}		&	-	&	\ref{ssec:generation},\ref{sssec:generation_function},\ref{sssec:beer_lambert} 	\\
\texttt{xmesh$\_$type}		&	Spatial mesh type									&	\texttt{linear}, \texttt{erf-linear}		&	-	&	\ref{ssec:spatial_mesh}			\\
\texttt{side}				&	Illumination side									&	\texttt{left}, \texttt{right}			&	-			&	-			\\
\texttt{N$\_$ionic$\_$species}		&	Number of mobile ionic species				&	$1$ (cations), $2$ (cations \& anions)				&	-	&	-		\\
\noalign{\smallskip}\hline
\end{tabular}
\end{table*}

\subsubsection{Layer types}
\label{sssec:layer_types}
Layer types, set using the \texttt{layer$\_$type} property, flag how each layer should be treated. \df\ currently uses three layer types:
\begin{enumerate}
\item \texttt{`electrode'}: A pseudo-layer which defines the boundary properties of the system. These are not discrete layers and do not appear in visualised outputs.
\item \texttt{`layer'}: A slab of semiconductor for which all properties are spatially constant.
\item \texttt{`active'}: As \texttt{layer} but flags the active layer of the device. The number of the first layer designated `active' is stored in the \texttt{active$\_$layer} property and is used for calculating further properties such as the active layer thickness, \texttt{d$\_$active}. Flagging the active layer proves particularly useful when automating parameter explorations as the active layer can easily be indexed.
\item \texttt{`interface'}: An interfacial region between two different material layers. The properties of the interface are varied according to the specific choice of grading method as defined in \texttt{build$\_$device} (see below). \textit{It is critical that interfacial layers are included between material layers with different energy levels and eDOS values i.e. at heterojunctions. See the included default input files for examples of how to set up devices with heterojunctions.} 
\end{enumerate}


\subsubsection{Spatial mesh}
\label{ssec:spatial_mesh}
\begin{figure}
\centering
	\includegraphics[width=0.48\textwidth]{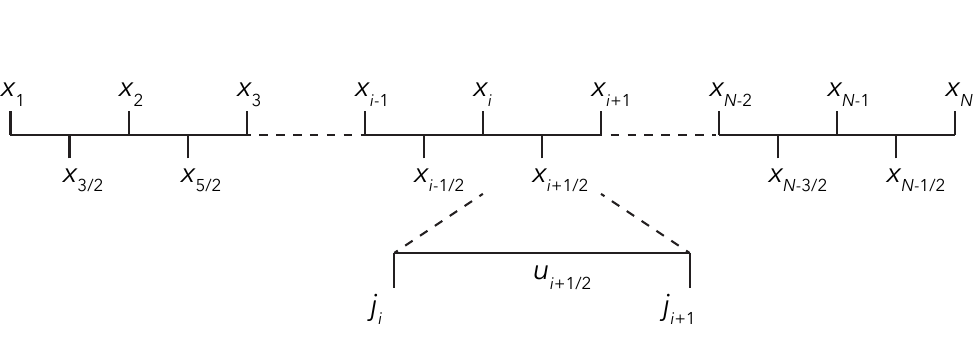}
	\caption[The computational spatial grid.]{\textbf{The computational spatial grid.} Variables are solved for on the subintervals while flux densities are calculated on the integer intervals. Figure concept taken from Ref \cite{Courtier2018c}.}
	\label{fig:computational_grid}
\end{figure}

The computational grid is divided into $N$ intervals with $N-1$ subintervals, where the position of the subintervals is defined by $x_{i +1/2} = (x_{i+1} + x_{i})/2$ for $i = 1,2,3,...,N-1$. \texttt{pdepe} solves for the variable values $u_{i +1/2}$ on the subintervals ($x_{i+1/2}$) and their associated flux densities $j_{i}$ on the integer intervals ($x_{i}$) as illustrated in Figure \ref{fig:computational_grid}.

Owing to the use of a finite element discretisation scheme the details of the spatial mesh in \df\ are of critical importance to ensure fast and reliable convergence. In this release of \df\ two types of spatial mesh are available:

\begin{enumerate}
\item \texttt{`linear'}: Linear piece-wise spacing
\item \texttt{`erf-linear'}: Mixed error function (bulk regions)- linear (interfacial regions) piece-wise spacing
\end{enumerate}

\texttt{meshgen$\_$x} generates integer and subinterval spatial meshes, \texttt{x} and \texttt{x$\_$sub} respectively (see Figure \ref{fig:computational_grid}), based on the layer \texttt{thickness}, the number of \texttt{layer$\_$points} defined in the device properties, and the \texttt{xmesh$\_$type}. The solution is interpolated for the integer grid points when generating the output solution matrix \texttt{sol.u} (see Section \ref{ssec:solution_structures}). The property \texttt{xmesh$\_$coeff} controls the spread of points for regions where an error function is used for point spacing i.e. \texttt{layer} and \texttt{active} layer types: higher values determine higher point densities close to the layer boundaries. In general we recommend using \texttt{`erf-linear'} for devices with relatively high ionic defect densities as, where ionic carriers are confined, high point densities are required at the layer boundaries to resolve the carrier distributions. Where the depletion of ionic carriers extends into the bulk \texttt{xmesh$\_$ceoff} can be reduced to increase the bulk point density. 

\subsubsection{Time mesh}
\label{ssec:time_mesh}
\texttt{pdepe} uses an adaptive time step for forward time integration and solution output is interpolated for the user-defined time mesh. Convergence of the solver is weakly-dependent of the user-defined time mesh interval spacing and strongly-dependent on the maximum time and the maximum time step. These can be adjusted by changing the \texttt{tmax} and \texttt{MaxStepFactor} properties of the parameters object (see Section \ref{ssec:parameters}). The options for different time mesh types (\texttt{tmesh$\_$type}) are:

\begin{enumerate}
\item \texttt{`linear'} or \texttt{1}
\item \texttt{`$\log 10$'} or \texttt{2}
\end{enumerate}

Typically, the time mesh is changed frequently for intermediate solutions within protocols to accommodate the different timescales on which carriers move. For example, in some cases the ionic carriers may be frozen to obtain a stable short timescale solution for the electronic carriers, before a second, longer timescale, solution is calculated with the ionic carriers mobile. Similarly the \texttt{tmesh$\_$type} is adjusted dependent on the voltage and light conditions. For example during a $J-V$ scan a linear time mesh is used in keeping with the linear change in the applied voltage with time. In other instances a logarithmic mesh is more appropriate in order to resolve time periods for which the carrier time derivatives are larger.

\subsubsection{\textrm{The device structures}}
\label{sssec:device_structures}
\texttt{build$\_$device} and \texttt{build$\_$property} are called during creation of the parameters object to build two important data structures, which we call the `device structures': \texttt{dev}, defined on the integer grid intervals ($x_i$), is used to determine the initial conditions only, while \texttt{dev$\_$sub}, defined on the subintervals ($x_{i+\frac{1}{2}}$), is used by the \texttt{pdepe} solver function. \texttt{dev} and \texttt{dev$\_$sub} contain arrays defining all spatially varying properties at every location within the device including the interfacial regions. \texttt{build$\_$property} enables the user to specify different types of interface grading for each property listed in \texttt{build$\_$device}. At present there are four generic grading option types:

\begin{enumerate}
\item \texttt{`zeroed'}: The value of the property is set to zero throughout the interfacial region.
\item \texttt{`constant'}: The value of the property is set to be constant throughout the interfacial region. Note that the user must define this value in the \texttt{.csv} file.
\item \texttt{`lin\_graded'}: The property is linearly graded using the property values of the adjoining layers.
\item \texttt{`exp\_graded'}: The property is exponentially graded using the property values of the adjoining layers.
\end{enumerate}

For the volumetric surface recombination scheme described in Section \ref{ssec:recombination_interfaces}, we also introduce parameter-specific grading schemes for the VSR time constants:
\begin{enumerate}
\item \texttt{`taun\_vsr'}: Sets $\tau_{n,\mathrm{vsr}}$ according to Equation \ref{eq:taun_int}.
\item \texttt{`taup\_vsr'}: Sets $\tau_{p,\mathrm{vsr}}$ according to Equation \ref{eq:taup_int}.
\end{enumerate}

The interface grading type for each property is set in the \texttt{build$\_$device} function e.g. the default grading type for the electron affinity, $\Phi_\mathrm{EA}$ is \texttt{`lin\_graded'}.

Dependent on the choice of grading option, input values may, or may not, be required for a given material property in the interfacial layers. For example, when using the \texttt{`lin\_graded'} option a value is not needed because the interface property values are calculated from those of the adjacent layers. By contrast, when using the \texttt{`constant'} option, a property value does need to be specified for the interfacial layer to avoid an error. Since property values that are not required are ignored, \textit{we recommend that users specify all property values for all layers} to future-proof against problems arising when experimenting with grading options.

\subsubsection{\textrm{The electronic carrier probability distribution function}}
The class \texttt{distro\_fun} defines the electronic carrier probability distribution function for the model and calculates equilibrium boundary, and initial carrier densities when building the parameters object. At present there are two available options for the choice of probability distribution function:

\begin{enumerate}
\item \texttt{`Boltz'} - The Boltzmann approximation ($\gamma = 0$)
\item \texttt{`Blakemore'} - The Blakemore approximation ($\gamma > 0$ see Section \ref{sssec:prob_dist_fun})
\end{enumerate} 

While the Boltzmann approximation results in marginally faster calculations owing to the absence of a diffusion enhancement we recommend using Blakemore statistics for their extended domain of validity.

\subsubsection{\textrm{The electronic carrier generation profile function}}
\label{sssec:generation_function}
\texttt{generation} calculates the two generation profiles \texttt{gx1} and \texttt{gx2}, which can be used for a constant bias light and a pulse source for example, at each spatial location in the device for the chosen \texttt{optical$\_$model} and light sources (see Section \ref{ssec:generation} and the flow diagram in the \SI\ Figure \ref{fig:generation_flow}). The light sources can be set using the \texttt{light\_source1} and \texttt{light\_source2} properties. There are two options for the \texttt{optical$\_$model}:

\begin{enumerate}
\item \texttt{`uniform'}: a uniform volumetric generation rate defined by the property \texttt{g0} is applied layers with \texttt{layer} and \texttt{active} layer types. The multiplier properties \texttt{int1} and \texttt{int2} define the intensities for light source $1$ and $2$ respectively.
\item \texttt{`Beer\_Lambert'}: The generation profile follows the Beer-Lambert law as detailed in Section \ref{sssec:beer_lambert} with material layer optical properties taken from \texttt{./Libraries/Index$\_$of$\_$Refraction\_library.xls} for the corresponding materials defined in the \texttt{par.material} cell array. As with uniform generation, the generation rate profile is multiplied by the intensity properties \texttt{int1} and \texttt{int2} before being applied.
\end{enumerate} 

A arbitrary generation profile calculated using an external program (such as Solcore\cite{Alonso-Alvarez2018a} or the McGeeHee Group's Transfer Matrix code\cite{Burkhard2010}) can be inserted into the parameters object \texttt{par} by overwriting the generation profile properties \texttt{gx1} or \texttt{gx2} following creation of the object. The profile must be interpolated for the subinterval ($x_{i+1/2}$) grid points defined by \texttt{x\_sub} (see Section \ref{ssec:spatial_mesh}). We also recommend that the generation rate is set to zero within the interfacial regions to avoid stability issues.

The light source time-dependencies are controlled using the \texttt{g1\_fun\_type} and \texttt{g2\_fun\_type} properties, which define the function type (e.g. sine wave) and the \texttt{g1\_fun\_arg} and \texttt{g2\_fun\_arg} properties, which are coefficient arrays for the function generator (see Section \ref{ssec:fun_gen}) e.g. the frequency, amplitude, etc.

\begin{figure}[t]
\centering
\includegraphics[width=0.48\textwidth]{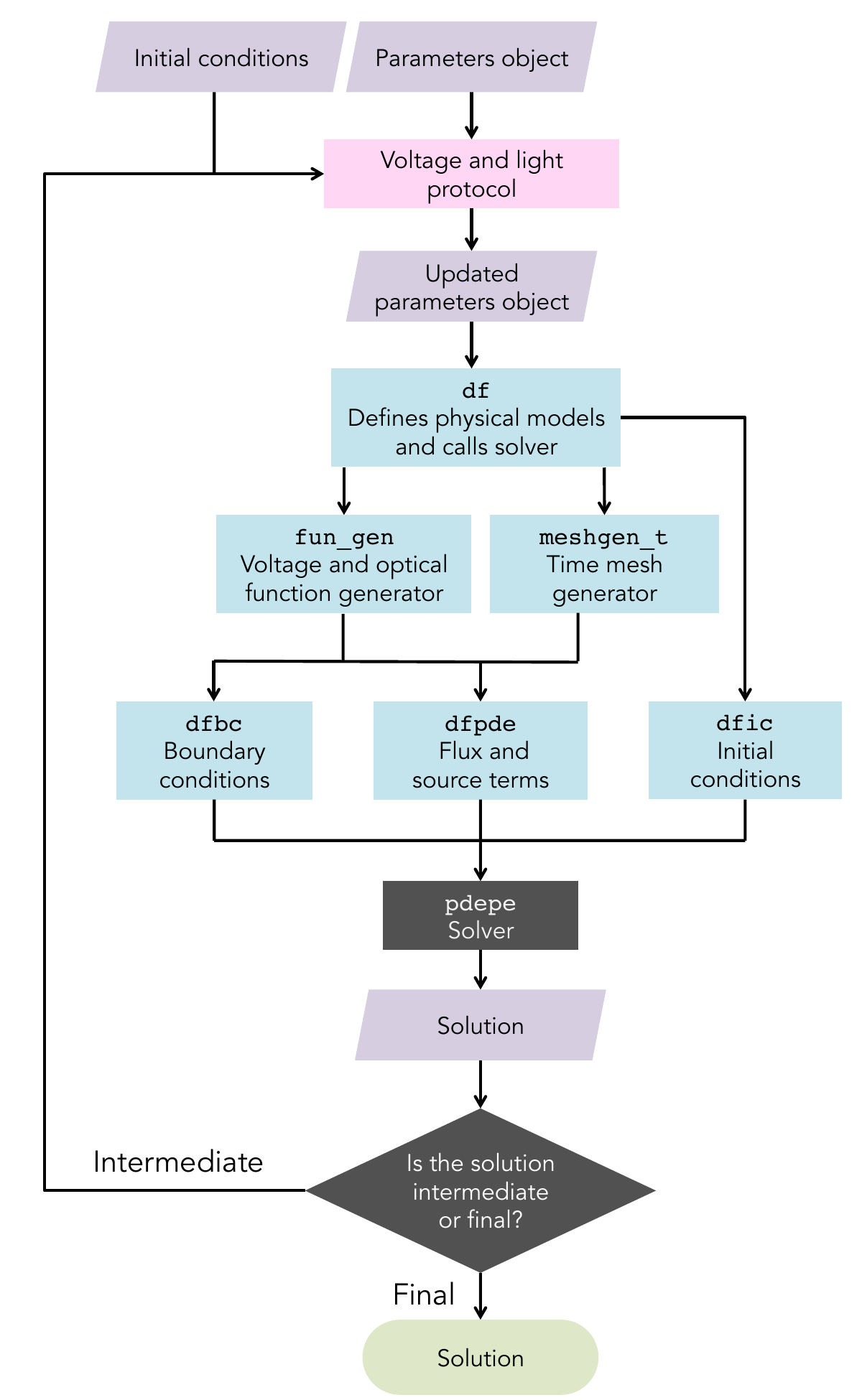}
\caption[Flow diagram illustrating execution of a \df\ protocol.]{\textbf{Flow diagram illustrating execution of a \df\ protocol.}}	
\label{fig:protocols_flow_diagram}
\end{figure}

\subsection{Protocols: \texttt{equilibrate}}
\label{ssec:protocols_equilibrate}
Once a device has been created and stored in the \texttt{MATLAB} workspace as a parameters object the next step is to find the equilibrium solution for the device. The function \texttt{equilibrate} starts with the initial conditions described in Section \ref{ssec:initial_condish} and runs through a number of steps to find the equilibrium solutions, with and without mobile ionic carriers, for the device described by \texttt{par}. The output structure \texttt{soleq} contains two solutions (see Section \ref{ssec:solution_structures});
\begin{enumerate}
\item \texttt{soleq.el}: only the electronic charge carriers are mobile and are at equilibrium.
\item \texttt{soleq.ion}: electronic and ionic carriers are mobile and are at equilibrium.
\end{enumerate}
Storing both solutions in this way allows devices with and without mobile ionic charge to be compared easily.

\subsection{Protocols: General}
\label{ssec:protocols_general}
A \df\ protocol is defined as a function that contains a series of instructions that takes an input solution (initial conditions) and produces an output solution. For many of the existing protocols, listed in Table \ref{tbl:protocols}, with the exception of \texttt{equilibrate}, the input is one of the equilibrium solutions, \texttt{soleq.el} or \texttt{soleq.ion}. Figure \ref{fig:protocols_flow_diagram} is a flow diagram illustrating the key functions called during execution of a protocol.

Protocols typically start by creating a temporary parameters object that is a duplicate of the input solution parameters object. This temporary object can then be used to write new voltage and light parameters that will be used by the function generator to define the generation rate at each point in space and time and the potential at the boundary at each point in time. Additional parameters, for example those defining the output time mesh or carrier transport are also frequently adjusted. The approach is to split a complex experimental protocol into a series of intermediate steps that facilitate convergence of the solver. For example, where ionic mobilities are separated by many orders of magnitude from electronic mobilities, a steady-state solution is easiest found by accelerating the ions to similar values to the electronic carriers using the \texttt{K\_a} and \texttt{K\_c} properties. The simulation can then be run with an appropriate time step and checked to confirm that a steady-state has been reached. The possibilities are too numerous to list here and users are encouraged to investigate the existing protocols listed in Table \ref{tbl:protocols} in preparation of writing their own. 

\begin{table*}
\caption{List of protocols available in \df\ at the time of publication.}
\label{tbl:protocols}
\begin{tabular}{p{3cm}p{13cm}}
\hline\noalign{\smallskip}
Command		&	Description		\\
\noalign{\smallskip}\hline\noalign{\smallskip}
\texttt{changeLight}	& Switches light intensity using incremental intensity steps. \\
\texttt{doCV}			& Cyclic Voltammogram (CV) simulation using triangle wave function generator	.\\
\texttt{doIMPS}			& Intensity Modulated Photocurrent Spectroscopy (IMPS) simulation at a specified frequency and light intensities.	\\
\texttt{doIMVS}			& Switches to open circuit (OC), runs to steady-state OC, then performs Intensity Modulated Photovoltage Spectroscopy (IMVS) measurement simulation at a specified frequency and light intensities.\\	
\texttt{doJV}			& Forward and reverse current-voltage (JV) scan with options for dark and constant illumination conditions.\\	
\texttt{doLightPulse}	& Uses light source 2 with square wave generator superimposed on light source 1 (determined by the initial conditions) to optically pulse the device.\\
\texttt{doSDP}			& Step, Dwell, Probe (SDP) measurement protocol: Jump to an applied potential, remain at the applied potential for a specified dwell time and then perform optically pulsed current transient. See Ref. \cite{Belisle2016} for further details of the experimental protocol. \\
\texttt{doSPV}			& Surface PhotoVoltage (SPV) simulation: Switches light on with high series resistance. \\
\texttt{doTPV}			& Transient PhotoVoltage (TPV) simulation: Switches to open circuit with bias light and optically pulses the device.\\
\texttt{equilibrate}	& Start from base initial conditions and find equilibrium solutions without mobile ionic charge (\texttt{soleq.el}) and with mobile ionic charge (\texttt{soleq.ion}).\\	
\texttt{findVoc}		& Obtains steady-state open circuit condition using Newton-Rhaphson minimisation.\\	
\texttt{findVocDirect}	& Obtains an approximate steady-state open circuit condition using 1 M$\Omega$ cm$^2$ series resistance.\\	
\texttt{genIntStructs}	& Generates solutions at various light intensities.\\
\texttt{genIntStructsRealVoc}	& Generates open circuit solutions at various light intensities.\\
\texttt{genVappStructs}	& Generates solutions at various applied voltages.\\
\texttt{jumptoV}	& Jumps to a new applied voltage and stabilises the cell at the new voltage for a user-defined time period.\\
\texttt{lightonRs}	& Switches on the light for a specified period of time with a user-defined series resistance.\\
\texttt{stabilize}	& Runs a set of initial conditions to a steady-state.\\
\texttt{sweepLight}	& Linear sweep of the light intensity over a user-defined time period.\\
\texttt{transient\_nid}	& Simulates a transient ideality factor ($n_{id}$) measurement protocol.\cite{Calado2019_ideality}\\
\texttt{VappFunction}	& Applies a user-defined voltage function to a set of initial conditions.\\
\noalign{\smallskip}\hline
\end{tabular}
\end{table*}

\subsection{The \df\ master function \texttt{df}}
\label{ssec:df_master_function}
\texttt{df} is the core of \df . The function takes the device parameters, and voltage and light conditions, calls the solver and outputs the solution (Figure \ref{fig:protocols_flow_diagram}). The underlying equations in \df\ can be customised for specific models and applications using the Equation Editor in the \texttt{dfpde} subfunction. \texttt{df} contains three important subfunctions: \texttt{dfpde}, \texttt{dfic}, and \texttt{dfbc} which we now proceed to describe.

\subsubsection{\df\ Partial Differential Equation function \texttt{dfpde} and the Equation Editor}
\label{sssec:df_pde_equation_editor}
\texttt{dfpde} defines the equations to be solved by \texttt{MATLAB}'s Partial Differential Equation: Parabolic and Elliptic solver toolbox (\texttt{pdepe}).\cite{pdepe2013} For one-dimensional Cartesian co-ordinates, \texttt{pdepe} solves equations of the form:

\begin{equation} 	\label{eq:pdepe}
C \left( x,t,u,\dfrac{\partial u}{\partial x} \right) \dfrac{\partial u}{\partial t} = 
\dfrac{\partial u}{\partial x} \left( F \left( x,t,u,\dfrac{\partial u}{\partial x} \right) \right)
+ S\left( x,t,u,\dfrac{\partial u}{\partial x} \right)
\end{equation}

Here, $u$ is a vector containing the variables $V$, $n$, $p$, $c$, and $a$ at each position in space $x$ and time $t$. $C$ is a vector defining the prefactor for the time derivative, $F$ is a vector determining the flux density terms (note: $F$ does \textit{not} denote the electric field in this section), and $S$ is a vector containing the source/sink terms for the components of $u$. By default $C = 1$ for charge carriers but could, for example, be used to change the active volume fraction of a layer in a mesoporous structure. The equations can be easily reviewed and edited in the \texttt{dfpde} Equation Editor as shown in Listing \ref{lst:equation_editor}. Here, device properties that have a spatial dependence are indexed with the variable $i$ to obtain the corresponding value at the location given by \texttt{x\_sub(i)}. A step-by-step example of how to change the physical model using the Equation Editor is given in the \SI , Section \ref{sec:edit_equations}. %

\begin{lstlisting}[float,language=Matlab, 
caption=\textbf{The Equation Editor.} Coefficients that are defined at every position are indexed for the current $x$ position using the index $i$. Gradient coefficients (prefixed with `grad') are equal to zero outside of the interfacial regions. Location: ./Core/df - dfpde subfunction. ,
firstnumber=247,
firstline=247,
lastline=279,
float=*,
basicstyle=\ttfamily\scriptsize,
label={lst:equation_editor}]
%% Equation editor
% Time-dependence prefactor term
C_V = 0; C_n = 1; C_p = 1; C_c = 1; C_a = 1;
C = [C_V; C_n; C_p; C_c; C_a];

% Flux terms
F_V = (epp(i)/eppmax)*dVdx;
F_n = mu_n(i)*n*(-dVdx + gradEA(i)) + (Dn(i)*(dndx - ((n/Nc(i))*gradNc(i))));
F_p = mu_p(i)*p*(dVdx - gradIP(i)) + (Dp(i)*(dpdx - ((p/Nv(i))*gradNv(i))));
F_c = mu_c(i)*(z_c*c*dVdx + kB*T*(dcdx + (c*(dcdx/(c_max(i) - c)))));
F_a = mu_a(i)*(z_a*a*dVdx + kB*T*(dadx + (a*(dadx/(a_max(i) - a)))));
F = [F_V; mobset*F_n; mobset*F_p; mobseti*K_c*F_c; mobseti*K_a*F_a];

% Electron and hole recombination
% Radiative
r_rad = radset*B(i)*(n*p - ni(i)^2);
% Bulk SRH
r_srh = SRHset*srh_zone(i)*((n*p - ni(i)^2)/(taun(i)*(p + pt(i)) + taup(i)*(n + nt(i))));
% Volumetric surface recombination
alpha = sign_xn(i)*q*dVdx/(kB*T) + alpha0_xn(i);
beta = sign_xp(i)*q*-dVdx/(kB*T) + beta0_xp(i);
r_vsr = SRHset*vsr_zone(i)*((n*exp(-alpha*xprime_n(i))*p*exp(-beta*xprime_p(i)) - ni(i)^2)...
    /(taun_vsr(i)*(p*exp(-beta*xprime_p(i)) + pt(i)) + taup_vsr(i)*(n*exp(-alpha*xprime_n(i)) + nt(i))));
% Total electron and hole recombination
r_np = r_rad + r_srh + r_vsr;

% Source terms
S_V = (q/(eppmax*epp0))*(-n + p - NA(i) + ND(i) + z_a*a + z_c*c - z_a*Nani(i) - z_c*Ncat(i));
S_n = g - r_np;
S_p = g - r_np;
S_c = 0;
S_a = 0;
S = [S_V; S_n; S_p; S_c; S_a];
\end{lstlisting}

\subsubsection{\df\ Initial Conditions \texttt{dfic}} \texttt{dfic} defines the initial conditions to be used by \texttt{pedpe}. If running \texttt{equilibrate} to obtain the equilibrium solution or running \texttt{df} with an empty input solution, the first set of initial conditions are as described in Section \ref{ssec:initial_condish}. Otherwise, the final time point of the input solution is used.

\subsubsection{\df\ Boundary Conditions \texttt{dfbc}} 
\begin{lstlisting}[float,language=Matlab,
caption=\textbf{Default boundary condition expressions for electrons, holes, cations and the electrostatic potential.} Location: ./Core/df - dfbc subfunction. ,
firstnumber=407,
firstline=407,
lastline=421,
float,
basicstyle=\ttfamily\scriptsize,
label={lst:dfbc}]
Pl = [-V_l;
  mobset*(-sn_l*(n_l - n0_l));
  mobset*(-sp_l*(p_l - p0_l));
  0;
  0;];

Ql = [0; 1; 1; 1; 1;];

Pr = [-V_r+Vbi-Vapp-Vres;
  mobset*(sn_r*(n_r - n0_r));
  mobset*(sp_r*(p_r - p0_r));
  0;
  0;];

Qr = [0; 1; 1; 1; 1;];
\end{lstlisting}

\texttt{dfbc} defines the system boundary conditions. The boundary condition expressions are passed to \texttt{pdepe} using two coefficients $P$ and $Q$, with $N_u$ elements, where $N_u$ is the number of independent variables being solved for. The boundary conditions are expressed in the form:

\begin{equation} 	\label{eq:pdepe_BCs}
P(x,t,u) + Q(x,t)F \left( x,t,u,\dfrac{\partial u}{\partial x} \right) = 0.
\end{equation}

For Dirichlet conditions $P$ must be non-zero to define the variable values, whereas for Neumann conditions $Q$ must be non-zero to define the flux density. Listing \ref{lst:dfbc} shows how the default \df\ boundary conditions (described in Section \ref{ssec:boundary_conditions}) are implemented in the current version of the code.\\

In addition to these subfunctions, \texttt{df} also calls two important external functions: the time mesh generator, \texttt{meshgen\_t} and the function generator, \texttt{fun\_gen}.

\subsubsection{The time mesh generator \texttt{meshgen\_t}.}
\label{ssec:meshgen_t}
\texttt{df} calls the time mesh generator \texttt{meshgen\_t} at the start of the code. As discussed in Section \ref{ssec:time_mesh}, the solver uses an adaptive time step and interpolates the solution to the user-defined mesh. Hence the choice of mesh is not critical to time integration convergence. The values of the mesh should be chosen such as to resolve the solution properly on the appropriate timescales. The total time step of the solution \texttt{tmax} and the maximum time step (controlled using the \texttt{MaxStepFactor} property) do however influence convergence strongly. For this reason, where convergence is proving problematic it is recommended that either \texttt{tmax} or \texttt{MaxStepFactor} is reduced and the solution obtained in multiple stages.

\subsubsection{The function generator \texttt{fun\_gen}.}
\label{ssec:fun_gen}
\texttt{df} calls \texttt{fun\_gen} to generate time-dependent algebraic functions that define the applied voltage and light intensity conditions. \texttt{df} includes the ability to call two different light intensity functions with different light sources, enabling users to simulate a constant bias light and additional pump pulse using the square wave generator for example. Each function type requires a coefficients array with a number of elements determined by the function type and detailed in the comments of \texttt{fun\_gen}. Listing \ref{lst:VappFunction} is an example from the \texttt{./Scripts/Vapp\_function\_script} script showing how to define a sine wave function for the applied voltage:

\begin{lstlisting}[float,language=Matlab,
float,
basicstyle=\ttfamily\scriptsize,
caption=\textbf{Setting the function type and coefficients for the applied voltage function.} Location:  ./Scripts/Vapp\_function\_script. ,
label={lst:VappFunction}]
% Voltage function type
par.V_fun_type = 'sin';
% DC offset voltage (V)
par.V_fun_arg(1) = 0;
% AC voltage amplitude (V)
par.V_fun_arg(2) = 20e-3;
% Frequency (Hz)
par.V_fun_arg(3) = 1e3;
% Phase (Rads)
par.V_fun_arg(4) = 0;
\end{lstlisting}


\subsection{Solution structures}
\label{ssec:solution_structures}
\texttt{df} outputs a solution structure \texttt{sol} with the following components:

\begin{itemize}
\item The solution matrix \texttt{u}: This is a three dimensional matrix for which the dimensions are (time, space, variable). The order of the variables are as follows: 
\begin{enumerate}
\item Electrostatic potential
\item Electron density
\item Hole density
\item Cation density (where 1 mobile ionic carrier is stipulated)
\item Anion density (where 2 mobile ionic carriers are stipulated)
\end{enumerate}
\item The spatial mesh \texttt{x}.
\item The time mesh \texttt{t}.
\item The parameters object \texttt{par}.
\end{itemize} 

All other outputs can be calculated from the above by calling methods from \texttt{dfana}.

\subsection{Calculating outputs: \texttt{dfana}}
\label{ssec:dfana}
\texttt{dfana} is a collection of functions (methods) that enable the user to calculate outputs such as carrier currents, quasi-Fermi levels, recombination rates etc. from the solution matrix \texttt{u}, the parameters object \texttt{par} and the specified physical models. The use of a class in this instance enables the package syntax \texttt{dfana.my\_calculation(sol)} to be used. For example the command:

\begin{lstlisting}[numbers=none]
rho = dfana.calcrho(sol, "whole"),
\end{lstlisting}

\noindent outputs a two dimensional matrix containing the space charge density as function of time and position. Many other examples of how \texttt{dfana} methods can be used to calculate outputs are given in the highlighted boxes of Section \ref{sec:theory}. The full list of the available analysis methods can be viewed and easily navigated by selecting the \texttt{dfana.m} in the \texttt{Current Folder} window and opening the functions browser sub-window.

\begin{tcolorbox}
Due to the computational cost of calling functions external to \texttt{pdepe}, \textit{the physical model described in the Equation Editor is not coupled to that used in the analysis functions}. Users should, therefore, take great care when adapting the physics of the simulation to make certain that the models defined in \texttt{dfana} and \texttt{df} are consistent.
\end{tcolorbox}

\subsection{Plotting the output: \texttt{dfplot}}
\label{ssec:dfplot}
\texttt{dfplot} is a class containing a collection of plotting methods. Similar to \texttt{dfana}, this enables the package syntax \texttt{dfplot.my\_plot(sol)} to be used. 
For variables plotted as a function of position, an optional vector argument $[t_1, t_2, t_3,... t_m]$ can be included to plot the solution at $t = t_1, t_2, t_3,... t_m$, where $m$ is the $m{th}$ time point to be plotted. For example the command:

\begin{lstlisting}[numbers=none]
dfplot.Vx(sol, [0, 0.2, 0.4, 0.6, 0.8]);
\end{lstlisting}

\noindent plots the electrostatic potential component of the solution as a function of position at $t = 0$, $0.2$, $0.4$, $0.6$, and $0.8$ s (an example is shown below in Figure \ref{fig:DF_vs_ASA_Vx_px_m1_1S}a). If no second argument is given then only the final time point is plotted. \texttt{dfplot} also includes the generic property plotting function \texttt{dfplot.x2d} to allow users to easily create new two-dimensional plots.

For variables plotted as a function of time the second argument defines the position. For example the command:

\begin{lstlisting}[numbers=none]
dfplot.Jt(sol, 1e-5);
\end{lstlisting}

\noindent plots the current density for each carrier as a function of position at $x = 10^{-5}$ cm. For plots where variables are integrated over a region of space, the second argument is a vector containing the limits $[x_1 , x_2]$. Further details can be found in the comments of \texttt{dfplot}.

\subsection{Getting started and the example scripts}
\label{sec:how_to}
While the underlying system may appear complex, \df\ has been designed such that with a few simple commands, users can simulate complex devices and transient optoelectronic experiment protocols. Table \ref{tbl:protocols} is a complete list of protocols available at the time of writing. In addition to the brief guide below, a quick start with up-to-date instructions can be found in the \texttt{README.md} file in the \df\ GitHub repository,\cite{Calado2017} and a series of example scripts for running specific protocols are also presented in the \texttt{Scripts} folder. New users are advised to study these scripts and adapt them to their own purposes. In addition, we have written an introductory workshop to guide students through the process of building basic semiconductor devices and applying optical and voltage biases to them. This can be found in the \texttt{Getting-started-workshop-2021} branch of the \df\ GitHub repository.

\subsubsection{How to build a device object, find the equilibrium solution and run a cyclic voltammogram}
\label{sssec:how_to_cv}
In this section some commonly-used commands are put together to show new users how to create a device object, obtain the device equilibrium solutions, and run a protocol, which in this example simulates a cyclic voltammogram (CV). The \texttt{doCV} protocol applies a triangular wave voltage function to the device, with optional constant illumination, for a set number of cycles enabling the device current-voltage characteristics at a given scan rate to be calculated.

At the start of each session, the system must be initialised by typing the command:

\begin{lstlisting}[numbers=none]
initialise_df;
\end{lstlisting}

To create a parameters object using the default parameters for a \Spiro /perovskite/\Ti\ perovskite solar cell the parameters class, \texttt{pc} is called with the file path to the appropriate \texttt{.csv} file as the input argument:

\begin{lstlisting}[numbers=none]
par = pc(`Input_files/spiro_mapi_tio2.csv');
\end{lstlisting}

The equilibrium solutions with and without mobile ionic carriers for the device can now be obtained by calling the \texttt{equilibrate} protocol:

\begin{lstlisting}[numbers=none]
soleq = equilibrate(par);
\end{lstlisting}

As discussed in Section \ref{ssec:protocols_equilibrate}, the output structure \texttt{soleq} contains two solutions: \texttt{soleq.el} and \texttt{soleq.ion}. In this example we are interested in seeing how mobile ionic carriers influence the device currents so we will use the solution including mobile ionic charge carriers, \texttt{soleq.ion}.

To perform a cyclic voltammogram simulation from $0$ to $1.2$ to $-0.2$ to $0$ V at $50$ mVs$^{-1}$, under $1$ sun illumination we call the \texttt{doCV} protocol with the appropriate argument values as detailed in the protocol comments shown in Listing \ref{lst:do_CV}.

\begin{lstlisting}[float,language=Matlab, 
caption=\textbf{First lines of the doCV protocol function.} The input arguments for \df\ protocols are detailed in the comments at the start of each function. Location: ./Protocols/doCV .,
firstline=1,
lastline=11,
firstnumber=1,
float=*,
basicstyle=\ttfamily\scriptsize,
label={lst:do_CV}]
function sol_CV = doCV(sol_ini, light_intensity, V0, Vmax, Vmin, scan_rate, cycles, tpoints)
% Performs a cyclic voltammogram (CV) simulation
% Input arguments:
% SOL_INI = solution containing initial conditions
% LIGHT_INTENSITY = Light intensity for bias light (Suns)
% V0 = Starting voltage (V)
% VMAX = Maximum voltage point (V)
% VMIN = Minimum voltage point (V)
% SCAN_RATE = Scan rate (Vs-1)
% CYCLES = No. of scan cycles
% TPOINTS = No. of points in output time array
\end{lstlisting}

\begin{lstlisting}[numbers=none]
solcv =...
doCV(soleq.ion,1,0,1.2,-0.2,50e-3,2,400);
\end{lstlisting}

\begin{figure}
\includegraphics[width=0.48\textwidth]{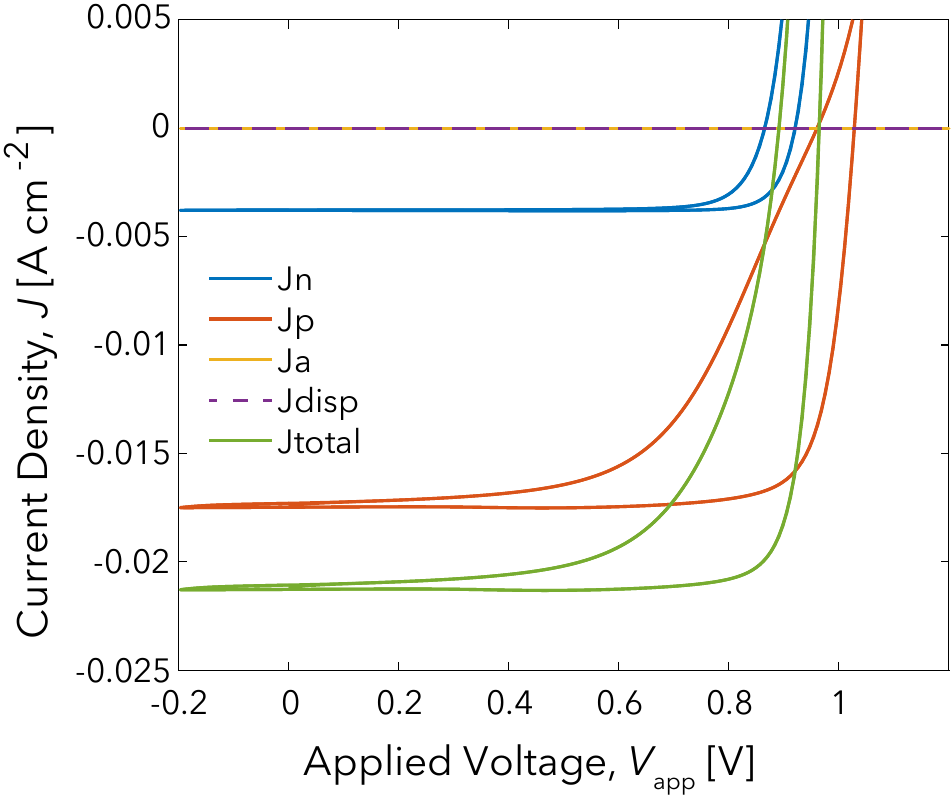}
\caption[Current-voltage scan results obtained from the cyclic voltammogram protocol (\texttt{doCV}) applied to the default \Spiro /perovskite /\Ti\ solar cell parameters.]{\textbf{Current-voltage scan results obtained from the cyclic voltammogram protocol (\texttt{doCV}) applied to the default \Spiro /perovskite /\Ti\ solar cell parameters.} See the corresponding guide in Section \ref{sssec:how_to_cv} for step-by-step instructions on how to obtain these results and plot the figure.}	
\label{fig:ExampleCV}
\end{figure}

Once a solution has been calculated the different components of the currents can be plotted as a function of voltage using the command:

\begin{lstlisting}[numbers=none]
dfplot.JVapp(solcv, par.d_midactive);
ylim([-25e-3, 5e-3]);
\end{lstlisting}

The second argument of \texttt{dfplot.JVapp} is the pre-calculated dependent property \texttt{d\_midactive}, the value of which is equal to the position at the midpoint of the active layer of the device. The resulting plot is given in Figure \ref{fig:ExampleCV} for reference.

\subsubsection{How to change the physical model}
A detailed step-by-step example of how to modify the physical model to account for the possible effects of photogenerated mobile ionic charge carriers is described in Section \ref{sec:edit_equations} of the \SI .

\subsection{Advanced features}
\label{ssec:advanced_features}

\subsubsection{Rebuilding device structures and spatial meshes: \texttt{refresh\_device}}
In some situations where device properties, such as layer widths, are changed in a user-defined script or function users may need to rebuild the device structures \texttt{dev} and \texttt{dev\_sub}, and spatial meshes \texttt{x} and \texttt{x\_sub}. To maintain code performance this is not performed automatically since non-device-related parameters are frequently changed within protocols. \texttt{meshgen\_x} and \texttt{build\_device} can be rerun and stored using the new parameter set (`refreshed') using the syntax:

\begin{lstlisting}[numbers=none]
par = refresh_device(par);
\end{lstlisting}

To further illustrate how to use \texttt{refresh\_device}, \texttt{refresh\_device\_script}, a script describing the necessary steps to change the interfacial recombination parameters, is provided in the \texttt{Scripts} folder of the \df\ repository.
\subsubsection{Parallel computing and parameter exploration: \texttt{explore}}
\df 's parameter exploration class \texttt{explore}, takes advantage of \texttt{MATLAB}'s parallel computing toolbox, to enable multiple simulations to be calculated using a parallel pool. \texttt{explore\_script} is an example script demonstrating how to use \texttt{explore} to run an active layer thickness versus light intensity parameter exploration and plot the outputs using \texttt{explore}'s embedded plotting tools. 

\section{Validation against existing models}
\label{sec:comparisons}
To verify the numerical accuracy of the simulation, results from \df\ were compared against those from two analytical and two numerical models. In Section \ref{ssec:Depletion_approx} current-voltage characteristics obtained using analytical and numerical solutions for a p-n junction solar cell are compared. In Section \ref{ssec:TPV} the simulation's time integration is verified by calculating the transient photovoltage response of a single, field-free layer and comparing it to the solution obtained using a zero-dimensional kinetic model. In Section \ref{sec:3_layer_verification}, numerical solutions for three-layer, dual heterojunction devices obtained using \df\ are compared with those from the Advanced Semiconductor Analysis (ASA) simulation tool, an established, commercially available package.\cite{Zeman2011}. Finally, in Section \ref{ssec:comparison_ionmonger}, \JV\ characteristics calculated using \df\ for devices dominated by bulk and interfacial recombination processes are compared with those of \im\ \cite{Courtier2019a}, a recently published, free-to-use, mixed ionic-electronic carrier semiconductor device simulator.

The location of the \texttt{MATLAB} scripts and the parameter sets used to obtain the results in this section can be found in the \SI , Section \ref{sec:SI_Comparisons}.

\subsection{The depletion approximation for a p-n junction}
\label{ssec:Depletion_approx}

\begin{figure}[t!]
\centering
	\includegraphics[width=0.48\textwidth]{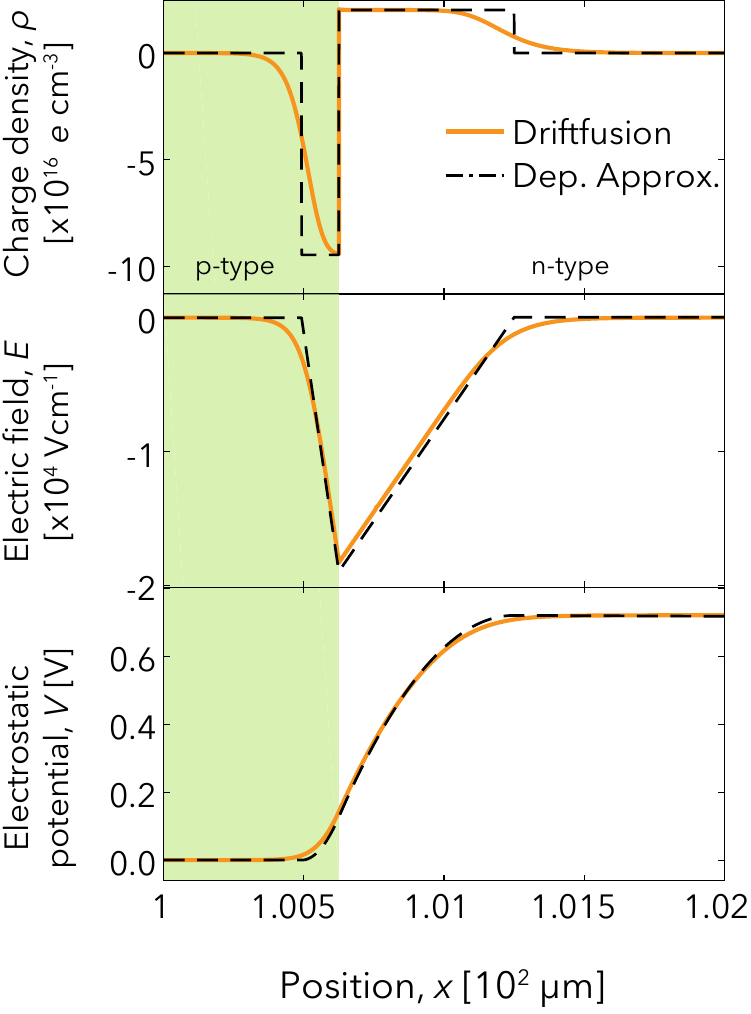}
	\caption[Analytical and numerical solutions of a p-n junction]{\textbf{Analytical and numerical solutions of a p-n junction}. Solutions obtained from \df\ (solid orange curve) and the depletion approximation (Dep. Approx., dashed black curve) for a p-n junction with $E_g = 1.12$ eV and $N_\mathrm{A} = 9.47 \times 10^{15}$ cm$^{-3}$, $N_\mathrm{D} = 2.01 \times 10^{15}$ cm$^{-3}$. Green and white regions indicate the n-type and p-type layers respectively. The complete parameter sets for the simulations are given in Tables \ref{tbl:Par_dep_approx_layers} and \ref{tbl:Par_dep_approx_device_wide}.}
	\label{fig:pn_rho_F_Phi}
\end{figure}

\paragraph{The p-n junction depletion approximation} The Depletion Approximation (DA) allows the continuity equations and Poisson's Equation (Equations \ref{eq:cont_n}, \ref{eq:cont_p} and \ref{eq:Poisson}) to be solved analytically for a p-n homojunction.\cite{Shockley1949} The depletion region at the junction of the device is assumed to have zero free carriers such that the space charge density $\rho$ can be described using a step function with magnitude equal to the background doping density (see Figure \ref{fig:pn_rho_F_Phi}, top panel). Transport and recombination of free carriers in the depletion region is also ignored
. Poisson's equation can then be solved by applying fixed carrier density ($p(x=-\infty) = p_0$ and $n(x=\infty) = n_0$) and zero-field boundary conditions to obtain the depletion widths for n- and p-type regions, $w_{\mathrm{n}}$ and $w_{\mathrm{p}}$ respectively:\cite{Sze1981}

\begin{equation} \label{eq:wn}
w_{\mathrm{n}} = \dfrac{N_{\mathrm{A}}}{N_{\mathrm{A}} + N_{\mathrm{D}}}\sqrt{\dfrac{2\varepsilon_r \varepsilon_\mathrm{0} V_{\mathrm{bi}}}{q\left( \dfrac{1}{N_{\mathrm{A}}} +\dfrac{1}{N_{\mathrm{D}}} \right)}}
\end{equation}

\begin{equation} \label{eq:wp}
w_{\mathrm{p}} = \dfrac{N_{\mathrm{D}}}{N_{\mathrm{A}} + N_{\mathrm{D}}}\sqrt{\dfrac{2\varepsilon_r \varepsilon_\mathrm{0} V_{\mathrm{bi}}}{q\left( \dfrac{1}{N_{\mathrm{A}}} +\dfrac{1}{N_{\mathrm{D}}} \right)}}
\end{equation}

Solving the DA for the current flowing across the junction under the assumption that the diffusion length of both carriers is significantly greater than the device thickness ($L_{n,p} >> d)$ yields the \textit{Shockley diode equation}:

\begin{equation} \label{eq:Ideal_diode}
J = J_0\left(\exp \left( {\frac{qV_\mathrm{app}}{k_\mathrm{B}T}} \right) -1\right) - J_{\mathrm{SC}},
\end{equation}

\noindent where \Jsc\ and $J_0$ are the short circuit and dark saturation current densities respectively. Here we use the convention that a positive applied (forward) bias generates a positive current flowing across the junction.

To make the comparison between numerical and analytical current-voltage (\JV) characteristics, values for $J_0$ and \Jsc\ need to be related to input parameters for the simulation. $J_0$ embodies the recombination characteristics of the device. For the contribution to recombination in the quasi-neutral region, $J_0$ can be related to the electron and hole diffusion lengths $L_n$ and $L_p$, minority carrier lifetimes $\tau_n$ and $\tau_p$, and diffusion coefficients $D_n$ and $D_p$, according to:\cite{Sze1981}

\begin{equation} \label{eq:J0}
J_0 = \dfrac{q D_p p_{0,n-type}}{L_p} + \dfrac{q D_n n_{0,p-type}}{L_n}			
\end{equation}
\\
where $L_p = \sqrt{\tau_p D_p}$ and $L_n = \sqrt{\tau_n D_n}$.

The material band gap and AM$1.5$ solar spectrum were used to calculate the theoretical maximum current density $J_{\mathrm{SC,max}}$ and corresponding uniform generation rate throughout the depletion region $g_0$. Figure \ref{fig:AM1p5_Jsc} of the \SI\ shows the AM1.5 Global Tilt solar spectrum obtained from Ref.\cite{Burkhard2010a} used for the calculation. 

The limiting short circuit photocurrent for a perfectly absorbing semiconductor of band gap $E_{\mathrm{g}}$ is given by:\cite{Nelson2003}

\begin{equation} \label{eq:Jscmax}
J_{\mathrm{SC}}(E_{\mathrm{g}}) =  q \int_{0}^{\infty} \eta(E_\gamma) \phi _0(E_\gamma) dE_\gamma,
\end{equation}

\noindent where $\eta$ is the external quantum efficiency. If $\eta = 1$ for photon energies $E_\gamma \geq E_\mathrm{g}$, and $\eta = 0$ for $E_\gamma < E_\mathrm{g}$, the maximum theoretically achievable short circuit current for a single junction $J_{\mathrm{SC,max}}$ is given by the integral of $\phi _0$ from the bandgap energy to infinity. Figure \ref{fig:AM1p5_Jsc} of the \SI\ shows the maximum achievable current density $J_{\mathrm{SC,max}}$ over a range of band gap energies. For the comparison we use the bandgap of silicon $E_\mathrm{g} = 1.12$ eV resulting in $J_{\mathrm{SC,max}} = 42.7$ mA cm$^{-2}$.

\paragraph{Simulation methods} A p-n junction was devised in \df\ with very thick n- and p-type layers ($\approx 100\ \mu$m, $p_0 \approx N_\mathrm{A} = 9.47 \times 10^{15}$ cm$^{-3}$, $n_0 \approx N_\mathrm{D} = 2.01 \times 10^{15}$ cm$^{-3}$) to approximate the assumptions and boundary conditions used in the DA. Furthermore, ionic carriers and trapped electronic charges are not included in the simulations in this section. The surface recombination velocity was set to zero for minority carriers at both boundaries. A special recombination scheme was implemented using the following simplified first order expressions:
\begin{equation} \label{eq:pn_recntype}
r = \dfrac{n-n_0}{\tau _n} \quad	\mathrm{for} \quad x < d/2 - w_p
\end{equation}
\begin{equation} \label{eq:pn_recptype}
r = \dfrac{p-p_0}{\tau _p} \quad	\mathrm{for} \quad x > d/2 + w_n,
\end{equation}
\\
where $\tau _n$ and $\tau _p$ are the electron and hole lifetimes respectively. For simplicity, $\tau _n$ and  $\tau _p$ were set equal to one another and, for consistency with the DA, recombination was switched off in the depletion region.

To convert the value obtained for $J_\mathrm{SC,max}$ into a uniform carrier generation rate, the short circuit flux density ($j_\mathrm{SC,max} = J_\mathrm{SC,max}/q$) was divided by the depletion region thickness $d_\mathrm{DR}$, yielding $g_0 =  j_\mathrm{SC,max}/d_\mathrm{DR}$. The complete parameter sets for the simulations in this section are given in Tables \ref{tbl:Par_dep_approx_layers} and \ref{tbl:Par_dep_approx_device_wide}.

\paragraph{Results} Both the analytical and numerical solutions for the space charge density, electric field, and electric potential are shown in Figure \ref{fig:pn_rho_F_Phi}: The space charge widths, field strength and potential profiles all show good agreement.

\begin{figure}[t]
\centering
	\includegraphics[width=0.48\textwidth]{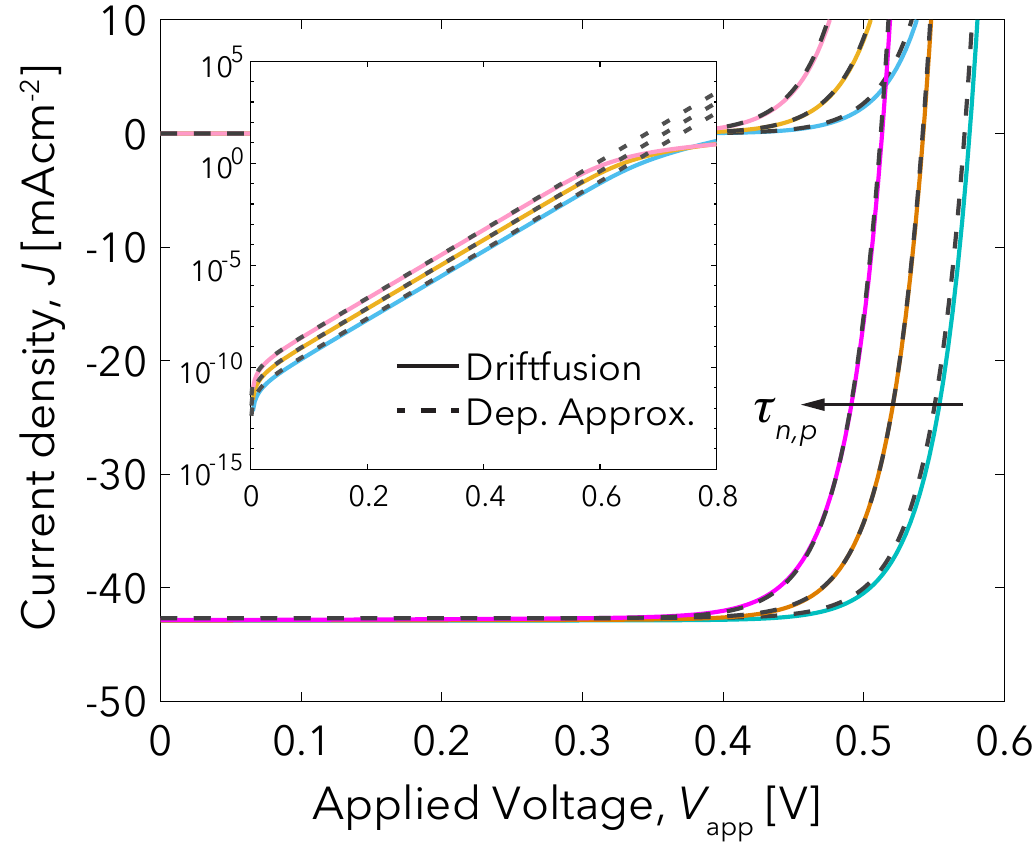}
	\caption[Comparison of current-voltage characteristics obtained using \df\ and the Depletion Approximation for a p-n junction]{\textbf{Comparison of current-voltage characteristics obtained using \df\ and the Depletion Approximation for a p-n junction.} Current-voltage characteristics for numerical and analytical solutions for a p-n junction with $\tau _n = \tau _p = 10^{-6}, 10^{-7}$, and $10^{-8}$ s. Results from \df\ and the Depletion Approximation (Dep. Approx.) are denoted by solid and dashed lines respectively. The complete parameter sets for the simulations are given in Tables \ref{tbl:Par_dep_approx_layers} and \ref{tbl:Par_dep_approx_device_wide}. The dark currents are shown on a logarithmic scale in the inset.}
	\label{fig:pn_JV_Voc}
\end{figure}

Figure \ref{fig:pn_JV_Voc} shows the light and dark current-voltage curves for the analytical solution obtained using equation \ref{eq:Ideal_diode}, as compared to the numerical solutions from \df\ for three values of $\tau_{n,p}$. For $\tau_{n,p} = 10^{-6}$ and $\tau_{n,p} = 10^{-7}$ the agreement is very good, even at current densities as low as $10^{-14}$ A cm$^{-2}$. The solutions begin to diverge at $\tau_{n,p} = 10^{-8}$ s, for which $L_{n,p} = 2.3$ $\mu$m such that $L_{n,p} << d$ and the underlying assumptions of the DA are no longer valid. The deviation from the ideal model in \df\ at higher current densities (Figure \ref{fig:pn_JV_Voc}, inset) is expected due to the absence of series resistance in the DA.

\subsection{Transient photovoltage response of a single layer field-free device}
\label{ssec:TPV}

\paragraph{Analytical methods} To verify the time-dependence of the solution from \df , the transient photovoltage (TPV) response for a field-free slab of intrinsic semiconductor was calculated numerically and compared to results from a zero-dimensional (0-D) kinetic model. During a TPV experiment the device is illuminated at open circuit with a constant bias light and pulsed with an optical excitation source such as to produce a small additional photovoltage $\Delta V$. As detailed in \SI\ Section \ref{sec:TPV_derivation}, it can be shown using a kinetic model that $\Delta V$ is given by:

\begin{multline} 	\label{eq:Delta_V_pulse}
\Delta V_\mathrm{OC} = \dfrac{2k_\mathrm{B}T}{q n_\mathrm{OC}}\dfrac{\Delta g}{k_\mathrm{TPV}} (1 - \exp(-k_\mathrm{TPV}(t+t_\mathrm{pulse})))		\\
\mathrm{for}\ -t_\mathrm{pulse} < t \leqslant 0,
\end{multline}

\begin{multline} 	\label{eq:Voc_kt}
\Delta V_\mathrm{OC}= \dfrac{2k_\mathrm{B}T}{q n_\mathrm{OC}}\exp(-k_\mathrm{TPV}t) \\
\mathrm{for}\ t >0,
\end{multline}

\noindent where $t_\mathrm{pulse}$ is the length of the laser pulse, $\Delta g$ is the additional volumetric generation rate due to the excitation pulse, $n_\mathrm{OC}$ is the steady-state open circuit carrier density, and $k_\mathrm{TPV}$ is the decay rate constant of the TPV signal. 

\paragraph{Simulation methods} A $100$ nm field-free single layer of semiconductor with $E_g = 1.6$ eV was simulated in \df\ with the zero flux density boundary conditions for electronic carriers representing perfect blocking contacts. The constant volumetric generation rate was set to $g_0 = 1.89\times10^{21}$ \rateunit\ $ = 1$ sun equivalent based on the integrated photon flux density for the AM$1.5$G solar spectrum and a step function absorption. In this special case the splitting of the quasi-Fermi levels in the simulation is solely attributable to changes in chemical potential, such that no series resistance term is needed for the potential boundary conditions. It should be noted that for instances where an electric field is present in the device these boundary conditions will not represent an open circuit condition; a high value for $R_s$ or a mirrored cell approach (see Ref. \cite{Calado2016} for details) should be used instead. The second order band-to-band recombination coefficient was set to $B = 10^{-10}$ cm$^3$s$^{-1}$ and SRH recombination was switched off. Since the underlying kinetic theory demands that the TPV perturbation must be small such that the additional carrier density $\Delta n << n_\mathrm{OC}$, we used $t_\mathrm{pulse} = 1\ \mu$s and set the pulse intensity equivalent to be $20 \%$ of the bias light intensity. The complete parameter sets for the simulations are given in Tables \ref{tbl:Par_tpv_layers} and \ref{tbl:Par_tpv_device_wide}. 

\begin{figure}[t]
\centering
\includegraphics[width=0.46\textwidth]{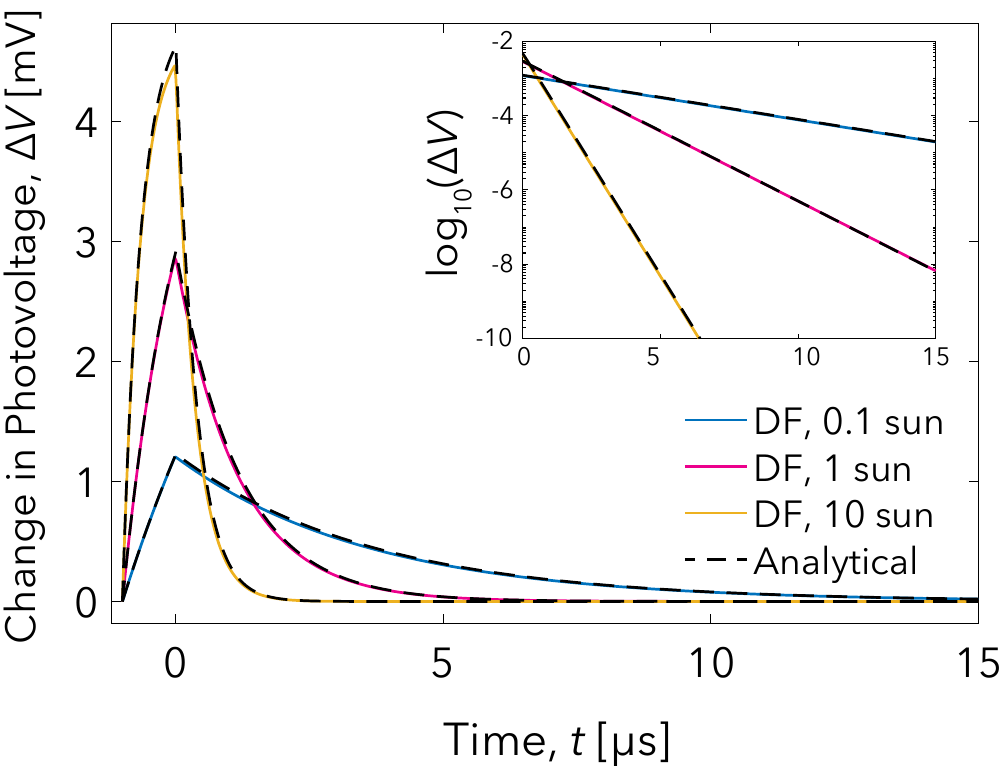}
\caption[Zero-dimensional theoretical prediction versus 1-D numerical drift-diffusion simulations for the transient photovoltage response of a single layer of semiconductor.]{\textbf{Zero-dimensional theoretical prediction versus 1-D numerical drift-diffusion simulations for the transient photovoltage response of a single layer of semiconductor.}  Change in photovoltage as a function of time calculated using \df\ (DF, solid curves) and a zero-dimensional kinetic model (Analytical, black dashed curves) at bias light intensities of $0.1$, $1$, and $10$ sun equivalent. The pulse length was $1\ \mu$s and the pulse intensity was set to be $20 \%$ of the bias intensity in each instance. The inset shows the results on a $\log_{10}$ scale. The complete parameter sets for the simulations are given in Tables \ref{tbl:Par_tpv_layers} and \ref{tbl:Par_tpv_device_wide}.}
\label{fig:0D_DD_TPV_Comp}
\end{figure}

\paragraph{Results} A comparison of the results from the analytical and simulation models for bias light intensities of $0.1$, $1$, and $10$ sun equivalent is shown in Figure \ref{fig:0D_DD_TPV_Comp}. The steady-state charge carrier densities and open circuit voltage obtained using \df\ agreed to within $10$ decimal places with the analytical values calculated using Equations \ref{eq:n_btb} and \ref{eq:Voc_zero_D} ($n = 4.35\times10^{15} $ cm$^{-3}$,  $V_\mathrm{OC} = 1.08$ V at 1 sun equivalent). The transient photovoltage perturbations also behaved as predicted, with the rate constants extracted from fitting the TPV decays correct to within $3$ significant figures of the values calculated using the analytical expression in Equation \ref{eq:Voc_kt}.

\FloatBarrier
\subsection{Numerical solution for three-layer devices with electronic carriers}	\label{sec:3_layer_verification}
\label{sec:ASA_comparison}

\paragraph{Simulation methods} To verify that the graded treatment of the interfaces in \df\ produces similar results to models that use abrupt interfaces, a HTL/absorber/ETL device was simulated using both \df\ and the Advanced Semiconductor Analysis (ASA) simulation tool.\cite{Zeman2011} To maintain consistent layer dimensions the linear grid spacing for the ASA simulations and the interface thickness in \df\ were set to be $1$ nm.

The base material parameters for the active layer were based loosely on those for a perovskite material excluding mobile ionic charge. The parameters for the contact layers were not chosen to simulate real materials but rather to produce a large built-in potential and to vary as many properties as possible including the layer thickness, dielectric constants, recombination coefficients, mobilities etc. For optical generation the Beer-Lambert option (without back contact reflection) was chosen and the same optical constant and incident photon flux density spectrum data were used in both simulators. Four different parameter sets (PS) were compared, the key differences for which are summarised in the first four columns of Table \ref{tab:DF_vs_ASA_params}.

\begin{table*}
\centering
\caption{Summary of the key simulation parameters for comparison of \df\ with ASA. CB and VB denote the conduction and valence bands respectively.}
\label{tab:DF_vs_ASA_params}       
\begin{tabular}{p{1cm}p{5cm}p{1.4cm}p{1.8cm}p{1.4cm}p{1.4cm}p{1.4cm}}
\hline\noalign{\smallskip}
Parameter set	&	Description		&	Built-in voltage (V) 	&	Active layer thickness (nm) &  \multicolumn{3}{c}{CB and VB effective density of states (\densunit	)} \\
&&&& HTL  & Absorber  &  ETL \\
\noalign{\smallskip}\hline\noalign{\smallskip}
1a 	& 3-layer device based on MAPI active layer 	& 1.05		&	400		&	$10^{19}$ 	&	$10^{18}$ 	&	$10^{19}$	\\
1b 	& As 1a with thinner active layer 				& 1.05		&	200		&	$10^{19}$ 	&	$10^{18}$ 	&	$10^{19}$	\\
2a 	& Randomised layer properties					& 0.6		&	200	 	& 	$10^{19}$ 	&	$10^{18}$ 	&	$10^{20}$ 	\\
2b 	& As 2a with uniform eDOS all layers 			& 0.6		&	200		&	$10^{18}$	& 	$10^{18}$	&	$10^{18}$	\\
\noalign{\smallskip}\hline
\end{tabular}
\end{table*}

The complete parameter sets for the simulated devices are given in Tables \ref{tbl:Par_ASA1_layer} -  \ref{tbl:Par_ASA2_device_wide}.

\paragraph{Results: Beer-Lambert optical model} The results for the integrated generation rate profiles are shown in Figure \ref{fig:DF_vs_ASA_gen}. Despite the wavelength-dependent generation rates appearing to be very closely matched between the two simulators (Figure \ref{fig:DF_vs_ASA_beer_lambert}), the integration across photon energies resulted in marginally different generation profiles in the two simulations corresponding to a total difference in generation current of $1.24$ \Junit . As a means to ensure that the input generation profiles were identical, the generation profile from ASA was inserted into the \df\ parameters objects using the method described in Section \ref{sssec:generation_function}.

\paragraph{Results: Current-voltage characteristics} To compare the current outputs from the different simulation tools current-voltage scans were performed from $V_\mathrm{app} = 0$ to $1.3$ V. Since ASA solves for the steady-state current, the \JV\ scan rate was set to be low ($k_\mathrm{scan} = 10^{-10}$ V s$^{-1}$) in \df\ to minimise contributions from the displacement current.

\begin{figure}
\centering
	\includegraphics[width=0.48\textwidth]{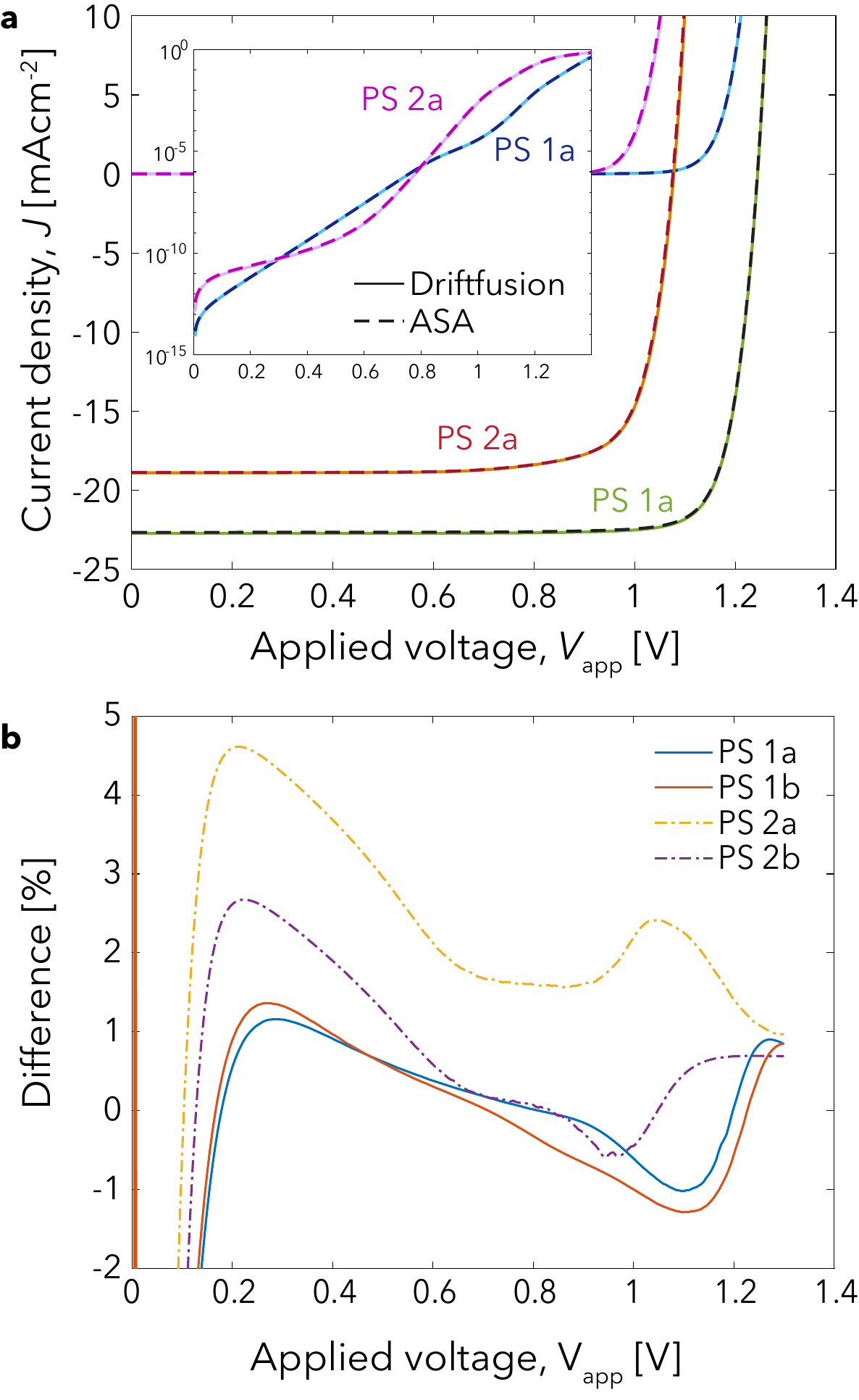}
	\caption[Comparison of dark and light current-voltage characteristics calculated by \df\ and ASA]{\textbf{Comparison of dark and light current-voltage characteristics calculated by \df\ and ASA} \textbf{a} Current-voltage characteristics for Parameter Sets (PS) $1$a and $2$a. Inset dark currents shown on log scale. Dashed and solid lines indicate the current calculated using \df\ and ASA respectively. \textbf{b} Percentage difference between dark current calculated using \df\ and ASA for the 4 different parameter sets investigated. The complete parameter sets for the simulations are given in Tables \ref{tbl:Par_ASA1_layer} -  \ref{tbl:Par_ASA2_device_wide}.}
	\label{fig:DF_vs_ASA_JV_error_v2}
\end{figure}

Figure \ref{fig:DF_vs_ASA_JV_error_v2}a shows a comparison of the \JV\ characteristics obtained from the two simulation tools for Parameter Sets (PS) 1a and 2a. While the different parameter sets result in distinctly different characteristics for the two devices, the data show excellent agreement between the two simulators despite the significant difference in discretisation schemes and interface treatment.

\begin{figure}
\centering
	\includegraphics[width=0.46\textwidth]{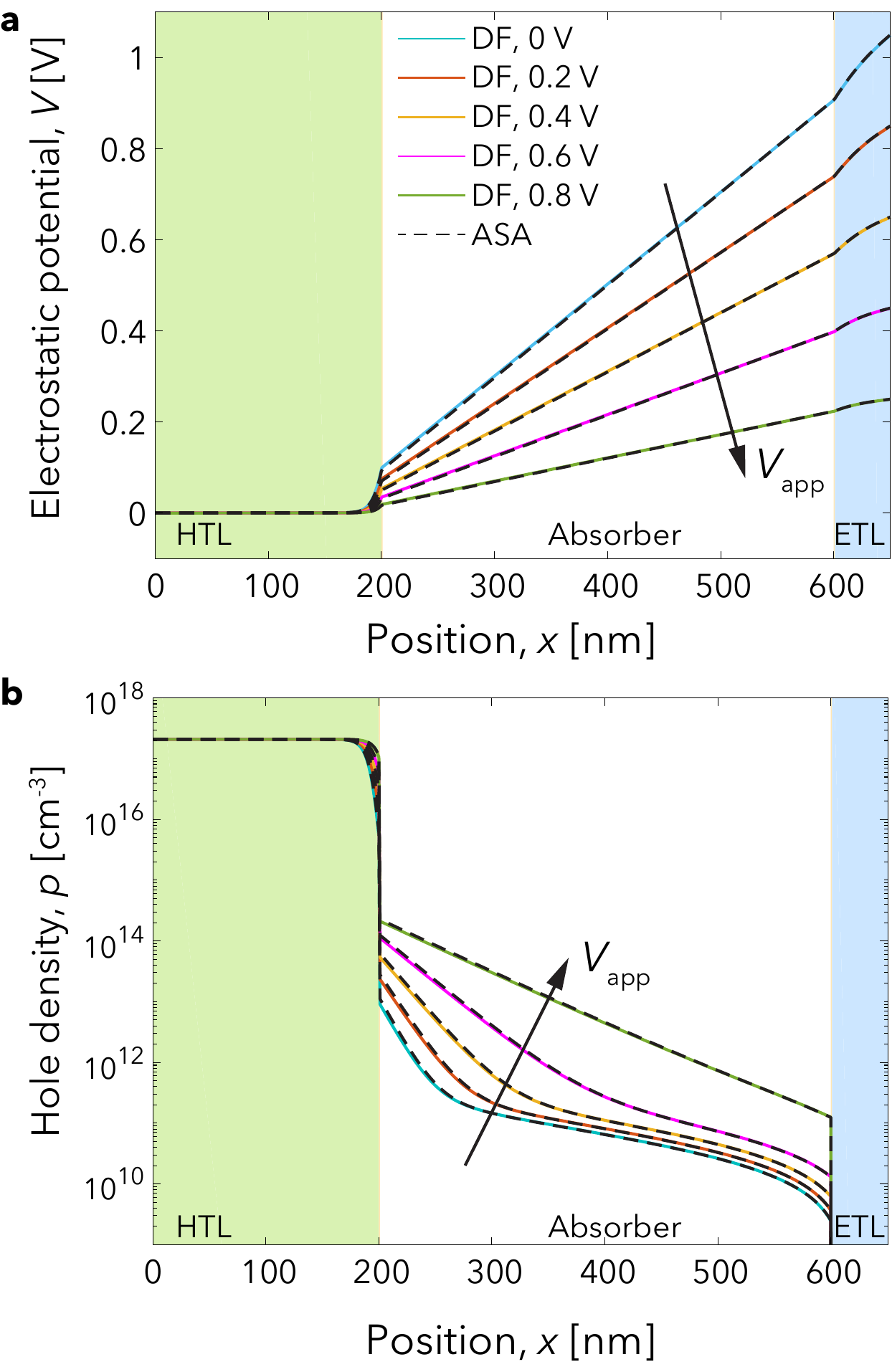}
	\caption[Comparison of the electrostatic potential and hole density profiles during a \JV\ scan for a three-layer device calculated by \df\ and ASA.]{\textbf{Comparison of the electrostatic potential and hole density profiles during a \JV\ scan for a three-layer device calculated by \df\ and ASA.} Results from \df\ indicated by solid lines whereas results from ASA are indicated by dashed lines. Corresponding electron densities are given in Supplemental Information Figure \ref{fig:DF_vs_ASA_nx_m1_1S}.}
	\label{fig:DF_vs_ASA_Vx_px_m1_1S}
\end{figure}

Closer examination of the results from the two simulators (Figure \ref{fig:DF_vs_ASA_JV_error_v2}b) reveals that, the percentage difference (calculated as $100 \times (J_\mathrm{ASA} - J_\mathrm{DF})/J_\mathrm{ASA}$) for PS 1a is on the order of $1$ \% for current densities beyond $J = 10^{-12}$ \Jun . Halving the active layer thickness has little impact on this difference (PS 1b). The percentage difference calculated for PS 2a, however, is much greater, with a maximum of $\approx 5$ \% for current densities $J > 10^{-12}$ \Jun . 
The larger difference between the two simulators in this instance can be attributed to a change of over $7$ orders of magnitude in the electron density at the absorber-ETL interface (Figure \ref{fig:npx_DF_vs_ASA_ps2_0V}) due to both a transition in the conduction band eDOS from $N_\mathrm{CB} = 10^{18}$ to $10^{20}$ \densunit\ and a change in the conduction band energy of $0.3$ eV. Under these circumstances the difference in discretisation schemes between the two tools becomes apparent: the linear discretisation method used in the PDEPE and \df\ cannot calculate the change in carrier densities within the interfaces to as high a degree of accuracy as the internal boundary conditions used in ASA (see Section \ref{ssec:linear_discretisation} for further details). The deviation can be reduced significantly, however, by using a uniform eDOS of $N_\mathrm{CB} = 10^{18}$ \densunit\ across all layers as the results for PS 2b show. With respect to device characteristics, despite the percentage difference in results for PS 2a, the key metrics of the \JV\ curve such as the \Voc , ideality factor and fill-factor are all preserved.
Figure \ref{fig:DF_vs_ASA_Vx_px_m1_1S} shows a comparison of the electrostatic potential and hole density profiles calculated using \df\ and ASA for PS 1a under illumination at increasing applied bias. The electron density is given in the \SI , Figure \ref{fig:DF_vs_ASA_nx_m1_1S}. The agreement here between the solutions is excellent taking into consideration the difference in treatment of the interfaces between the two simulators.

\FloatBarrier
\subsection{Numerical solution for three-layer devices with electronic and mobile ionic carriers}	
\label{ssec:comparison_ionmonger}

Courtier et al. recently published \im ,\cite{Courtier2019a} a three-layer (HTL/absorber/ETL) drift-diffusion simulation tool for modelling perovskite solar cells which solves for coupled electron, hole and cation carrier distributions. In contrast to \df , \im\ uses abrupt interfaces and solves carriers and the electrostatic potential for each of the three device layers simultaneously. The advantage to this approach is that boundary conditions can be established between the absorber and transport layers, such that interfacial recombination between electrons from one material and holes from another can be evaluated at the same grid point. The disadvantage is that $8$ variables are solved for simultaneously (minority carriers are not included in the transport regions) as opposed to $4$ in \df , for an equivalent three-layer device with a single mobile ionic species. Here we compare results obtained from \df\ with those from \im\ for devices with similar parameter sets. 

\paragraph{Methods} We modelled three layer perovskite HTL/absorber/ETL solar cells with electronic and mobile ionic carriers in the absorber layer and electronic carriers only in the HTL and ETL using \im\ and \df . For the \im\ simulations only holes were solved for within the HTL and electrons within the ETL. For the \df\ simulations all carriers were solved for in all regions with the ionic carrier mobility set to zero in the HTL, ETL and interfacial regions. Since the ionic carriers are inert and their charge is compensated by a static background charge density (see Equation \ref{eq:Poisson}) they have no effect in these regions. For the comparison we ran \JV\ scan simulations at a variety of scan speeds using similar parameter sets to those published in Ref. \cite{Courtier2019}: a device dominated by bulk recombination and a device dominated by interfacial recombination using the volumetric surface recombination scheme described in Section \ref{ssec:recombination_interfaces}.

\begin{figure}
\centering
\includegraphics[width=0.48\textwidth]{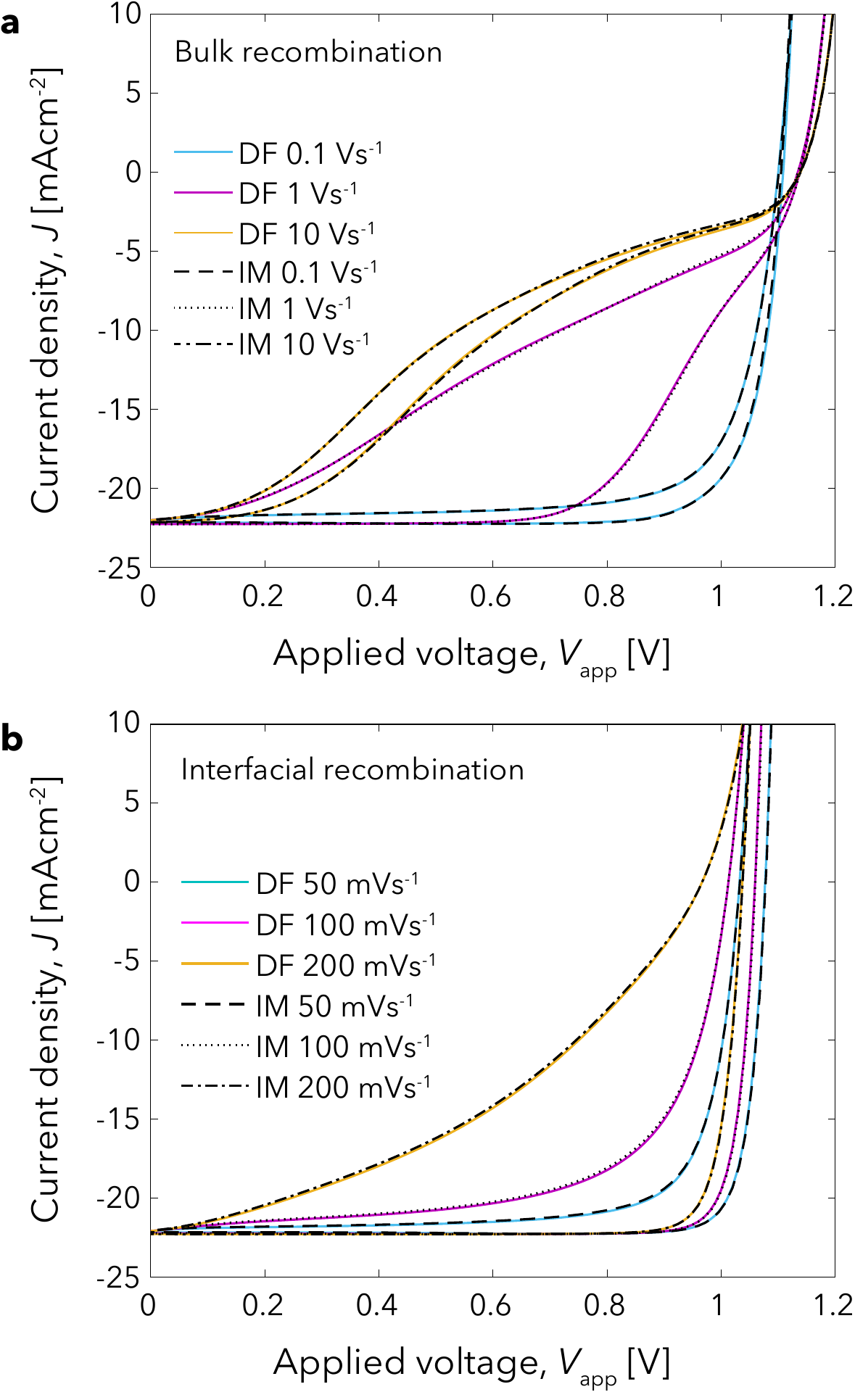}
\caption[Comparison of results calculated using \df\ and \im\ for three-layer solar cells including mobile ionic carriers in the absorber layer.]{\textbf{Comparison of results calculated using \df\ and \im\ for three-layer solar cells including mobile ionic carriers in the absorber layer.} (\textbf{a}) Current-voltage scans at $k_\mathrm{scan}=0.1$, $1$, and $10$ V s$^{-1}$ for a device dominated by bulk recombination. (\textbf{b}) Current-voltage scans at $k_\mathrm{scan}=50$, $100$, and $200$ mV s$^{-1}$ for a device dominated by interfacial recombination using the scheme described in Section \ref{ssec:recombination_interfaces}. The complete parameter sets for the simulations are given in Tables \ref{tbl:Par_ion_monger_layers} - \ref{tbl:Par_ion_monger_device_wide}.}
\label{fig:DF_vs_IM}
\end{figure}

\paragraph{Results: Current-voltage characteristics} Figure \ref{fig:DF_vs_IM}a and \ref{fig:DF_vs_IM}b show the \JV\ results from \df\ (solid coloured curves) and \im\ (black dashed curves) for the bulk and interfacial recombination dominated devices at varying voltage scan rates. The electrostatic potential profile, ionic carrier accumulation and electronic carrier profiles during the $1$ V s$^{-1}$ forward scan for the bulk recombination device, and at $100$ mV s$^{-1}$ for in the interfacial recombination dominated device, at increasing applied bias are given in the \SI\ Figures \ref{fig:DF_vs_IM_solx} and \ref{fig:DF_vs_IM_solx_IR} respectively. For both sets of parameters the results obtained from the two simulations are very similar with marginal differences in the calculated current outputs. This variance arises principally from the treatment of electronic currents across interfaces, differences in the spatial mesh, and calculation of ionic carrier densities throughout all layers of the device in \df\ as opposed to \im , which introduces small additional integration errors into the space charge density (\SI\ Figure \ref{fig:Ion_interfaces}). For the device dominated by interfacial recombination (Figure \ref{fig:DF_vs_IM}b), additional errors are also introduced by the volumetric surface recombination scheme. Figure \ref{fig:DF_vs_IM_rec_flux_compare} shows that, while the scheme is self-consistent, differences arise from the surface carrier densities (specifically the electron density at the active layer-HTL interface). These differences are accentuated by increasing the energetic barriers to minority carriers from $0.4$ to $0.8$ eV (Equations \ref{eq:alpha_interface} and \ref{eq:beta_interface}) as shown in \SI\ Figure \ref{fig:DF_vs_IM_interface_points}. The differences can however be reduced by increasing the interface thickness and using a larger number of interface mesh points (Figure \ref{fig:DF_vs_IM_interface_points}). To this end, consistency with established analytical models that use abrupt interfaces is somewhat sacrificed in \df\ in favour of greater flexibility, which enables the physical models to be easily edited, devices with any number of material layers to be simulated and a range of interface-specific properties and grading functions to be specified.\\

We have verified \df\ against two analytical and two existing numerical models and found that in all cases the calculated results are in reasonably good agreement. The results show that the discrete interface approach produces results comparable to models with abrupt interfaces, albeit with marginal errors introduced owing to the combination of the interface treatment and the linear discretisation scheme used in \df .

\section{Conclusions}
\label{sec:conclusions}
We have developed an efficient and powerful one-dimensional drift-diffusion simulation tool for modelling semiconductor devices with mixed ionic-electronic conducting layers based on \texttt{MATLAB}'s \texttt{pdepe} toolbox. Distinct from existing codes, in addition to electronic carriers, \df\ can include up to two ionic carrier species and virtually any number of material layers. This flexibility is made possible by a discrete interlayer interface approach whereby the material properties can be graded between two adjoining semiconductor layers. This method has the added advantage that interface-specific properties can easily be specified.

The default physical models underlying the simulation were described and analytical approximations to the carrier densities and fluxes within the interfacial regions were given. Using these solutions, a method for approximating surface recombination within interfacial regions using a volumetric recombination scheme was derived.

The system architecture was presented and the processes by which users can define device properties and change the simulation's physical model were outlined. Protocol functions, determining time-dependent voltage and light conditions, were described as well as the simulation solution structures and how to use built-in analysis and plotting functions to calculate and visualise outputs. A step-by-step guide was presented detailing how to create a device, find an equilibrium solution, and calculate current-voltage characteristics using a cyclic voltammogram protocol. Lastly, advanced features, including how to calculate multiple solutions in parallel, were briefly introduced.

\df\ was verified by comparing calculated solutions with two analytical and two existing numerical models which tested different aspects of the simulation. In all cases the agreement was good and the general device behaviour was reproduced. The discrete treatment of the interfaces resulted in small variations in the calculated currents as compared to other simulators using abrupt layer boundaries.

The ease with which the underlying models can be changed and new material layers can be introduced places \df\ in a unique space compared to other free-to-use simulation codes for which an intricate knowledge of the numerical mechanics is required to adapt the physical models and device architecture. By making \df\ both accessible and free-to-use our hope is that this work will advance our collective understanding of mixed ionic-electronic conducting materials and devices.

\FloatBarrier

\begin{acknowledgements}
The authors would like to thank the UK Engineering and Physical Sciences Research Council for funding this work (Grant No. EP/J002305/1, No. EP/M025020/1, No. EP/M014797/1, No. EP/N020863/1, No. EP/R020574/1, No. EP/R023581/ 1, and No. EP/L016702/1). J.N. thanks the European Research Council for support under the European Union's Horizon 2020 research and innovation program (grant agreement No. 742708). We would also like to thank: Emilio Palomares and Davide Moia for lending the expertise of their students to work on this project, and Diego Alonso \' Alvarez for his invaluable advice on preparing and improving the code.
\end{acknowledgements}

%
\section*{Funding}
The authors would like to thank the UK Engineering and Physical Sciences Research Council (grant No. EP/J002305/1, EP/M025020/1, EP/M014797/1, EP/R020574/1, EP/R023581/1, and EP/L016702/1) and the European Union's Horizon 2020 research and innovation program grant agreement (grant No. 742708) for their support in funding this work.

\section*{Conflicts of interest}
The authors declare that they have no conflicts of interest.

\section*{Availability of data and material} 
The data presented in this work is available at reasonable request from the authors.

\section*{Code availability}
The simulation code is freely available at:\\
 \url{https://github.com/barnesgroupICL/Driftfusion}

\section*{Authors' contributions}
P.C., P.R.F.B., I.G., M.A., and B.H. contributed to the development of the simulation code. P.C., P.R.F.B., J.N., I.G., and M.A. wrote the manuscript. P.R.F.B. and J.N. supervised the development and publication of this work.

\bibliographystyle{spphys}       
\bibliography{library}   

\begin{thebibliography}{10}
\providecommand{\url}[1]{{#1}}
\providecommand{\urlprefix}{URL }
\expandafter\ifx\csname urlstyle\endcsname\relax
  \providecommand{\doi}[1]{DOI \discretionary{}{}{}#1}\else
  \providecommand{\doi}{DOI \discretionary{}{}{}\begingroup
  \urlstyle{rm}\Url}\fi

\bibitem{VanRoosbroeck1950}
W.~{Van Roosbroeck}, Bell System Technical Journal \textbf{29}(4), 560 (1950).
\newblock \doi{10.1002/j.1538-7305.1950.tb03653.x}

\bibitem{Foster2014}
J.M. Foster, H.J. Snaith, T.~Leijtens, G.~Richardson, Siam J. Appl. Math
  \textbf{74}(6), 1935 (2014).
\newblock \doi{10.1137/130934258}

\bibitem{VanReenen2015a}
S.~van Reenen, M.~Kemerink, H.J. Snaith, The Journal of Physical Chemistry
  Letters \textbf{6}, 3808 (2015).
\newblock \doi{10.1021/acs.jpclett.5b01645}

\bibitem{Richardson2016}
G.~Richardson, S.E. O'Kane, R.G. Niemann, T.A. Peltola, J.M. Foster, P.J.
  Cameron, A.B. Walker, Energy {\&} Environmental Science \textbf{9}, 1476
  (2016).
\newblock \doi{10.1039/C5EE02740C}

\bibitem{Neukom2017}
M.T. Neukom, S.~Z{\"{u}}fle, E.~Knapp, M.~Makha, R.~Hany, B.~Ruhstaller, Solar
  Energy Materials and Solar Cells \textbf{169}, 159 (2017).
\newblock \doi{10.1016/j.solmat.2017.05.021}

\bibitem{Sherkar2017a}
T.S. Sherkar, C.~Momblona, L.~Gil-Escrig, J.~{\'{A}}vila, M.~Sessolo, H.J.
  Bolink, L.J.A. Koster, ACS Energy Letters \textbf{2}(5), 1214 (2017).
\newblock \doi{10.1021/acsenergylett.7b00236}

\bibitem{Garcia-Rosell2018b}
M.~Garc{\'{i}}a-Rosell, A.~Bou, J.A. Jim{\'{e}}nez-Tejada, J.~Bisquert,
  P.~Lopez-Varo, Journal of Physical Chemistry C \textbf{122}(25), 13920
  (2018).
\newblock \doi{10.1021/acs.jpcc.8b01070}

\bibitem{Courtier2019}
N.E. Courtier, J.M. Cave, J.M. Foster, A.B. Walker, G.~Richardson, Energy {\&}
  Environmental Science \textbf{12}(1), 396 (2019).
\newblock \doi{10.1039/C8EE01576G}

\bibitem{Bertoluzzi2019}
L.~Bertoluzzi, C.C. Boyd, N.~Rolston, J.~Xu, R.~Prasanna, B.C. O'Regan, M.D.
  McGehee, Joule \textbf{4}(1), 109 (2019).
\newblock \doi{10.1016/j.joule.2019.10.003}

\bibitem{Huang2020}
Y.~Huang, P.~Lopez-Varo, B.~Geffroy, H.~Lee, J.E. Bour{\'{e}}e, A.~Mishra,
  P.~Baranek, A.~Rolland, L.~Pedesseau, J.M. Jancu, J.~Even, J.B. Puel,
  M.~Gueunier-Farret, Journal of Photonics for Energy \textbf{10}(02), 1
  (2020).
\newblock \doi{10.1117/1.jpe.10.024502}

\bibitem{Xiao2014a}
Z.~Xiao, Y.~Yuan, Y.~Shao, Q.~Wang, Q.~Dong, C.~Bi, P.~Sharma, A.~Gruverman,
  J.~Huang, Nature Materials \textbf{14}(February), 193  (2015).
\newblock \doi{10.1038/nmat4150}

\bibitem{Eames2015}
C.~Eames, J.M. Frost, P.R.F. Barnes, B.C. O'Regan, A.~Walsh, M.S. Islam, Nature
  Communications \textbf{6}, 7497 (2015).
\newblock \doi{10.1038/ncomms8497}

\bibitem{Yang2015}
T.Y. Yang, G.~Gregori, N.~Pellet, M.~Gr{\"{a}}tzel, J.~Maier, Angewandte Chemie
  \textbf{127}, 8016  (2015).
\newblock \doi{10.1002/ange.201500014}

\bibitem{Haruyama2015}
J.~Haruyama, K.~Sodeyama, L.~Han, Y.~Tateyama, Journal of the American Chemical
  Society \textbf{137}(32), 10048 (2015).
\newblock \doi{10.1021/jacs.5b03615}

\bibitem{Moia2019}
D.~Moia, I.~Gelmetti, P.~Calado, W.~Fisher, M.~Stringer, O.~Game, Y.~Hu,
  P.~Docampo, D.~Lidzey, E.~Palomares, J.~Nelson, P.R.F. Barnes, Energy
  Environ. Sci. \textbf{12}(4), 1296 (2019).
\newblock \doi{10.1039/C8EE02362J}

\bibitem{Calado2019_ideality}
P.~Calado, D.~Burkitt, J.~Yao, J.~Troughton, T.M. Watson, M.J. Carnie, A.M.
  Telford, B.C. O'Regan, J.~Nelson, P.R. Barnes, Physical Review Applied
  \textbf{11}(4), 044005 (2019).
\newblock \doi{10.1103/PhysRevApplied.11.044005}

\bibitem{OKane2017}
S.E.J. O'Kane, G.~Richardson, A.~Pockett, R.G. Niemann, J.M. Cave, N.~Sakai,
  G.E. Eperon, H.J. Snaith, J.M. Foster, P.J. Cameron, A.B. Walker, J. Mater.
  Chem. C \textbf{5}(2), 452 (2017).
\newblock \doi{10.1039/C6TC04964H}

\bibitem{Calado2016}
P.~Calado, A.M. Telford, D.~Bryant, X.~Li, J.~Nelson, B.C. O'Regan, P.R.
  Barnes, Nature Communications \textbf{7}, 13831 (2016).
\newblock \doi{10.1038/ncomms13831}

\bibitem{fluxim2019semiconducting}
{FLUXiM AG}.
\newblock {SETFOS: Semiconducting emissive thin film optics simulator} (2019)

\bibitem{Courtier2019a}
N.E. Courtier, J.M. Cave, A.B. Walker, G.~Richardson, J.M. Foster, Journal of
  Computational Electronics \textbf{18}(4), 1435 (2019).
\newblock \doi{10.1007/s10825-019-01396-2}

\bibitem{COMSOL2019}
{COMSOL AB}.
\newblock {COMSOL Multiphysics} (2019)

\bibitem{MATLAB2017}
{The Mathworks}.
\newblock {MATLAB} (2017)

\bibitem{Jacobs2018}
D.A. Jacobs, H.~Shen, F.~Pfeffer, J.~Peng, T.P. White, F.J. Beck, K.R.
  Catchpole, The Journal of Applied Physics \textbf{124}, 225702 (2018).
\newblock \doi{10.1063/1.5063259}

\bibitem{Tessler2020}
N.~Tessler, Y.~Vaynzof, ACS Energy Letters \textbf{5}, 1260 (2020).
\newblock \doi{10.1021/acsenergylett.0c00172}

\bibitem{Singh2021}
A.~Singh, W.~Kaiser, A.~Gagliardi, Solar Energy Materials and Solar Cells
  \textbf{221}(October 2020), 110912 (2021).
\newblock \doi{10.1016/j.solmat.2020.110912}

\bibitem{Calado2017}
P.~Calado, P.R.F. Barnes, I.~Gelmetti, M.~Azzouzi, B.~Hilton.
\newblock {Driftfusion} (2017).
\newblock \doi{10.5281/zenodo.3670155}.
\newblock \urlprefix\url{https://github.com/barnesgroupICL/Driftfusion}

\bibitem{Sze1981}
S.M. Sze, \emph{{Physics of semiconductor devices}}, 2nd edn. (John Wiley {\&}
  Sons, Inc., New York, 1981)

\bibitem{Nelson2003}
J.~Nelson, \emph{{The physics of solar cells}} (Imperial College Press, 2003)

\bibitem{calado2017transient}
P.~Calado, {Transient optoelectronic characterisation and simulation of
  perovskite solar cells}.
\newblock Ph.D. thesis, Imperial College London (2017).
\newblock \urlprefix\url{https://spiral.imperial.ac.uk/handle/10044/1/66894}

\bibitem{Barnes2011}
P.R.F. Barnes, A.Y. Anderson, J.R. Durrant, B.C. O'Regan, Phys. Chem. Chem.
  Phys. \textbf{13}, 5798 (2011).
\newblock \doi{10.1039/c0cp01554g}

\bibitem{pdepe2013}
L.F. Shampine, J.~Kierzenka.
\newblock {PDEPE} (2013)

\bibitem{Skeel1990}
R.D. Skeel, M.~Berzins, SIAM Journal on Scientific and Statistical Computing
  \textbf{11}(1), 1 (1990).
\newblock \doi{10.1137/0911001}

\bibitem{Walsh2015}
A.~Walsh, D.O. Scanlon, S.~Chen, X.G. Gong, S.H. Wei, Angewandte Chemie
  International Edition \textbf{54}, 1791 (2015).
\newblock \doi{10.1002/anie.201409740}

\bibitem{Kilic2007}
M.S. Kilic, M.Z. Bazant, A.~Ajdari, Physical Review E - Statistical, Nonlinear,
  and Soft Matter Physics \textbf{75}(2), 1 (2007).
\newblock \doi{10.1103/PhysRevE.75.021502}

\bibitem{Zeman2011}
M.~Zeman, J.~van~den Heuvel, M.~Kroom, J.~Willemen, B.~Pieters, J.~Krc,
  S.~Solntsev.
\newblock {Advanced Semiconductor Analysis (ASA)} (2011).
\newblock
  \urlprefix\url{https://www.tudelft.nl/en/ewi/over-de-faculteit/afdelingen/electrical-sustainable-energy/photovoltaic-materials-and-devices/software-platform/asa-software}

\bibitem{Zeghbroeck2011}
B.V. Zeghbroeck, \emph{{Principles of semiconductor devices}}, online edn.
  (University of Colorado Boulder, Colorado Boulder, 2011)

\bibitem{Shockley1952}
W.~Shockley, W.T. Read, Physical Review \textbf{87}(46), 835 (1952).
\newblock \doi{10.1103/PhysRev.87.835}

\bibitem{Tress2011}
W.~Tress, {Device physics of organic solar cells}.
\newblock Ph.D. thesis, Dresden University of Technology (2012)

\bibitem{Courtier2018c}
N.E. Courtier, G.~Richardson, J.M. Foster, Applied Mathematical Modelling
  \textbf{63}, 329 (2018).
\newblock \doi{10.1016/j.apm.2018.06.051}

\bibitem{Blakemore1982}
J.S. Blakemore, Solid State Electronics \textbf{25}(11), 1067 (1982).
\newblock \doi{10.1016/0038-1101(82)90143-5}

\bibitem{Farrell2017a}
P.~Farrell, T.~Koprucki, J.~Fuhrmann, Journal of Computational Physics
  \textbf{346}, 497 (2017).
\newblock \doi{10.1016/j.jcp.2017.06.023}

\bibitem{Alonso-Alvarez2018a}
D.~Alonso-{\'{A}}lvarez, T.~Wilson, P.~Pearce, M.~F{\"{u}}hrer, D.~Farrell,
  N.~Ekins-Daukes, Journal of Computational Electronics \textbf{17}, 1099
  (2018).
\newblock \doi{10.1007/s10825-018-1171-3}

\bibitem{Burkhard2010}
G.F. Burkhard, E.T. Hoke, M.D. McGehee, Advanced Materials \textbf{22}(30),
  3293 (2010).
\newblock \doi{10.1002/adma.201000883}

\bibitem{Belisle2016}
R.A. Belisle, W.H. Nguyen, A.R. Bowring, P.~Calado, X.~Li, S.J.C. Irvine, M.D.
  Mcgehee, P.R.F. Barnes, B.C. O'Regan, Energy {\&} Environmental Science
  \textbf{10}(1), 192 (2016).
\newblock \doi{10.1039/C6EE02914K}

\bibitem{Shockley1949}
W.~Shockley, Bell Labs Technical Journal \textbf{28}(3), 435 (1949).
\newblock \doi{10.1002/j.1538-7305.1949.tb03645.x}

\bibitem{Burkhard2010a}
G.F. Burkhard, E.T. Hoke, M.D. McGehee.
\newblock {Transfer Matrix Optical Modeling} (2011).
\newblock
  \urlprefix\url{https://web.stanford.edu/group/mcgehee/transfermatrix/}

\end{thebibliography}


\begin{thebibliography}{1}
\providecommand{\url}[1]{{#1}}
\providecommand{\urlprefix}{URL }
\expandafter\ifx\csname urlstyle\endcsname\relax
  \providecommand{\doi}[1]{DOI \discretionary{}{}{}#1}\else
  \providecommand{\doi}{DOI \discretionary{}{}{}\begingroup
  \urlstyle{rm}\Url}\fi

\bibitem{Calado2017}
P.~Calado, P.R.F. Barnes, I.~Gelmetti, M.~Azzouzi, B.~Hilton.
\newblock {Driftfusion} (2017).
\newblock \doi{10.5281/zenodo.3670155}.
\newblock \urlprefix\url{https://github.com/barnesgroupICL/Driftfusion}

\bibitem{Scharfetter1969}
D.L. Scharfetter, H.K. Gummel, IEEE Transactions on Electron Devices
  \textbf{ED-16}(1) (1969).
\newblock \doi{10.1109/T-ED.1969.16566}

\bibitem{Farrell2016a}
P.~Farrell, N.~Rotundo, D.H. Doan, M.~Kantner, J.~Fuhrmann, T.~Koprucki,
  {Numerical methods for drift-diffusion models}.
\newblock Tech. rep., Leibniz-Institut im Forschungsverbund Berlin, Berlin
  (2016)

\bibitem{NREL2017}
NREL.
\newblock {Reference Solar Spectral Irradiance: ASTM G-173} (2017).
\newblock \urlprefix\url{https://www.nrel.gov/grid/solar-resource/spectra.html}

\bibitem{calado2017transient}
P.~Calado, {Transient optoelectronic characterisation and simulation of
  perovskite solar cells}.
\newblock Ph.D. thesis, Imperial College London (2017).
\newblock \urlprefix\url{https://spiral.imperial.ac.uk/handle/10044/1/66894}

\bibitem{Nelson2003}
J.~Nelson, \emph{{The physics of solar cells}} (Imperial College Press, 2003)

\bibitem{Shuttle2008b}
C.G. Shuttle, {Recombination dynamics in polythiophene: fullerene solar cells}.
\newblock Ph.D. thesis, Imperial College London (2008)

\bibitem{Courtier2019a}
N.E. Courtier, J.M. Cave, A.B. Walker, G.~Richardson, J.M. Foster, Journal of
  Computational Electronics \textbf{18}(4), 1435 (2019).
\newblock \doi{10.1007/s10825-019-01396-2}

\bibitem{Courtier2019}
N.E. Courtier, J.M. Cave, J.M. Foster, A.B. Walker, G.~Richardson, Energy {\&}
  Environmental Science \textbf{12}(1), 396 (2019).
\newblock \doi{10.1039/C8EE01576G}

\end{thebibliography}

\end{document}


\title{Supplemental information for Driftfusion \thanks{UK Engineering and Physical Sciences Research Council grant No. EP/J002305/1, EP/M025020/1, EP/M014797/1, EP/R020574/1, EP/R023581/1, EP/L016702/1, EP/T028513/1 (ATIP) and European Union`s Horizon 2020 research and innovation program grant agreement No. 742708.}}
\subtitle{An open source code for simulating ordered semiconductor devices with mixed ionic-electronic conducting materials in one dimension}

\titlerunning{Driftfusion: An open source code for simulating ordered semiconductor devices}

\author{Philip Calado$^1$ \and Ilario Gelmetti$^2$ \and Benjamin Hilton$^1$ \and Mohammed Azzouzi$^1$
\and Jenny Nelson$^1$ \and Piers R. F. Barnes$^1$}


\institute{P. Calado	\\
              \email{p.calado13@imperial.ac.uk}             \\
$^1$ Department of Physics, Imperial College London, London SW7 2AZ, UK.	\\
$^2$ Institute of Chemical Research of Catalonia (ICIQ), Barcelona Institute of Science and Technology (BIST), Avda. Paisos Catalans 16, 43007 Tarragona, Spain.
}

\date{}
\maketitle

\section{Universal constants, abbreviations and symbols}
\begin{table}[h]
\caption{\textbf{Universal constants.}}
\label{tbl:constants}
\begin{tabular}{llll}
\textbf{Constant Name}		& \textbf{Symbol}  	& 	\textbf{\df\ Property Name}	 & \textbf{Value} \\
\hline\noalign{\smallskip}
Boltzmann constant 		& $k_\mathrm{B}$ 	& 	\texttt{kB}		& $8.62\times10^{-5}$ eV K$^{-1}$ \\
Charge of an electron 		& $q$ 				&	\texttt{e}	 	& $1.60\times10^{-19}$ C \\
Permittivity of free space & $\varepsilon_0$ 	& 	\texttt{epp0}	& $8.85\times10^{-12}$ m$^{-3}$kg$^{-1}$s$^4$A$^{2}$ \\ 
\end{tabular}
\end{table}

\clearpage
\begin{table}[h]
\caption{\textbf{Table of symbols.} $\dagger$The absorption and photon flux data are loaded directly from libraries according to the layer names defined in \texttt{stack} and the choice of light source using the properties \texttt{light\_source1} and \texttt{light\_source2}. $^*$Solely for the \texttt{Methods-depletion-approximation} branch. $^{**}$For the .csv file the name \texttt{thickness} can also be used. For a complete description of all user-defined and dependent properties please see the comments in the parameters class \texttt{pc}. For information on calculated outputs see the analysis class \texttt{dfana}. VSR denotes auto-generated coefficients used in the volumetric surface recombination scheme.}
\label{tbl:symbols}
\begin{tabular}{p{1.4cm}p{1.4cm}p{2.3cm}p{4.2cm}p{2cm}}
\textbf{Symbol}			& 	\textbf{\df\ Variable Name} &	\textbf{Variable type}	&	\textbf{Name}				& \textbf{Unit}	\\
\hline\noalign{\smallskip}
$\alpha _\mathrm{abs}$	& 	$\dagger$				&		-						&	Absorption coefficient			& cm$^{-1}$	\\
$\gamma$				& 	-						&		-						&	Reaction order					&	\\
$\varepsilon_\mathrm{r}$& 	\texttt{epp}			&	Property					&	Relative permittivity			& 	\\
$\eta$					& 	-						&		-						&	External quantum efficiency	&	\\
$\kappa$				&	-						&		-						&	Reflectance						&	-	\\
$\lambda$				& 	-						&		-						&	Wavelength						& nm	\\
$\mu_n$					& 	\texttt{mu\_n}			&	Property					&	Electron mobility				& \mobunit	\\
$\mu_p$					& 	\texttt{mu\_p}			&	Property					&	Hole mobility					& \mobunit	\\
$\mu_a$					& 	\texttt{mu\_a}			&	Property					&	Anion mobility					& \mobunit	\\
$\mu_c$					& 	\texttt{mu\_c}			&	Property					&	Cation mobility					& \mobunit	\\
$\rho$					& 	\texttt{rho}			&	Calculated output			&	Charge density					& cm$^{-3}$		\\
$\sigma$				& 	-						&		-						&	Conductivity					& S cm$^{-1}$		\\
$\tau_n,\mathrm{SRH}$	& 	\texttt{taun}			&	Property					&	SRH electron lifetime			& s			\\
$\tau_p,\mathrm{SRH}$	& 	\texttt{taup}			&	Property					&	SRH hole lifetime				& s			\\
$\tau_n,\mathrm{vsr}$	& 	\texttt{taun\_vsr}		&	Dependent property			&	VSR electron lifetime, 		& s			\\
$\tau_p,\mathrm{vsr}$	& 	\texttt{taup\_vsr}		&	Dependent property			&	VSR hole lifetime				& s			\\
$\tau_n$				& 	\texttt{taun}$^*$		&	Property					&	Electron first order lifetime	& s			\\
$\tau_p$				& 	\texttt{taup}$^*$		&	Property					&	Hole first order lifetime		& s			\\
$\phi$					& 	$\dagger$				&		-						&	Photon flux						& cm$^{-2}$s$^{-1}$\\
$\Phi_\mathrm{EA}$		& 	\texttt{EA}				&	Property					&	Electron affinity				& eV	\\
$\Phi_\mathrm{IP}$		& 	\texttt{IP}				&	Property					&	Ionisation potential			& eV	\\
$\Phi_l$				& 	\texttt{Phi\_l}			&	Property					&	Left-hand electrode workfunction			& eV	\\
$\Phi_r$				& 	\texttt{Phi\_r}			&	Property					&	Right-hand electrode workfunction			& eV	\\
$a$ 					& 	\texttt{a}				&	Solution variable			&	Mobile anion carrier density 	& cm$^{-3}$	\\
$a_\mathrm{max}$		&	\texttt{a\_max}			&	Property					&	Limiting anion density			& cm$^{-3}$	\\
$c$						&	\texttt{c}				&	Solution variable			&	Mobile cation carrier density	& cm$^{-3}$	\\
$c_\mathrm{max}$		&	\texttt{c\_max}			&	Property					&	Limiting cation density		& cm$^{-3}$	\\
$d$						& 	\texttt{d}$^{**}$			&	Property					&	Layer thickness					& cm	\\
$d_\mathrm{dev}$		& 	\texttt{dcum(end)}		&	Dependent property			&	Device thickness				& cm	\\
$d_\mathrm{DR}$			& 	-						&		-						&	Depletion region thickness		& cm 	\\
$g$						& 	\texttt{gx1},\texttt{gx2}, \texttt{gx} &	Property	&	Volumetric generation rate		& cm$^{-3}$ s$^{-1}$	\\
$g_0$					& 	\texttt{g0}				&	Property					&	Uniform generation rate		& cm$^{-3}$ s$^{-1}$	\\
$j$ 					& 	\texttt{j}				&	Calculated output			&	Flux density 					& cm$^{-2}$ s$^{-1}$	\\
$k_{\gamma}$			& 	-						&		-						&	Rate coefficient for reaction order $\gamma$	& Variable 	\\
$B$ 					& 	\texttt{B}				&	Property					&	Band-to-band recombination rate coefficient		& \rateunit	\\
$k_\mathrm{scan}$		& 	-						&		-						&	Current-voltage scan rate		& Vs$^{-1}$		\\
\kTPV\					& 	-						&		-						&	TPV decay rate coefficient		& s$^{-1}$		\\
$n$ 					& 	\texttt{n}				&	Solution variable			&	Electron carrier density 		& cm$^{-3}$	\\
$n_0$ 					& 	\texttt{n0}				&	Dependent property			&	Equilibrium electron density	& cm$^{-3}$	\\
$n_\mathrm{id}$			& 	-						&		-						&	Diode ideality factor			&	\\
$n_\mathrm{i}$			& 	\texttt{ni}				&	Dependent property			&	Intrinsic carrier density		& cm$^{-3}$	\\
$n_\mathrm{t}$ 			& 	\texttt{nt}				&	Dependent property			&	SRH electron trap parameter	& cm$^{-3}$	\\
$p$ 					& 	\texttt{p}				&	Solution variable			&	Hole carrier density 			& cm$^{-3}$	\\
$p_0$ 					& 	\texttt{p0}				&	Dependent property			&	Equilibrium hole density			& cm$^{-3}$	\\
$p_\mathrm{t}$ 			& 	\texttt{pt}				&	Dependent property			&	SRH hole trap parameter			& cm$^{-3}$	\\
$r$						& 	\texttt{r}				&	Calculated output			&	Volumetric recombination rate		& cm$^{-3}$ s$^{-1}$	\\
$r_\mathrm{btb}$		& 	\texttt{r.btb}			&	Calculated output			&	Band-to-band recombination rate			& cm$^{-3}$ s$^{-1}$	\\
$r_\mathrm{SRH}$		& 	\texttt{r.srh}			&	Calculated output			&	SRH recombination rate				& cm$^{-3}$ s$^{-1}$		\\
$r_\mathrm{vsr}$		& 	\texttt{r.vsr}			&	Calculated output			&	Volumetric surface recombination rate		& cm$^{-3}$ s$^{-1}$		\\
$s_n$					& 	\texttt{sn} 	&	Property			&	Electron surface recombination velocity at interfaces &	cm s$^{-1}$	\\
$s_{n,l}$, $s_{n,r}$		& 	\texttt{sn\_l}, \texttt{sn\_r} 	&	Property			&	Electron surface recombination velocity at left ($l$) and right-hand ($r$) system boundaries &	cm s$^{-1}$	\\
$s_p$					& 	\texttt{sp} 	&	Property			&	Hole surface recombination velocity &	cm s$^{-1}$	\\
$s_{p,l}$, $s_{p,r}$		& 	\texttt{sp\_l}, \texttt{sp\_r}	&	Property			&	Hole surface recombination velocity at left ($l$) and right-hand ($r$) system boundaries &	cm s$^{-1}$	\\
-						& 	\texttt{stack}			&	Property					&	Materials layers names			&	-		\\

\end{tabular}
\end{table}

\clearpage
\begin{table}[h]
\caption{\textbf{Table of symbols continued.} For a complete description of all user-defined and dependent properties please see the comments in the parameters class \texttt{pc}. For information on calculated outputs see the analysis class \texttt{dfana}.}
\label{tbl:symbols2}
\begin{tabular}{p{1.4cm}p{1.4cm}p{2.4cm}p{4cm}p{2cm}}
\textbf{Symbol}			& \textbf{\df\ Property Name}	&	\textbf{Variable type}	&	\textbf{Name}	& \textbf{Unit}	\\
\hline\noalign{\smallskip}
$t$ 					& 	\texttt{t}				&	Property					&	Time 							& s		\\
$w_\mathrm{n}$			& 	-						&		-						&	$n$-type depletion width		& cm			\\
$w_\mathrm{p}$			& 	-						&		-						&	$p$-type depletion width		& cm			\\
$x$ 					& 	\texttt{x}				&	Property					&	Position 						& cm	\\
$x_n$ 					& 	-				&	-					&	Translated spatial co-ordinate for electron distribution within the interfacial regions 						& cm	\\
$x_p$ 					& 	-				&	-					&	Translated spatial co-ordinate for hole distribution within the interfacial regions 									& cm	\\
$y$						& 	-						&		-						&	Generic carrier concentration	& cm$^{-3}$	\\	
$z$						& 	\texttt{z}				&	Property					&	Integer charge state	& - 	\\	
$D_n$					& 	\texttt{Dn}				&	Solver variable				&	Electron diffusion coefficient		& cm$^2$ s$^{-1}$\\	
$D_p$					& 	\texttt{Dp}				&	Solver variable				&	Hole diffusion coefficient			& cm$^2$ s$^{-1}$\\	
$E$						& 	-						&		-						&	Energy							& eV	\\
$E_\gamma$				& 	-						&		-						&	Photon energy					& eV	\\
$E_\mathrm{CB}$			& 	\texttt{Ec}			&	Property					&	Conduction band energy			& eV	\\
$E_\mathrm{VB}$			& 	\texttt{Ev}			&	Property					&	Valence band energy			& eV	\\
$E_\mathrm{Fi}$			& 	\texttt{Efi}			&	Dependent property			&	Intrinsic Fermi energy			& eV	\\
$E_\mathrm{F0}$			& 	\texttt{E0}				&	Property					&	Equilibrium Fermi level		& eV	\\
$E_\mathrm{Fn}$			& 	\texttt{Efn}			&	Calculated 	output			&	Electron quasi Fermi level		& eV	\\
$E_\mathrm{Fp}$			& 	\texttt{Efp}			&	Calculated 	output			&	Hole quasi Fermi level			& eV	\\
$E_\mathrm{vac}$		& 	\texttt{Evac}			&	Calculated 	output			&	Vacuum energy						& eV	\\
$E_\mathrm{g}$			& 	\texttt{Eg}				&	Dependent property			&	Bandgap	energy						& eV	\\
$E_\mathrm{t}$			& 	\texttt{Et}				&	Property					&	SRH trap energy					& eV	\\
$F$						& 	\texttt{F}				&	Calculated 	output			&	Electric field 					& V cm$^{-1}$		\\
$J$			 			& 	\texttt{J}				&	Calculated 	output			&	Current density 					& A cm$^{-2}$		\\
$J_\mathrm{SC}$			& 	\texttt{Jsc}			&	Calculated 	output			&	Short circuit current density	& A cm$^{-2}$	\\
$J_0$					& 	-						&		-						&	Dark saturation current			& A cm$^{-2}$	\\
$J_{0\mathrm{,rad}}$	& 	-						&		-						&	Black body recombination current & A cm$^{-2}$	\\
$L$						& 	-						&		-						&	Diffusion length					& cm	\\
$\bar{L}$				& 	-						&		-						&	Mean free path length				& cm	\\
$N_\mathrm{ani}$		& 	\texttt{Nani}			&	Property					&	Intrinsic anion  density		& cm$^{-3}$			\\
$N_\mathrm{cat}$		& 	\texttt{Ncat}			&	Property					&	Intrinsic cation  density		& cm$^{-3}$			\\
$N_\mathrm{A}$ 			& 	\texttt{NA}				&	Dependent property			&	Acceptor density 				& cm$^{-3}$			\\
$N_\mathrm{D}$ 			& 	\texttt{ND}				&	Dependent property			&	Donor density 					& cm$^{-3}$			\\
$N_\mathrm{CB}$ 		& 	\texttt{Ncb}			&	Property					&	Conduction band effective density of states	& cm$^{-3}$			\\
$N_\mathrm{VB}$ 		& 	\texttt{Nvb}			&	Property					&	Valence band effective density of states		& cm$^{-3}$			\\
$P$						& 	\texttt{P}				&	Calculated 	output			&	Power density					& W cm$^{-2}$	\\
$Q$						& 	\texttt{sigma}			&	Calculated 	output			&	Integrated charge density		& C	cm$^{-2}$		\\
$R_S$					& 	\texttt{Rs}				&	Property					&	Series resistance (area normalised)	& $\mathrm{\Omega}$cm$^{2}$\\
$S$						& 	-						&			-					&	Source/sink rate				& cm$^{-3}$ s$^{-1}$	\\
$T$						& 	\texttt{T}				&	Property					&	Temperature						&	K	\\
$T_\mathrm{S}$			& 	-						&			-					&	Black body temperature			&	K	\\
$U_0$					& 	-						&			-					&	Black body recombination rate	& cm$^{-3}$ s$^{-1}$\\
$V$						& 	\texttt{V}				&	Solution variable			&	Electrostatic potential		& V		\\
$V_\mathrm{app}$ 		& 	\texttt{Vapp}			&	Property					& 	Applied potential				&	V		\\
$V_\mathrm{bi}$ 		& 	\texttt{Vbi}			&	Dependent property			&	Built-in potential 			& V		\\
$V_\mathrm{OC}$ 		& 	\texttt{Voc}			&	Calculated 	output			&	Open circuit voltage 			& V		\\
$V_\mathrm{Rs}$ 		& 	\texttt{Vres}			&	Solver variable				&	Voltage drop across external circuit		& V		\\
\end{tabular}
\end{table}

\clearpage
\begin{table}
\caption{\textbf{Abbreviations.}}
\label{tbl:abbreviations}
\begin{tabular}{ll}
\textbf{General}	&						\\
ASA				&	Advanced Semiconductor Analysis	\\
AM				& 	Air Mass						\\
CB				&	Conduction Band					\\
DD				&	Drift-Diffusion					\\
DSSC 			& 	Dye Sensitised Solar Cell		\\
eDOS			&	Effective density of states		\\
EQE				&	External Quantum Efficiency		\\
ETL				&	Electron Transport Layer		\\
FF				&	Fill Factor						\\
HTL				&	Hole transport Layer			\\
\JV\ 			& 	Current density-Voltage Scan	\\
LED				&	Light Emitting Diode			\\
OPV 			&	Organic Photovoltaic 			\\
PCE				&	Power Conversion Efficiency		\\
PDEPE			&	Partial Differential Equation solver for Parabolic and Elliptic equations	\\
PSC 			& 	Perovskite Solar Cell			\\
PV 				& 	Photovoltaics					\\
QFL 			& 	Quasi Fermi Level				\\
SRH				&	Shockley-Read-Hall				\\
VB				&	Valence Band					\\
				&									\\
\textbf{Materials}	&		\\
ITO				&	Indium tin oxide				\\
MAPI			&	Methylammonium lead iodide		\\
\mpTi			&	Mesoporous \Ti\					\\
PCBM			&	Phenyl-C61-butyric acid methyl ester				\\
Spiro-OMeTAD	&	2,2',7,7'-Tetrakis[N,N-di(4-methoxyphenyl)amino]-9,9'	\\
				&	-spirobifluorene					\\
\end{tabular}
\end{table}

\section{Physical principles}
\subsection{Carrier flux densities}

\begin{figure*}[h]
\centering
\includegraphics[width=\textwidth]{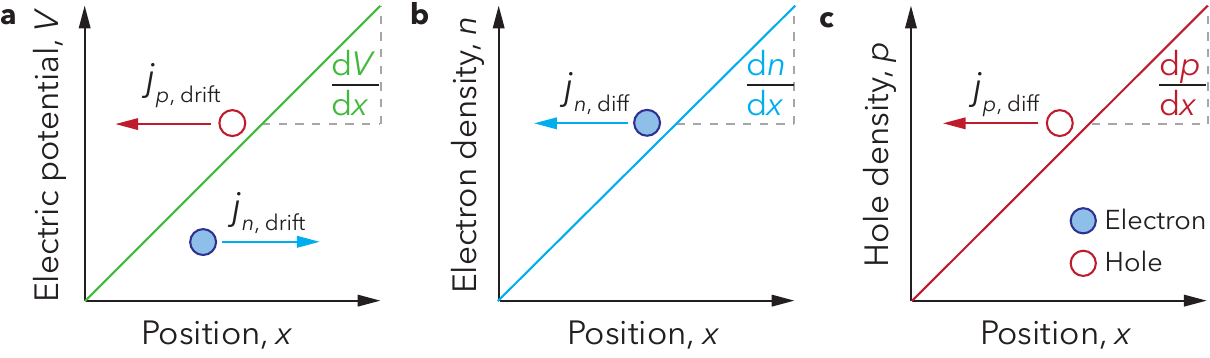}
\caption[Electron and hole carrier flux densities]{\textbf{Electron and hole carrier flux densities.} Direction of electron $j_n$ and hole $j_p$ carrier flux densities in response to positive gradients in (\textbf{a}) the electric field potential (note: $F = -\dv{V}{x}$), and (\textbf{b}) electron and (\textbf{c}) hole carrier densities. The subscripts `drift' and `diff' denote drift and diffusion flux densities respectively. Analogous flux densities can be drawn for mobile ionic species by substituting cations for holes and anions for electrons.}
\label{fig:Fluxes}
\end{figure*}

\clearpage
\FloatBarrier
\subsection{Energy level diagrams}	
\label{sec:EL_diagrams}

\begin{figure}[h]
\centering
\includegraphics[width=\textwidth]{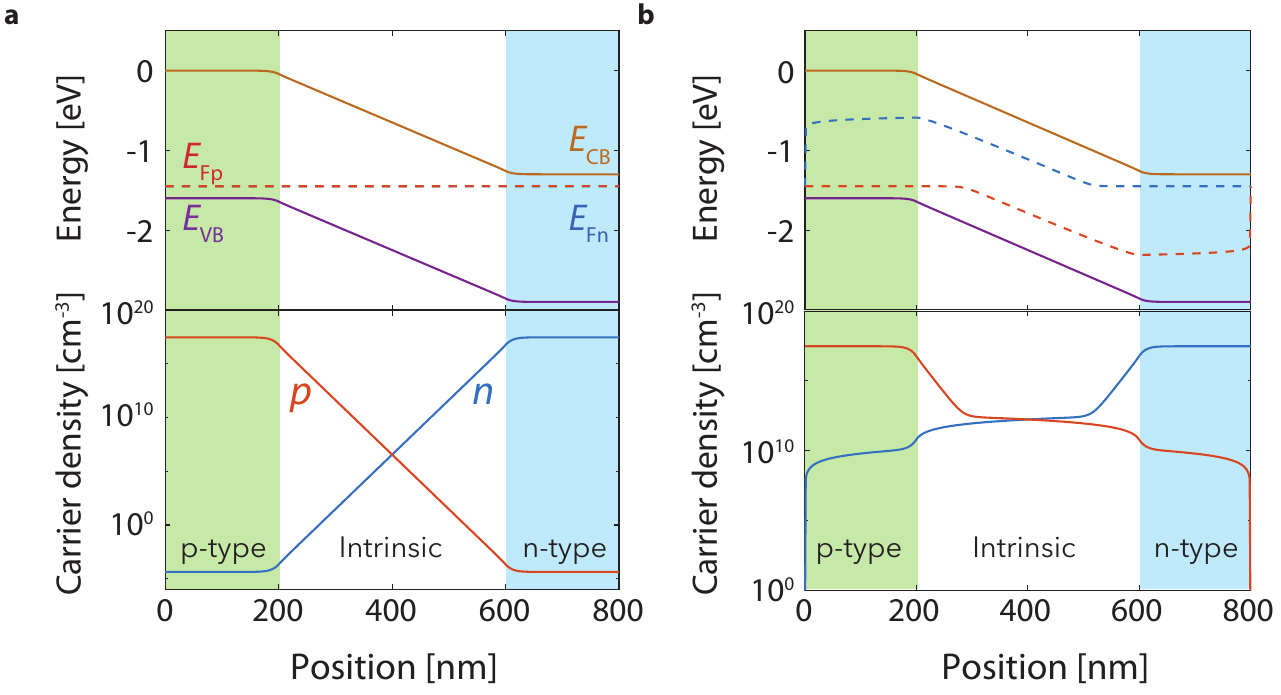}
	\caption[Short circuit equilibrium and illuminated \pin\ energy level diagrams and charge densities]{\textbf{Short circuit equilibrium and illuminated p-i-n energy level diagrams and charge densities}. (\textbf{a}) A simulated \pin\ structure at thermal equilibrium. p and n-type regions are shaded in green and blue respectively. The built-in field predominantly drops across the active layer (white region) while the chemical potential gradient is equal and opposite accounting for the same slope in the $n$ and $p$ profiles plotted on a logarithmic scale. This results in flat quasi-Fermi levels (QFLs) for both electrons and holes indicating that no current is flowing. (\textbf{b}) The same device at short circuit under optical bias: The QFLs split due to population of the valence and conduction bands with photoexcitated electrons and holes. Here, the Fermi level gradients indicate that net electron and hole currents are flowing in the device.}
	\label{fig:pin_band_diagrams}
\end{figure}

Energy level diagrams\footnote{The terms `band diagrams' and `energy level diagrams' are often used interchangeably although we avoid this terminology here to prevent confusion with band structure diagrams (plots of energy versus crystal momentum).} provide a convenient method for visualising the spatial solution of the simulation at a single time point. The electron and hole quasi-Fermi levels (QFLs), and conduction and valence band energies at each location within the device are plotted with respect to \Evac. The slope of the bands indicates the strength of the electric field. The position of the QFLs with respect to the bands gives an indication of the occupation state of the bands, while the slope of the QFLs determines the direction and strength of the overall electrochemical driving force. Figure \ref{fig:pin_band_diagrams}a and \ref{fig:pin_band_diagrams}b shows example energy level diagrams of a \pin\ structure at equilibrium and short circuit under illumination.

\clearpage
\FloatBarrier
\section{Steady-state analytical solution for electronic carrier densities within the interfacial regions}

\subsection{Solution}
\label{sec:ss_carriers_interfaces}
To examine how the carrier densities change within the interfacial regions we begin with the continuity equation for electrons and assume no generation such that:

\begin{equation} \label{eq:cont_n_inter}
\dfrac{\partial n(x,t)}{\partial t} = -\dfrac{\partial j_n(x,t)}{\partial x} - r_n(x,t)
\end{equation}

For simplicity, we will assume that electrons within the interfaces are in a steady-state condition such that $\partial n(x,t)/\partial t =0$ and

\begin{equation} \label{eq:cont_n_SS}
-\dfrac{\partial j_n(x,t)}{\partial x} = r(x,t),
\end{equation}

\noindent where:

\begin{equation} \label{eq:current_e_inter}
j_n(x,t) = \mu_n(x,t) n\left(-\dfrac{\partial \Phi_\mathrm{EA}(x)}{\partial x} - qF \right) \\- 
D_n \left(\dfrac{\partial n(x,t)}{\partial x}  -  \dfrac{n(x,t)}{N_{\mathrm{CB}}(x)}\dfrac{\partial N_{\mathrm{CB}}(x)}{\partial x}\right).
\end{equation}

\begin{equation} \label{eq:current_e_inter_alpha}
j_n(x,t) = k_\mathrm{B}T \mu_n(x,t) \left(\alpha n(x,t) - \dfrac{\partial n(x,t)}{\partial x} \right)
\end{equation}

\noindent where,

\begin{equation} \label{eq:alpha_interface_SI}
\alpha = -\frac{1}{k_\mathrm{B}T} \left( \dfrac{\partial \Phi_\mathrm{EA}(x)}{\partial x} + qF \right) + \dfrac{1}{N_{\mathrm{CB}}(x)}\dfrac{\partial N_{\mathrm{CB}}(x)}{\partial x},
\end{equation}

To solve Equation \ref{eq:cont_n_SS} we make the following approximations and assumptions;
\begin{enumerate}
\item Transport within the interface is fast with respect to the surrounding materials such that the carriers within the interface can be treated as begin at steady-state.
\item Given that the interfacial regions are largely depleted of carriers, the electric field can be treated as constant throughout the interfaces.
\item The recombination rate $r$ is uniform throughout the thickness of the interface such that it can be treated as constant for a given majority carrier boundary density.
\item The non-generalised Einstein relation ($D = k_\mathrm{B}T \mu$) can be applied (i.e. Boltzmann statistics are valid).
\end{enumerate}

Under these conditions the following general solution to Equation \ref{eq:cont_n_SS} can be obtained:

\begin{equation} \label{eq:general_interface_solution}
n(x') = \dfrac{C_I}{\alpha} \exp(\alpha x') - \dfrac{qrx'}{k_B T \alpha \mu_n } + C_{II},
\end{equation}

\noindent $C_I$, $C_{II}$ are constants, and $x'$ is the translated spatial co-ordinate such that $x' = x - x_1$, where $x_1$ is the position of the left-hand interface boundary as shown in Figure \ref{fig:Interface_schematic} of the Main Text.

Using the boundary conditions $n(0) = n_\mathrm{s}$, and flux $j_n(0) = j_{n,\mathrm{s}}$ the following expressions for the electron density and flux within the interfaces can be obtained:

\begin{equation} \label{eq:n_sol_n0ns_j0js}
n(x') = n_\mathrm{s} e^{\alpha x'}  + \dfrac{j_{n,s}}{ k_B T \alpha \mu_n}(1-e^{\alpha x'}) - \dfrac{r}{k_B T \alpha^2 \mu_n}(1 -e^{\alpha x'} + \alpha x'),
\end{equation}


\begin{equation} \label{eq:n_general_sol_jns_SI}
j_{n}(x') = j_{n,\mathrm{s}} - rx'.
\end{equation}

A similar method can be used to obtain analogous expressions for the hole carrier and flux densities:

\begin{equation} \label{eq:p_sol_p0s_j0js}
p(x') = p_\mathrm{s} e^{\beta x'}  + \dfrac{j_{p,s}}{k_B T \beta \mu_p}(1- e^{\beta x'}) - \dfrac{r}{k_B T \beta^2 \mu_p}(1 -e^{\beta x'} + \beta x'),
\end{equation}

\begin{equation} \label{eq:p_general_sol_jps_SI}
j_{p}(x') = j_{p,\mathrm{s}} - rx'.
\end{equation}

where,

\begin{equation} \label{eq:beta_interface_SI}
\beta = \frac{1}{k_\mathrm{B}T} \left( \dfrac{\partial \Phi_\mathrm{IP}(x)}{\partial x} +qF \right) + \dfrac{1}{N_{\mathrm{VB}}(x)}\dfrac{\partial N_{\mathrm{VB}}(x)}{\partial x}
\end{equation}


%

\subsection{Validation under different transport and recombination regimes}
\label{sec:interface_ana_validation}

Repository branch: \ 	\texttt{Methods-interface-solutions}\\
Parameters file:	\	\texttt{./Input\_files/3\_layer\_test\_vary.csv}\\
Script: 			\	\texttt{./Scripts/interface\_numerical\_analytical}\\

\begin{figure}[h!]
\centering
	\includegraphics[width=\textwidth]{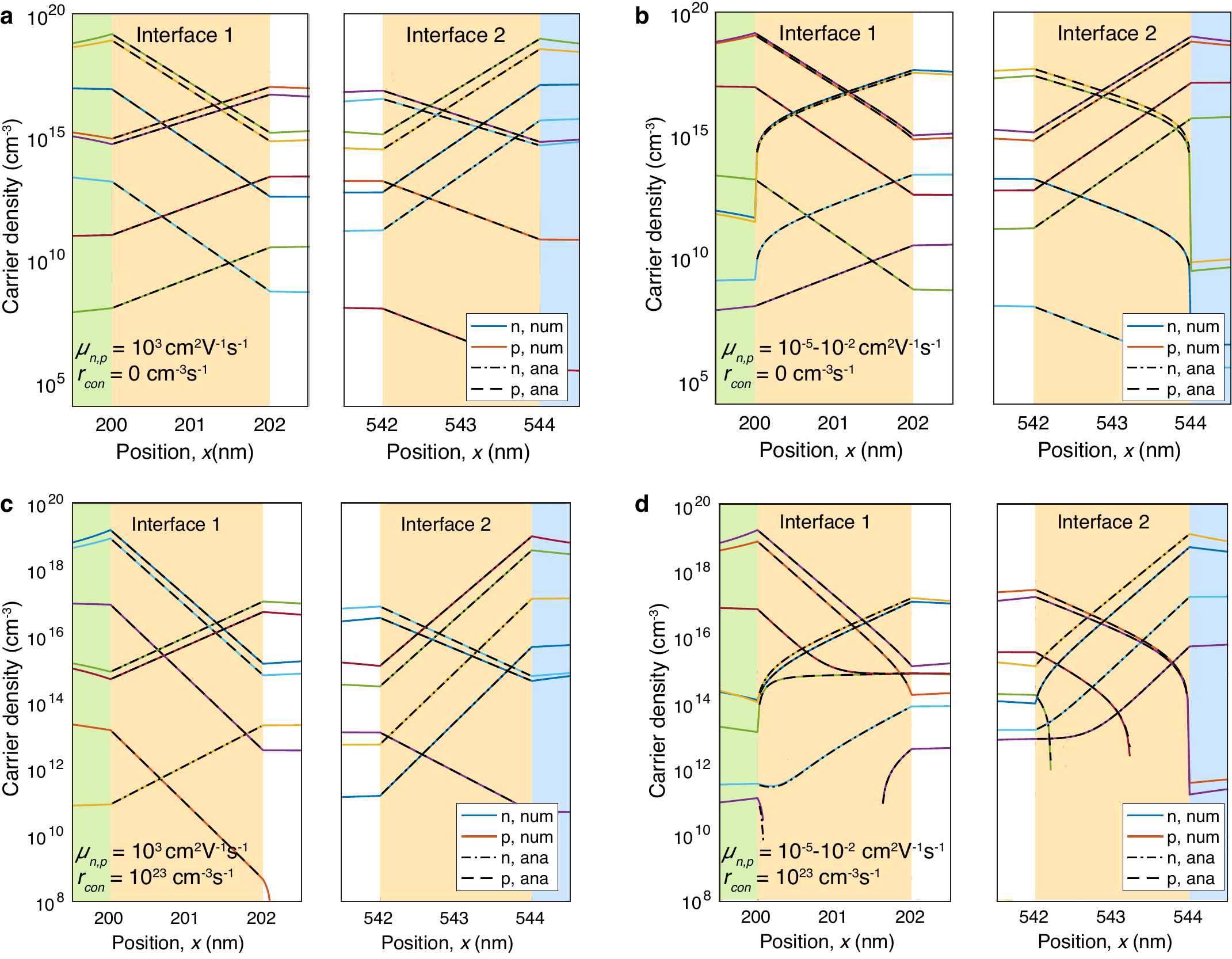}
	\caption[A comparison of numerical solutions and analytical approximations to the interfacial carrier density profiles within a three-layered solar cell with mobile ions during current-voltage scans.]{\textbf{A comparison of numerical solutions and analytical approximations to the interfacial carrier density profiles within a three-layered solar cell with mobile ions during current-voltage scans.} The following mobility and recombination regimes were tested; (\textbf{a}) High mobility, no recombination; (\textbf{b}) Low mobility, no recombination; (\textbf{c}) High mobility, high recombination and; (\textbf{d}) Low mobility, high recombination. Solutions are plotted at $t = 0, 4, 8$ and $12$ s, corresponding to $V_{app} = 0, 0.4, 0.8 $ (forward scan), and $0.8 $ (reverse scan) V. A constant and uniform recombination rate, $r_{con}$ was set throughout the interfacial regions. Numerical (num) and analytical (ana) solutions for electrons ($n$) and holes ($p$) are indicated by solid colours and black dashed curves respectively.}
	\label{fig:interface_solutions}
\end{figure}

To validate the analytical solutions derived in Section \ref{sec:ss_carriers_interfaces}, here we present results comparing analytical and numerical solutions of the carrier densities within the interfacial regions for a three-layer device including ionic carriers in the active layer. Solely for testing purposes a constant and uniform recombination rate, $r_{con}$ was set throughout the interfacial regions. Four different mobility and recombination regimes were tested; 1. High mobility, no recombination; 2. Low mobility, no recombination; 3. High mobility, high recombination; and 4. Low mobility, high recombination. Current-voltage scans from $0$ to $1$ to $0$ V at a scan rate of $100$ mV s$^{-1}$ were executed in each case. To calculate the analytical values the boundary values $n_s$, $p_s$, $j_{n,s}$ and $j_{p,s}$ were taken from the numerical solution (the nearest interior point on the subinterval mesh) for each interface. The solutions in Figure \ref{fig:interface_solutions} are plotted at $t = 0, 4, 8$ and $12$ s, corresponding to $V_{app} = 0, 0.4, 0.8 $ (forward scan), and $0.8 $ (reverse scan) V.

The results from the analytical and numerical solutions show close agreement. The small differences between the two can be accounted for by the fact that the fluxes are calculated on the subinterval mesh. In the low mobility, high recombination case, negative carrier densities result from the unphysical constant recombination rate. In practice this can be avoided by using relatively high mobilities within the interfacial regions such that the $r$ term in Equations \ref{eq:n_sol_n0ns_j0js} and \ref{eq:p_sol_p0s_j0js} is minimised. As discussed in Section \ref{ssec:recombination_interfaces}, the use of a recombination zone further minimises this issue by including recombination only in the region with the highest minority carrier density.

\clearpage
\section{Interfacial volumetric surface recombination model}
\label{sec:VSR}

\subsection{Recombination zone}
\begin{figure}[h]
\centering
\includegraphics[width=0.5\textwidth]{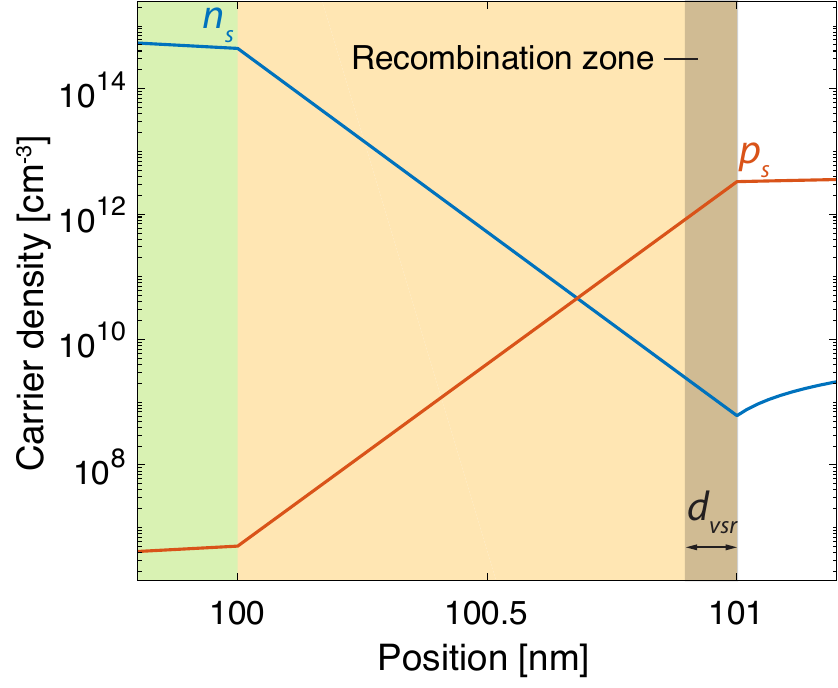}
	\caption[Recombination zone within an interface for the volumetric surface recombination scheme.]{\textbf{Recombination zone within an interface for the volumetric surface recombination scheme.} The recombination zone for the volumetric surface recombination model is located close to the boundary with the highest minority carrier density.}
	\label{fig:rec_zone}
\end{figure}
%
\FloatBarrier
\section{System architecture}
\subsection{Creating the generation profile}

\begin{figure}[h!]
\centering
\includegraphics[width=\textwidth]{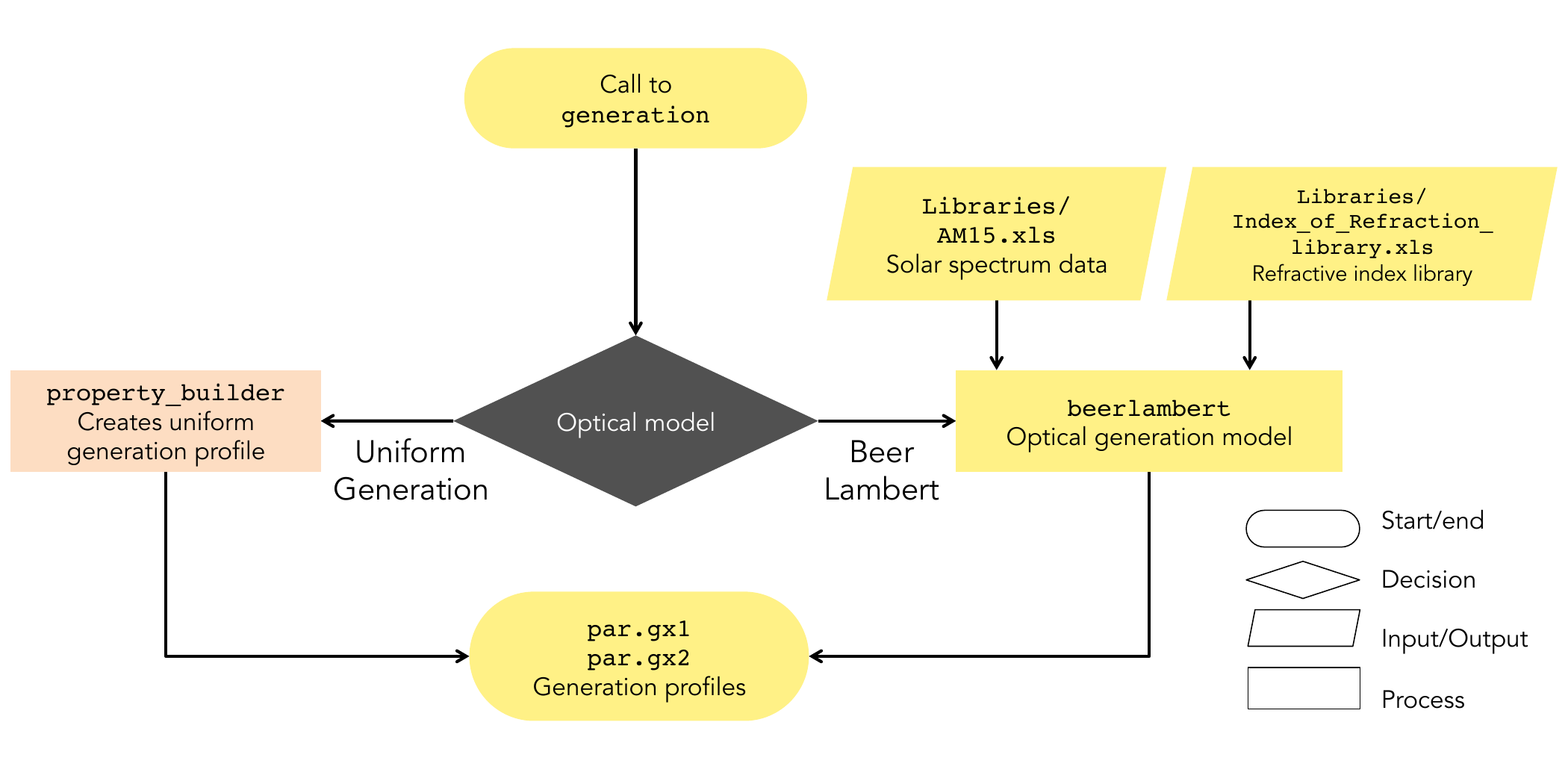}
\caption[Flow diagram for building the generation profile.]{\textbf{Flow diagram for building the generation profile.}}	
\label{fig:generation_flow}
\end{figure}

\section{Using the equation editor to adapt the physical model: Example}
\label{sec:edit_equations}
In the default version of \df\ ionic carriers do not have generation and recombination terms. In the following example we show how to change the underlying physical model by adding generation and recombination terms for ionic carriers. In general, the following steps are required:
\begin{enumerate}
\item Add terms to the equation editor \texttt{dfpde}, initial conditions \texttt{dfic} and boundary conditions \texttt{dfbc} subfunctions in \texttt{df} as required.
\item Ensure that changes in the equation editor and boundary conditions are reproduced comprehensively in \texttt{dfana}, with particular attention to \texttt{dfana.calcJ} (flux terms) and \texttt{dfana.calcr} (recombination terms). 
\item Add necessary properties to \texttt{pc}.
\item Define how the properties are graded in \texttt{build\_device}.
\end{enumerate}

\FloatBarrier
\subsection{Switching to two ionic carrier species}
\begin{figure}[h!]
	\includegraphics[width=\textwidth]{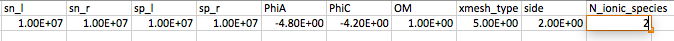}
	\caption[Setting the number of ionic carrier variables to 2.]{\textbf{Setting the number of ionic carrier variables to 2.} }
	\label{fig:N_ionic_species}
\end{figure}

We start with the default expressions shown in Listing \ref{lst:equation_editor} of the Main Text and the input properties defined in {\fontfamily{qcr}\selectfont Input\_files/spiro\_mapi\_tio2.csv}. To ensure charge conservation we maintain ionic carrier neutrality, two ionic carrier variables are required. As such we first need to set \texttt{N\_ionic\_species} to $2$ in the .csv file as shown in Figure \ref{fig:N_ionic_species}

\FloatBarrier
\subsection{Adding new device properties}
In order to be able to adjust our generation term we will introduce two new device properties \texttt{k\_iongen} and \texttt{k\_ionrec} that characterise the ionic generation and recombination rates. For simplicity we will assume that these terms are material specific and equal to zero in the transport layers of our three-layer device.
\texttt{k\_iongen} and \texttt{k\_ionrec} need to be added as arrays to the parameters class \texttt{pc} and the device builder before we can create a parameters object for the device. First we create the properties in \texttt{pc} with default values as shown in Listing S\ref{lst:adding_parameters}.

\begin{lstlisting}[language=Matlab, float, caption=\textbf{Creating property arrays in the parameters class.} New coefficients for ionic generation and recombination coefficients are added in lines 156 and 157.,
firstnumber=144,
firstline=144,
lastline=157,
basicstyle=\ttfamily\scriptsize,
label={lst:adding_parameters}]
%% Mobile ions
% Mobile ion defect density [cm-3]
N_ionic_species = 1;
K_anion = 1;         % Coefficients to easily accelerate ions
K_cation = 1;        % Coefficients to easily accelerate ions
Nani = [1e19];
Ncat = [1e19];
% Approximate density of iodide sites [cm-3]
% Limits the density of iodide vancancies
a_max = [1.21e22];
c_max = [1.21e22]; 
% Ionic generation and recombination rates
k_iongen = [0];
k_ionrec = [0];
\end{lstlisting}

Listing S\ref{lst:device_builder} shows how new properties can then be added to the device builder. The new coefficient arrays \texttt{k\_iongen} and \texttt{k\_ionrec} are used as the first input arguments. The device building code \texttt{build\_device} then uses these values with our chosen interface grading option to define the values of these properties at every point in the device and stores the resulting arrays in the device structures \texttt{par.dev} and \texttt{par.dev\_sub}. Here the \texttt{`lin\_graded'} option is chosen to linearly grade the new properties in the interface regions.

\begin{lstlisting}[float, language=Matlab,
caption=\textbf{Adding properties to the device builder.},
firstline=1,
lastline=28,
basicstyle=\ttfamily\scriptsize,
label={lst:device_builder}]
function dev = build_device(par, meshoption)
% BUILD_DEVICE calls BUILD_PROPERTY for each device property. BUILD_PROPERTY then defines the
% properties at each point on the grid defined by MESHOPTION

switch meshoption
    case 'whole'
        xmesh = par.xx;
    case 'sub'
        xmesh = par.x_sub;
end

% Properties with constant values in the interfacial regions
% Constant properties
dev.mu_c = build_property(par.mu_c, xmesh, par, 'constant', 0);
dev.mu_a = build_property(par.mu_a, xmesh, par, 'constant', 0);
dev.epp = build_property(par.epp, xmesh, par, 'constant', 0);
dev.k_iongen = build_property(par.k_iongen, xmesh, par, 'constant', 0);
dev.k_ionrec = build_property(par.k_ionrec, xmesh, par, 'constant', 0);

% Linearly graded properties
dev.EA = build_property(par.EA, xmesh, par, 'lin_graded', 0);
dev.IP = build_property(par.IP, xmesh, par, 'lin_graded', 0);
dev.E0 = build_property(par.E0, xmesh, par, 'lin_graded', 0);

% Logarithmically graded properties
dev.NA = build_property(par.NA, xmesh, par, 'zeroed', 0);
dev.ND = build_property(par.ND, xmesh, par, 'zeroed', 0);
dev.Nc = build_property(par.Nc, xmesh, par, 'log_graded', 0);
\end{lstlisting}

\begin{figure}[h!]
	\includegraphics[width=\textwidth]{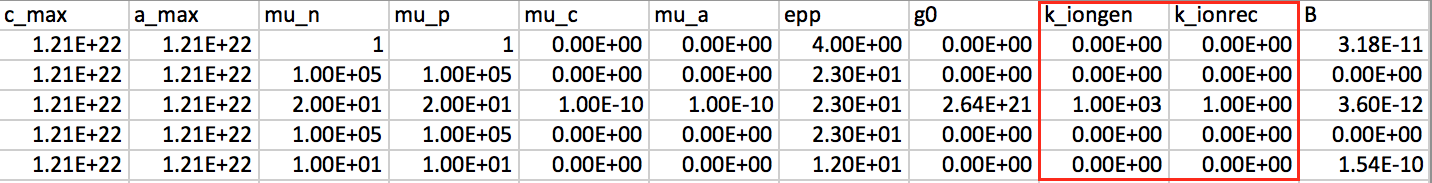}
	\caption[Adding new properties to the .csv file.]{\textbf{Adding new properties to the .csv file.} The new properties \texttt{k\_iongen} and \texttt{k\_ionrec} have been added to the final two columns.}
	\label{fig:Adding_new_properties_to_csv}
\end{figure}

Once the new properties have been added to the \texttt{pc} and \texttt{build\_device}, they can additionally be added to the \texttt{.csv} parameters file and \texttt{import\_properties} to enable them to be easily imported. Figure \ref{fig:Adding_new_properties_to_csv} shows two new columns added to the \texttt{.csv} file defining \texttt{k\_iongen} and \texttt{k\_ionrec} for each layer.

New properties can be inserted anywhere the user chooses into the table as \texttt{import\_properties} checks for matches between the column headers and listed property names. Lastly, the new properties are added to \texttt{import\_properties} using try statements (Listing \ref{lst:import_properties}). If a match to a column heading is found in the \texttt{.csv} file the the properties are imported, otherwise the default values in \texttt{pc} are used.

\begin{lstlisting}[float, language=Matlab,
caption=\textbf{Including new properties to be imported from the .csv file in import\_properties.},
firstnumber=182,
firstline=182,
lastline=191,
basicstyle=\ttfamily\scriptsize,
label={lst:import_properties}]
try
    par.k_iongen = T{1, 'k_iongen'};
catch
    warning('k_iongen undefined in .csv. Using default in PC')
end
try
    par.k_ionrec = T{1, 'k_ionrec'};
catch
    warning('k_ionrec undefined in .csv. Using default in PC')
end
\end{lstlisting}

This completes the steps necessary to add new properties to the parameters object and make them easily accessible from the .csv file. We will now look at how to adapt the physical model using the Equation Editor to include our new ionic carrier generation and recombination terms.

\FloatBarrier
\subsection{Using the equation editor}
For simplicity, we will assume that ionic carriers are generated at a rate proportional to the optical intensity and hence the electron and hole generation rate $g$. We will further assume that the ionic carriers recombine at a rate proportional to their respective densities. \textit{We stress here that there is no physical basis for this choice of models and they are only been used for illustrative purposes.} To adapt the \df\ master code \texttt{df}
to our purposes we first need to unpack the new variables from the parameters object by adding the to the list of variables at the start of the code as shown in Listing S\ref{lst:df_ion_gen}. 

\begin{lstlisting}[float, language=Matlab,
caption=\textbf{Unpacking the new coefficients at the start of df.},
firstnumber=93,
firstline=93,
lastline=102,
basicstyle=\ttfamily\scriptsize,
label={lst:df_ion_gen}]
epp = device.epp;           % Dielectric constant
eppmax = max(par.epp);      % Maximum dielectric constant (for normalisation)
B = device.B;               % Radiative recombination rate coefficient
ni = device.ni;             % Intrinsic carrier density
taun = device.taun;         % Electron SRH time constant
taup = device.taup;         % Electron SRH time constant
taun_vsr = device.taun_vsr; % Electron SRH time constant- volumetric interfacial surface recombination scheme
taup_vsr = device.taup_vsr; % Electron SRH time constant- volumetric interfacial surface recombination scheme
k_iongen = device.k_iongen; % Ion generation coefficient
k_ionrec = device.k_ionrec; % Ion recombination coefficient
\end{lstlisting}

While this approach is laborious, creating individual variables in the workspace in this way (as opposed to using \texttt{par.k\_iongen} directly in the Equation Editor) dramatically improves the performance of the code.

These terms are then simply added to the appropriate lines in the Equation Editor as shown in Listing S\ref{lst:equation_editor_ion_gen}.

\begin{lstlisting}[float, language=Matlab,
caption=\textbf{Adding source terms for mobile cations in the Equation Editor.} New terms for generation and recombination of cations are introduced on line 277.,
firstnumber=247,
firstline=247,
lastline=279,
basicstyle=\ttfamily\scriptsize,
label={lst:equation_editor_ion_gen}]
%% Equation editor
% Time-dependence prefactor term
C_V = 0; C_n = 1; C_p = 1; C_c = 1; C_a = 1;
C = [C_V; C_n; C_p; C_c; C_a];

% Flux terms
F_V = (epp(i)/eppmax)*dVdx;
F_n = mu_n(i)*n*(-dVdx + gradEA(i)) + (Dn(i)*(dndx - ((n/Nc(i))*gradNc(i))));
F_p = mu_p(i)*p*(dVdx - gradIP(i)) + (Dp(i)*(dpdx - ((p/Nv(i))*gradNv(i))));
F_c = mu_c(i)*(z_c*c*dVdx + kB*T*(dcdx + (c*(dcdx/(c_max(i) - c)))));
F_a = mu_a(i)*(z_a*a*dVdx + kB*T*(dadx + (a*(dadx/(a_max(i) - a)))));
F = [F_V; mobset*F_n; mobset*F_p; mobseti*K_c*F_c; mobseti*K_a*F_a];

% Electron and hole recombination
% Radiative
r_rad = radset*B(i)*(n*p - ni(i)^2);
% Bulk SRH
r_srh = SRHset*srh_zone(i)*((n*p - ni(i)^2)/(taun(i)*(p + pt(i)) + taup(i)*(n + nt(i))));
% Volumetric surface recombination
alpha = sign_xn(i)*q*dVdx/(kB*T) + alpha0_xn(i);
beta = sign_xp(i)*q*-dVdx/(kB*T) + beta0_xp(i);
r_vsr = SRHset*vsr_zone(i)*((n*exp(-alpha*xprime_n(i))*p*exp(-beta*xprime_p(i)) - ni(i)^2)...
    /(taun_vsr(i)*(p*exp(-beta*xprime_p(i)) + pt(i)) + taup_vsr(i)*(n*exp(-alpha*xprime_n(i)) + nt(i))));
% Total electron and hole recombination
r_np = r_rad + r_srh + r_vsr;

% Source terms
S_V = (q/(eppmax*epp0))*(-n + p - NA(i) + ND(i) + z_a*a + z_c*c - z_a*Nani(i) - z_c*Ncat(i));
S_n = g - r_np;
S_p = g - r_np;
S_c = k_iongen*g - k_ionrec*(c*a-Ncat(i)*Nani(i));
S_a = 0;
S = [S_V; S_n; S_p; S_c; S_a];
\end{lstlisting}

\subsection{Adapting analysis functions in \texttt{dfana} to the new physical model}
\label{ssec:adapting_analysis_functions}
\FloatBarrier
\textit{It is particularly important to bear in mind that the physical models in the master code and the analysis functions are not coupled.} Users therefore need to adapt any of the analysis functions in \texttt{dfana} to account for changes to the physical model. For example, if one were to change the recombination model, \texttt{dfana.calcr} would need to be updated in accordance with the changes made to the expressions in the Equation Editor.

\section{Troubleshooting}
We hope that your experience using \df\ will be a relatively trouble-free experience. Along with the possibility of minor bugs that come with experimental research software, you may encounter two common error messages: spatial discretization and time integration failures.

\subsection{Spatial discretization has failed}
This error typically occurs for one of two reasons:
\begin{enumerate}
\item The point density of spatial mesh is too high: Try reducing the number of points in individual layers or using a different spatial mesh type (see \texttt{meshgen\_x} for possible options).
\item The boundary conditions are not consistent with initial conditions: The initial value of a variable must be consistent with those set in the boundary conditions. If the initial or boundary conditions are changed from the default, ensure that they are consistent with one another.
\end{enumerate}

\subsection{Time integration failure}
\label{ssec:time_integration_failure}
This error occurs when the solver cannot converge and has a number of possible solutions:
\begin{enumerate}
\item Reduce the \texttt{par.tmax} property to reduce the total time of the simulation.
\item Reduce the \texttt{par.MaxStepFactor} property to reduce the solver's maximum time step. Please note that this may slow down solving time considerably.
\item Use the \texttt{par.K\_cation} and \texttt{par.K\_anion} properties to change the ionic carrier transport to be on a similar time scale to electronic carrier transport.
\item Split the protocol into a number of intermediate steps and associated solutions.
\item Relax (increase) the solver tolerances \texttt{par.RelTol} and \texttt{par.AbsTol}.
\item Stabilise the solution with the ionic carriers frozen (set \texttt{par.mobseti = 0}) then run a subsequent simulation with the ion transport switched on (set \texttt{par.mobseti = 1}).
\end{enumerate}
Time integration errors can also occur when inconsistencies are introduced to the physical model in the simulation.

\subsection{Unexpected values calculated using \texttt{dfana}}
\label{ssec:unexpected_calculate_values}
As discussed in the Main Text, owing to the computational cost of using functions external to \texttt{df} for the solving the equations, \textit{the physical model described in the Equation Editor is not coupled to that used in the analysis functions}. For greater consistency an obvious change to the programming architecture would be to use centralised flux and source functions to describe the physical model, which could then be called during both the solving and analysis stages. Multiple tests have shown, however, that this method comes with a significant increase in computation cost owing to the large number of calls made by the solver to such functions. Hence the present method of adapting the model in both \texttt{df} and \texttt{dfana} separately, while more prone to user error, is considerably more efficient.

\subsection{Bug reporting}
\label{ssec:bug_reporting}
If you find a bug with \df\ please raise an issue using the \texttt{Issues} tab on the \df\ GitHub repository page.\citep{Calado2017} This is the best way to ensure that other users can see which bugs have been addressed and understand how they have been fixed.

\section{Known issues}
\label{known_issues}

\subsection{Linear discretisation}
\label{ssec:linear_discretisation}
Many existing drift-diffusion models use Scharfetter-Gummel finite volume discretisation scheme\citep{Scharfetter1969}\cite{Farrell2016a} to account for the exponential change in carrier density in a constant electric field. In order to take advantage of MATLAB's Partial Differential Equation Parabolic and Elliptic (PDEPE) toolbox and the ease with which transport models can be altered with it, a major trade off is the use a simplified finite element discretisation scheme for which carrier densities are assumed to change linearly between neighbouring grid points. This can be compensated for somewhat with the use of a logarithmically-spaced spatial grid and a higher density of grid points. There are however limits to the allowed point density and while the comparison simulations that use the Scharfetter-Gummel scheme (Section \ref{sec:ASA_comparison} of the Main Text) is generally good, the user should be aware that the use of lower point densities will increase this error. One consequence of this simplified scheme is that the currents may not always be close to zero at equilibrium. Following exploratory work at lower point densities, users are encouraged to use as high a point density as possible for obtaining the final results, especially since the solver is highly efficient. Communication with Mathworks has suggested the maximum point density is a function of various factors and therefore cannot be distilled to a single value. A degree of trial and error is therefore required on the part of the user to find the maximum allowed point density for a given device configuration.

\subsection{Numerical errors with interfacial currents}
Following from the brief discussion in Subsection \ref{ssec:linear_discretisation}, and as noted in the comparison of \df\ and ASA results in Section \ref{sec:ASA_comparison}, the linear discretisation method and large gradients in the band energies and eDOS within the interfacial interlayers of \df\ makes these regions particularly prone to numerical errors. As shown in Section \ref{sec:IM_comp_SI}, errors can be minimised by reducing band energy and eDOS gradients as well as increasing the number of spatial mesh points within the interfaces. However, where accurate calculation of very low current densities ($<10^{-12}$ A cm$^{-2}$ for example, based on the ASA comparison calculations) across one or more heterojunction interfaces is required we presently do not recommend the use of \df\ for this purpose.

\subsection{Normalisation of the dielectric constant}
As may be noted from the Equation Editor (Listing \ref{lst:equation_editor}, Main Text) the dielectric constant must be normalised for the electrostatic potential flux and source terms to avoid a spatial discretisation error. The origin of this problem is currently unknown as normalisation has little effect on the magnitude of the input values.

\subsection{High extraction coefficients do not result in constant charge density at the boundaries}
Under some uncommon operating conditions e.g. large preconditioning voltages, the carrier densities at the boundaries do not tend to their limiting value when using high extraction coefficients. Under most circumstances this can be solved by switching to fixed carrier densities at the boundaries. This does however present a problem when calculating currents for a single carrier device as the minority carrier fluxes can no longer be used as the boundary value for the fluxes. This issue is currently ongoing and under investigation.

\subsection{Drift and diffusion currents do not sum to give the total current}
\label{ssec:known_issues_dd_currents}
Drift and diffusion currents calculated using \texttt{dfana.Jddxt} do not sum to give the correct current. This is related to the way in which fluxes are calculated in the solver but a solution has yet to present itself. Total carrier currents are correctly calculated using the continuity equations in \texttt{dfana.calcJ} and users are recommended to use this method instead, reserving \texttt{dfana.Jddxt} only for individual analysis of the approximate drift and diffusion currents.

\subsection{\texttt{lightOnRs} protocol instability}
\label{ssec:known_issues_lightOnRs_protocol}
At the time of writing \texttt{lightOnRs} protocol does not converge for all devices parameter sets at high values of series resistance.

\clearpage
\section{Validation against existing models}
\label{sec:SI_Comparisons}
The following sections provide references to the relevant Github \df\ repository branches, parameters files and scripts required to reproduce the results in Section \ref{sec:comparisons} of the Main Text.

\FloatBarrier
\subsection{The depletion approximation for a p-n junction}

\begin{tabular}{l l}
Repository branch: & 	\texttt{Methods-depletion-approximation-comparison}\\
Parameters file:	&	\texttt{./Input\_files/TPV\_test.csv}\\
Script: 			&	\texttt{df\_methods\_depletion\_approx\_pn\_junction.m}
\end{tabular}

\begin{table}[h]
\small
\caption{Key layer-specific parameters in the \df\ simulation comparison with the depletion approximation for a p-n junction.}
\label{tbl:Par_dep_approx_layers}
\begin{tabular}{p{3cm}lp{1.5cm}p{1.5cm}p{1.5cm}p{1.5cm}l}
\hline\noalign{\smallskip}
Variable 					& Symbol 			& p-type quasi-neutral & p-type depletion region	& n-type depletion region	&	n-type  quasi-neutral		&	Unit  \\
\noalign{\smallskip}\hline\noalign{\smallskip}
Thickness 					& $d$ 				& $10^{-2}$ 	& $1.38 \times 10^{-5}$ & $6.25 \times 10^{-5}$ 	& 	$10^{-2}$ &	cm \\
Electron affinity			& $\Phi _{EA}$		& $0$		& $0$		& $0$			&	$0$				&	eV	\\
Ionisation potential		& $\Phi _{IP}$		& $-1.12$	& $-1.12$		& $-1.12$		&	$-1.12$			&	eV	\\
Equilibrium Fermi energy	& $E_\mathrm{F0}$	& $-0.94$	& $-0.94$		&	$-0.22$		&	$-0.22$		&	eV	\\
Acceptor density			& $N_\mathrm{A}$				& $2.01 \times 10^{15}$	& $2.01 \times 10^{15}$ &	$0$	&	$0$	&	\densunit	\\
Donor density				& $N_\mathrm{D}$				& $0$	& $0$	&	$9.47 \times 10^{15}$	&	$9.47 \times 10^{15}$	&	\densunit	\\
SRH trap energy				& $E_\mathrm{trap}$				& $-0.56$	& $-0.56$	& $-0.56$	&	$-0.56$		&	eV	\\
eDOS conduction band		& $N_\mathrm{CB}$	& $10^{19}$		& $10^{19}$		& $10^{19}$		&	$10^{19}$	&	\densunit	\\
eDOS valence band			& $N_\mathrm{VB}$	& $10^{19}$		& $10^{19}$		& $10^{19}$		&	$10^{19}$	&	\densunit	\\
Electron mobility			& $\mu _e$	& $2000$	& $2000$	& $2000$	&	$2000$		&	\mobunit	\\
Hole mobility				& $\mu _h$	& $2000$	& $2000$	& $2000$	&	$2000$		&	\mobunit	\\
Relative dielectric constant	&	$\varepsilon _r$	&	$12$	&	$12$	&	$12$	&	$12$	& 	-	\\
Uniform generation rate	&	$g_0$	&	$0$	&	$3.49 \times 10^{21}$	&	$3.49 \times 10^{21}$	&	$0$	&	\rateunit \\
First order recombination electron lifetime	&	$\tau _n$	&	$10^{-6}, 10^{-7}, 10^{-8}$	&	$10^{100}$ &	$10^{100}$ &	$10^{100}$	& s \\
First order recombination hole lifetime	&	$\tau _p$	&	$10^{100}$	&	$10^{100}$ &	$10^{100}$ &	$10^{-6}, 10^{-7}, 10^{-8}$	& s \\
\noalign{\smallskip}\hline
\end{tabular}
\end{table}

\begin{table}[h]
\small
\caption{Key device-wide parameters in the \df\ simulation comparison with the depletion approximation for a p-n junction.}
\label{tbl:Par_dep_approx_device_wide}
\begin{tabular}{p{6cm}lp{1.5cm}l}
\hline\noalign{\smallskip}
Variable 					& Symbol 			& 	Unit  \\
\noalign{\smallskip}\hline\noalign{\smallskip}
Left-hand boundary Fermi energy						&	$\Phi _{l}$	&	$-0.94$		&	eV	\\
Right-hand boundary Fermi energy						&	$\Phi _{r}$	&	$-0.22$		&	eV	\\
Left-hand boundary electron extraction coefficient	&	$s_{n,l}$	&	$0$			&	cm s$^{-1}$	\\
Right-hand boundary electron extraction coefficient	&	$s_{n,r}$	&	$10^{10}$	&	cm s$^{-1}$	\\	
Left-hand boundary hole extraction coefficient		&	$s_{p,left}$	&	$10^{10}$	&	cm s$^{-1}$	\\	
Right-hand boundary hole extraction coefficient		&	$s_{p,r}$	&	$0$			&	cm s$^{-1}$	\\
\noalign{\smallskip}\hline
\end{tabular}
\end{table}

\clearpage

\subsubsection{Additional data}

\begin{figure}[h!]
\centering
	\includegraphics[width=0.8\textwidth]{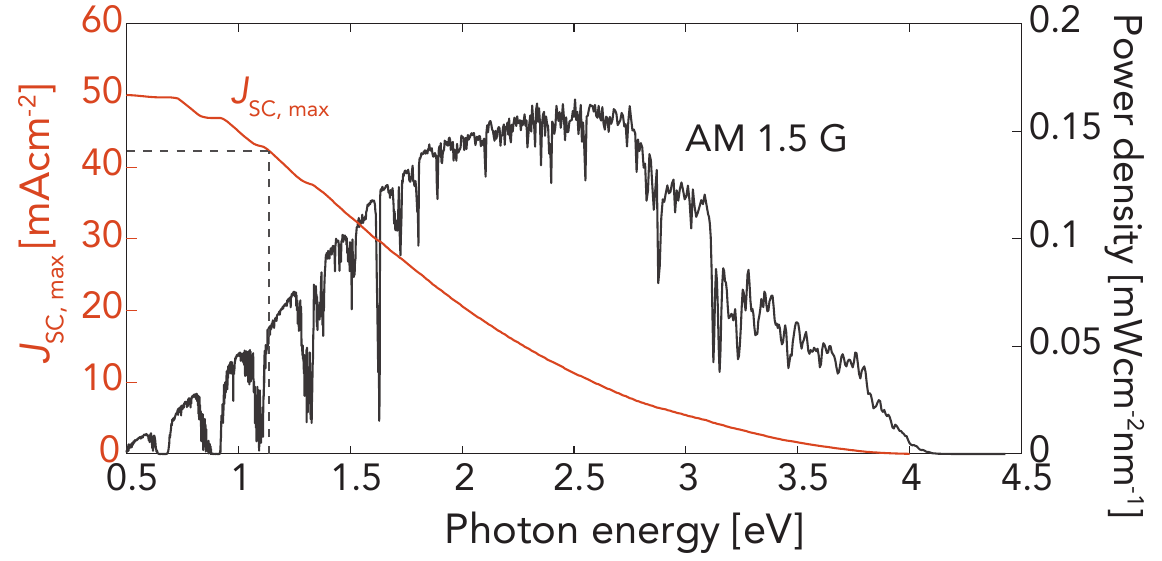}
	\caption[AM 1.5 solar spectrum and maximum theoretical short circuit current]{\textbf{AM 1.5 solar spectrum and maximum theoretical short circuit current.} AM 1.5 power density versus wavelength (right axis, black curve). Integrating the spectral photon flux above the band gap gives the maximum theoretical short circuit current $J_{\mathrm{SC,max}}$ (left axis, red curve). The dashed line indicates the position of $1.12$ eV. Note: 'Global Tilt' is for a south facing surface at an angle of $37$ degrees to the horizontal and includes sky diffuse and diffuse reflected light from the ground.\citep{NREL2017}}
	\label{fig:AM1p5_Jsc}
\end{figure}

\FloatBarrier
\subsection{Transient photovoltage response of a single field-free device}

\begin{tabular}{l l}
Repository branch:	& 	\texttt{Methods}\\
Parameters file:	&	\texttt{./Input\_files/TPV\_test.csv}\\
Script: 			&	\texttt{df\_methods\_TPV.m}
\end{tabular}

\begin{table}[h]
\small
\caption{Key layer-specific parameters in the \df\ simulation comparison with transient photovoltage kinetic model analytical solution.}
\label{tbl:Par_tpv_layers}
\begin{tabular}{p{3cm}lp{1.5cm}p{1.5cm}p{1.5cm}p{1.5cm}l}
\hline\noalign{\smallskip}
Variable 							& Symbol 			& 	Absorber layer 			&	Unit  \\
\noalign{\smallskip}\hline\noalign{\smallskip}
Thickness 							& $d$ 				& $10^{-5}$ 	 			&	cm \\
Electron affinity					& $\Phi _{EA}$		& $-3.8$						&	eV	\\
Ionisation potential				& $\Phi _{IP}$		& $-5.4$					&	eV	\\
Equilibrium Fermi energy			& $E_\mathrm{F0}$	& $-4.6$					&	eV	\\
Acceptor density					& $N_\mathrm{A}$				& $0$						&	\densunit	\\
Donor density						& $N_\mathrm{D}$				& $0$						&	\densunit	\\
eDOS conduction band				& $N_\mathrm{CB}$			& $10^{20}$					&	\densunit	\\
eDOS valence band					& $N_\mathrm{VB}$			& $10^{20}$					&	\densunit	\\
Electron mobility					& $\mu _e$			& $10$					&	\mobunit	\\
Hole mobility						& $\mu _h$			& $10$					&	\mobunit	\\
Relative dielectric constant		&	$\varepsilon _r$	&	$23$				& 	-	\\
Uniform generation rate			&	$g_0$			&	$1.89 \times 10^{21}$					&	\rateunit \\
Direct recombination coefficient	&	$k_{rad}$		&	$10^{-10}$	& s \\
\noalign{\smallskip}\hline
\end{tabular}
\end{table}

\begin{table}[h]
\small
\caption{Key device-wide parameters in the \df\ simulation comparison with transient photovoltage kinetic model analytical solution.}
\label{tbl:Par_tpv_device_wide}
\begin{tabular}{p{6cm}lp{1.5cm}l}
\hline\noalign{\smallskip}
Variable 					& Symbol 			& 	Unit  \\
\noalign{\smallskip}\hline\noalign{\smallskip}
Left-hand boundary Fermi energy						&	$\Phi _{l}$	&	$-4.6$		&	eV	\\
Right-hand boundary Fermi energy						&	$\Phi _{r}$	&	$-4.6$		&	eV	\\
Left-hand boundary electron extraction coefficient	&	$s_{n,l}$	&	$0$			&	cm s$^{-1}$	\\
Right-hand boundary electron extraction coefficient	&	$s_{n,r}$	&	$0$			&	cm s$^{-1}$	\\	
Left-hand boundary hole extraction coefficient		&	$s_{p,l}$	&	$0$			&	cm s$^{-1}$	\\	
Right-hand boundary hole extraction coefficient		&	$s_{p,r}$	&	$0$			&	cm s$^{-1}$	\\
\noalign{\smallskip}\hline
\end{tabular}
\end{table}

\newpage
\subsubsection{Derivation of a zero-dimensional kinetic model of recombination}	
\label{sec:TPV_derivation}
This section has been largely reproduced from reference \citep{calado2017transient}.

Under the assumption that the parabolic band and Boltzmann approximations are valid,\footnote{Criteria for validity: the band edges can be modelled as parabolic functions and the quasi Fermi levels should be $> 3k_\mathrm{B}T$ from their respective bands.\citep{Nelson2003}} the charge densities of electrons $n$ and holes $p$ for an intrinsic semiconductor with intrinsic carrier density $n_\mathrm{i}$ and Fermi energy $E_\mathrm{Fi}$ are given by:\citep{Nelson2003}

\begin{equation} 	\label{eq:n_Boltzmann}
n = n_\mathrm{i}\exp\left(\dfrac{E_\mathrm{Fn}-E_\mathrm{Fi}}{k_\mathrm{B}T}\right)
\end{equation}

\begin{equation} 	\label{eq:p_boltzmann}
p = n_\mathrm{i}\exp\left(\dfrac{E_\mathrm{Fi}-E_\mathrm{Fp}}{k_\mathrm{B}T}\right)
\end{equation}

Using Equations \ref{eq:n_Boltzmann} and \ref{eq:p_boltzmann}, the open circuit (OC) voltage \Voc\ in a field free, zero-dimensional material, where electron and hole charge carrier concentrations are equal ($n=p$), can be expressed as:

\begin{equation} 	\label{eq:Voc_zero_D}
V_\mathrm{OC} =  \dfrac{2k_\mathrm{B}T}{q}\ln\left(\dfrac{n_\mathrm{OC}}{n_\mathrm{i}}\right)
\end{equation}

Here, $n_\mathrm{OC}$ is carrier density at OC and is dependent on the generation rate $g$ and the recombination model. For example, in an idealised device with band-to-band recombination only, the recombination rate $r$ is given by $r = B(n^2-n_\mathrm{i}^2)$, where $B$ is the band-to-band recombination coefficient. At OC steady-state, generation is equal to recombination ($r=g$) leading to:

\begin{equation} 	\label{eq:n_btb}
n_\mathrm{OC} = \left (\dfrac{g}{B}+n_\mathrm{i}^2\right) ^{1/2}	 \approx	 \left (\dfrac{g}{B} \right) ^{1/2}
\end{equation}

\begin{figure}
\centering
\includegraphics[width=\textwidth]{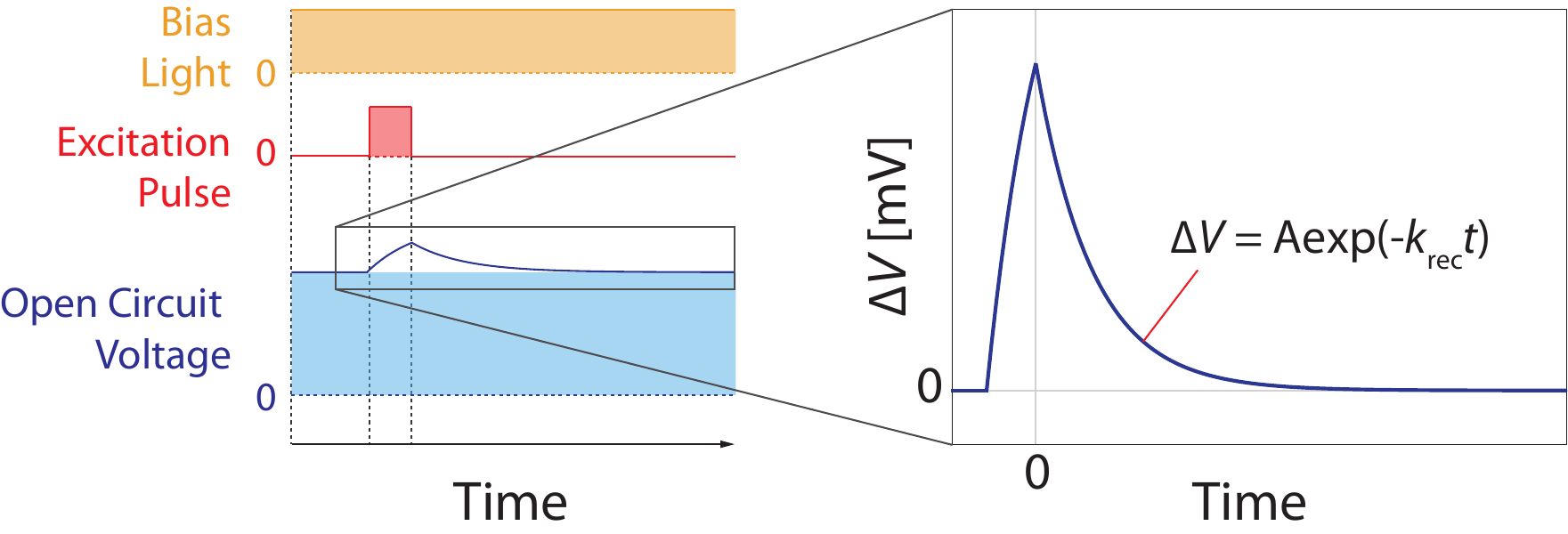}
\caption[Transient photovoltage experimental timeline]{\textbf{Transient photovoltage experimental timeline} The exponential decay rate constant of $\Delta V$ equates to $k_{rec}$ in the 0-D model.}	
\label{MT_fig:TPV_schematic}
\end{figure}

During a small perturbation measurement a small additional charge density $\Delta n << n_\mathrm{OC}$ is injected into the device such that the state of the system is not significantly altered. In a transient photovoltage measurement this charge is generated by an excitation light pulse imposed onto a background bias light (Figure \ref{MT_fig:TPV_schematic}a). 

Figure \ref{MT_fig:TPV_schematic}b is a schematic showing how the relaxation of the excess electron carrier density after a pulse results in a change in the electron quasi Fermi level (QFL). The recombination of excess charge with rate constant $k_{rec}$ , results in an associated decay of the QFLs. In this 0-D representation, the open circuit voltage is defined by the difference in electron and hole QFLs ($qV_\mathrm{OC} = E_\mathrm{Fn} - E_\mathrm{Fp}$) and decays with the same rate constant as the charge. In the most general case, following a small perturbation ending at $t=0$, the rate of change of addition charge $\dv{\Delta n}{t}$ can be expressed using a small perturbation rate constant, \kTPV\ (see reference \citep{Shuttle2008b} , pp. $101-103$ for full derivation):

\begin{equation} 	\label{eq:dDeltan_dt}
\dv{\Delta n}{t}\approx -k_\mathrm{TPV}\Delta n
\end{equation}

\begin{equation} 	\label{eq:k_pfo}
k_\mathrm{TPV} = \sum_{ij} \left(k_{ij} n^{\gamma_i} p^{\gamma_j} \left(\dfrac{\gamma_i}{n}+\dfrac{\gamma_j}{p}\right)\right)
\end{equation}

where $i$ and $j$ are summation indices and the exponents $\gamma_i$ and $\gamma_j$ can take any value (including non-integers). $k_{ij}$ is the associated rate coefficient for the reaction order. The solution to Equation \ref{eq:dDeltan_dt} is an exponential:

\begin{equation} 	\label{eq:Delta_n}
\Delta n = \exp(-k_\mathrm{TPV}t)	\qquad 	\mathrm{for}\ t >0
\end{equation}

Using Equation \ref{eq:Voc_zero_D}, the change in open circuit voltage $\Delta V_\mathrm{OC}$ produced by a light pulse introducing an additional charge $\Delta n$ can be expressed as:

\[\Delta V_\mathrm{OC} 	= \dfrac{2k_\mathrm{B}T}{q}\ln\left(1+\dfrac{\Delta n}{n_\mathrm{OC}}\right) \approx \dfrac{2k_\mathrm{B}T\Delta n}{n_\mathrm{OC}} \]

Substituting for $\Delta n$ using Equation \ref{eq:Delta_n} leads to:

\begin{equation} 	\label{eq:Voc_kt_SI}
\Delta V_\mathrm{OC}= \dfrac{2k_\mathrm{B}T}{qn_\mathrm{OC}}\exp(-k_\mathrm{TPV}t)		\qquad 	\mathrm{for}\ t >0
\end{equation}

A similar method can be used to find the change in charge carrier density and voltage rise during a pulse of duration $t_\mathrm{pulse}$:

\begin{equation} 	\label{eq:Delta_n_pulse}
\Delta n = \dfrac{\Delta g}{k_\mathrm{TPV}} (1 - \exp(-k_\mathrm{TPV}(t+t_\mathrm{pulse})))	\qquad 	\mathrm{for}\ -t_\mathrm{pulse} < t \leqslant 0
\end{equation}		

\begin{equation} 	\label{eq:Delta_V_pulse_SI}
\Delta V = \dfrac{2k_\mathrm{B}T}{qn_\mathrm{OC}}\dfrac{\Delta g}{k_\mathrm{TPV}} (1 - \exp(-k_\mathrm{TPV}(t+t_\mathrm{pulse})))		\qquad 	\mathrm{for}\ -t_\mathrm{pulse} < t \leqslant 0
\end{equation}

where $\Delta g$ is the additional generation rate from the pulse light.

In experimental measurements, a single empirical reaction order $\gamma$, with the corresponding rate constant $k_{\gamma}$, is typically assumed to dominate recombination such that:

\begin{equation} 	\label{eq:k_TPV_singleorder}
r = k_{\gamma}n^{\gamma}
\end{equation}

Accordingly, the slope of the $\log (k_\mathrm{TPV})$ vs. $\log (n_\mathrm{OC})$ plot can be used to determine $\gamma $:

\begin{equation} 	\label{eq:Reaction_order}
\log (k_\mathrm{TPV}) = (\gamma -1 ) \log(n_\mathrm{OC}) + \log(\gamma k_{\gamma})
\end{equation}	

\FloatBarrier
\subsection{Numerical solution for a three-layer device without mobile ions: comparison with Advanced Semiconductor Analysis (ASA) software}

\begin{tabular}{l l}
Repository branch:	& 	\texttt{Methods}\\
Parameters files: 	& \texttt{./Input\_files/3\_layer\_methods\_test1a.csv}\\
					& \texttt{./Input\_files/3\_layer\_methods\_test1b.csv}\\
					& \texttt{./Input\_files/3\_layer\_methods\_test2a.csv}\\
					& \texttt{./Input\_files/3\_layer\_methods\_test2b.csv}\\
Script:				&	 \texttt{df\_methods\_ASA\_comparison.m}	
\end{tabular}

\subsubsection{Parameter Sets 1a \& 1b}
\begin{table}[h]
\small
\caption{Key layer-specific parameters for Parameter Sets 1a and 1b in the \df\ simulation comparison with ASA. HTL and ETL denote hole and electron transport layers respectively.}
\label{tbl:Par_ASA1_layer}
\begin{tabular}{p{3cm}lp{1.5cm}p{1.5cm}p{1.5cm}p{1.5cm}l}
\hline\noalign{\smallskip}
Variable 					& Symbol 			&								&	Layer		&									&	Unit  \\
							&					& HTL							& Absorber		& ETL								&	\\
\noalign{\smallskip}\hline\noalign{\smallskip}
Optical parameters material reference			&	-	&	SiO$_2$	&	MAPICl	&	SiO$_2$	&	- \\
Thickness 					& $d$ 				& $2 \times 10^{-5}$	& (\textbf{a}) $4 \times 10^{-5}$ 	& $0.5 \times 10^{-5}$ &	cm \\
							&					&						& (\textbf{b}) $2 \times 10^{-5}$	&						&	\\
Electron affinity			& $\Phi _{EA}$		& $-3.3$				& $-3.8$				& $-4.0$				&	eV	\\
Ionisation potential		& $\Phi _{IP}$		& $-5.3$				& $-5.4$				& $-6.2$				&	eV	\\
Equilibrium Fermi energy	& $E_\mathrm{F0}$	& $-5.2$				& $-4.6$				& $-4.15$				&	eV	\\
Acceptor density			& $N_\mathrm{A}$				& $2.09 \times 10^{17}$& $0$ 					& $0$					&	\densunit	\\
Donor density				& $N_\mathrm{D}$				& $0$					& $0$					& $3.02 \times 10^{16}$ &	\densunit	\\
SRH trap energy				& $E_\mathrm{trap}$				& $-4.3$				& $-4.6$				& $-5.1$				&	eV	\\
eDOS conduction band		& $N_\mathrm{CB}$			& $10^{19}$				& $10^{18}$				& $10^{19}$				&	\densunit	\\
eDOS valence band			& $N_\mathrm{VB}$			& $10^{19}$				& $10^{18}$				& $10^{19}$				&	\densunit	\\
Electron mobility			& $\mu _e$			& $0.2$					& $20$					& $0.1$					&	\mobunit	\\
Hole mobility				& $\mu _h$			& $0.02$				& $20$					& $0.01$				&	\mobunit	\\
Relative dielectric constant	&	$\varepsilon _r$	&	$4$				& $23$					& $12$					&	-	\\
SRH recombination electron lifetime	&	$\tau _n$	&	$10^{-8}$	& $10^{-7}$ 			& $10^{-9}$ 			& s \\
SRH recombination hole lifetime		&	$\tau _p$	&	$10^{-8}$	& $10^{-7}$ 			& $10^{-9}$ 			& s \\
\noalign{\smallskip}\hline
\end{tabular}
\end{table}

\begin{table}[h]
\small
\caption{Key device-wide parameters for Parameter Sets 1a and 1b  in the \df\ simulation comparison with ASA.}
\label{tbl:Par_ASA1_device_wide}
\begin{tabular}{p{6cm}lp{1.5cm}l}
\hline\noalign{\smallskip}
Variable 					& Symbol 			& 	Unit  \\
\noalign{\smallskip}\hline\noalign{\smallskip}
Left-hand boundary Fermi energy						&	$\Phi _{l}$	&	$-5.2$		&	eV	\\
Right-hand boundary Fermi energy						&	$\Phi _{r}$	&	$-4.15$		&	eV	\\
Left-hand boundary electron extraction coefficient	&	$s_{n,l}$	&	$10^{8}$	&	cm s$^{-1}$	\\
Right-hand boundary electron extraction coefficient	&	$s_{n,r}$	&	$10^{8}$	&	cm s$^{-1}$	\\	
Left-hand boundary hole extraction coefficient		&	$s_{p,l}$	&	$10^{8}$	&	cm s$^{-1}$	\\	
Right-hand boundary hole extraction coefficient		&	$s_{p,r}$	&	$10^{8}$	&	cm s$^{-1}$	\\
\noalign{\smallskip}\hline
\end{tabular}
\end{table}

\newpage
\subsubsection{Parameter Sets 2a \& 2b}
\begin{table}[h]
\small
\caption{Key layer-specific parameters for Parameter Sets 2a and 2b in the \df\ simulation comparison with ASA. HTL and ETL denote hole and electron transport layers respectively.}
\label{tbl:Par_ASA2_layer}
\begin{tabular}{p{3cm}lp{1.5cm}p{1.5cm}p{1.5cm}p{1.5cm}l}
\hline\noalign{\smallskip}
Variable 					& Symbol 			&								&	Layer		&									&	Unit  \\
							&					& HTL							& Absorber		& ETL								&	\\
\noalign{\smallskip}\hline\noalign{\smallskip}
Optical parameters material reference			&	-	&	SiO$_2$	&	MAPICl	&	SiO$_2$	&	- \\
Thickness 					& $d$ 				& $1 \times 10^{-5}$	& $2 \times 10^{-5}$  	& $0.7 \times 10^{-5}$ &	cm \\
Electron affinity			& $\Phi _{EA}$		& $-3.0$				& $-3.8$				& $-4.1$				&	eV	\\
Ionisation potential		& $\Phi _{IP}$		& $-5.1$				& $-5.2$				& $-6.2$				&	eV	\\
Equilibrium Fermi energy	& $E_\mathrm{F0}$	& $-5.0$				& $-4.6$				& $-4.4$				&	eV	\\
Acceptor density			& $N_\mathrm{A}$				& $2.09 \times 10^{17}$& $8.32 \times 10^{7}$	& $0$					&	\densunit	\\
Donor density				& $N_\mathrm{D}$				& $0$					& $0$					& $9.12 \times 10^{14}$&	\densunit	\\
SRH trap energy				& $E_\mathrm{trap}$				& $-4.05$				& $-4.5$				& $-5.15$				&	eV	\\
eDOS conduction band		& $N_\mathrm{CB}$			& (\textbf{a}) $10^{19}$ 	& $10^{18}$	 		& (\textbf{a}) $10^{20}$ &	\densunit	\\
							&					& (\textbf{b}) $10^{18}$	&					& (\textbf{b}) $10^{18}$ &		\\
eDOS valence band			& $N_\mathrm{VB}$			& (\textbf{a}) $10^{19}$ & $10^{18}$			& (\textbf{a}) $10^{20}$ &	\densunit	\\
							&					& (\textbf{b}) $10^{18}$	&					& (\textbf{b}) $10^{18}$ &		\\
Electron mobility			& $\mu _e$			& $0.2$					& $20$					& $0.1$					&	\mobunit	\\
Hole mobility				& $\mu _h$			& $0.02$				& $2$					& $0.001$				&	\mobunit	\\
Relative dielectric constant	&	$\varepsilon _r$	&	$8$				& $10$					& $20$					&	-	\\
SRH recombination electron lifetime	&	$\tau _n$	&	$10^{-9}$	& $10^{-7}$ 			& $10^{-6}$ 			& s \\
SRH recombination hole lifetime		&	$\tau _p$	&	$10^{-9}$	& $10^{-7}$ 			& $10^{-6}$ 			& s \\
\noalign{\smallskip}\hline
\end{tabular}
\end{table}

\begin{table}[h]
\small
\caption{Key device-wide parameters for Parameter Sets 2a and 2b in the \df\ simulation comparison with ASA.}
\label{tbl:Par_ASA2_device_wide}
\begin{tabular}{p{6cm}lp{1.5cm}l}
\hline\noalign{\smallskip}
Variable 					& Symbol 			& 	Unit  \\
\noalign{\smallskip}\hline\noalign{\smallskip}
Left-hand boundary Fermi energy						&	$\Phi _{l}$	&	$-5.0$		&	eV	\\
Right-hand boundary Fermi energy						&	$\Phi _{r}$	&	$-4.4$		&	eV	\\
Left-hand boundary electron extraction coefficient	&	$s_{n,l}$	&	$10^{8}$	&	cm s$^{-1}$	\\
Right-hand boundary electron extraction coefficient	&	$s_{n,r}$	&	$10^{8}$	&	cm s$^{-1}$	\\	
Left-hand boundary hole extraction coefficient		&	$s_{p,l}$	&	$10^{8}$	&	cm s$^{-1}$	\\	
Right-hand boundary hole extraction coefficient		&	$s_{p,r}$	&	$10^{8}$	&	cm s$^{-1}$	\\
\noalign{\smallskip}\hline
\end{tabular}
\end{table}

\FloatBarrier

\subsubsection{Beer Lambert generation profile}

\begin{figure}[h!]
\centering
	\includegraphics[width=0.6\textwidth]{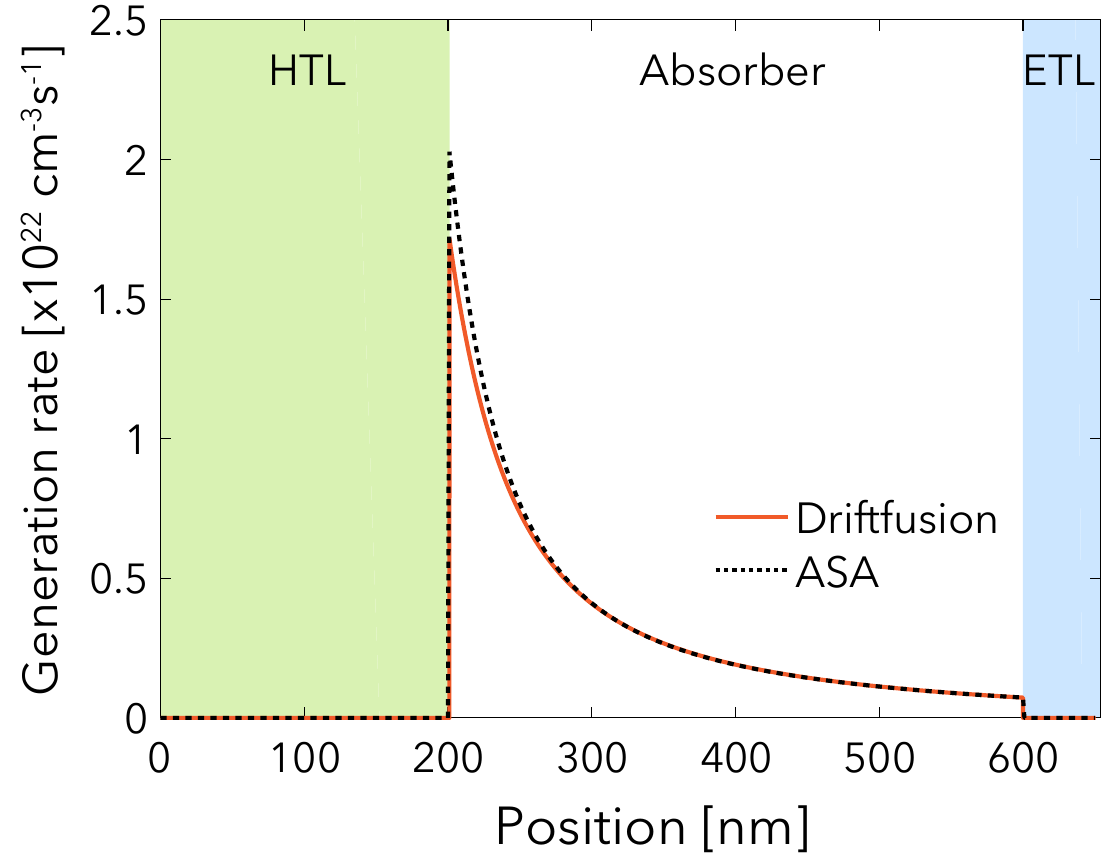}
	\caption[Integrated generation profile obtained using \df\ and ASA Beer Lambert models.]{\textbf{Integrated generation profile obtained using \df\ and ASA Beer Lambert models.}}
	\label{fig:DF_vs_ASA_gen}
\end{figure}

\begin{figure}[h!]
\centering
	\includegraphics[width=\textwidth]{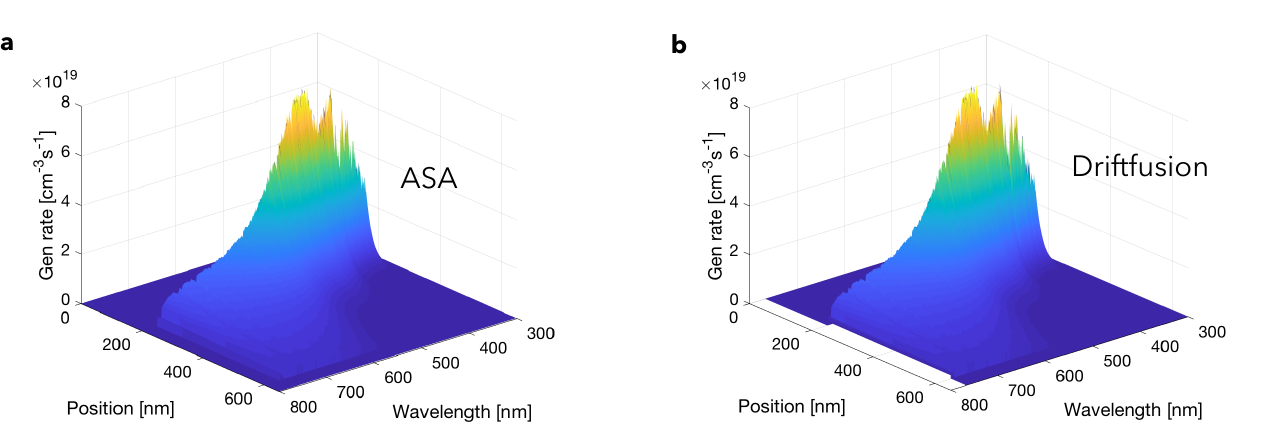}
	\caption[Comparison of optical generation rate as a function of wavelength and position using the Beer-Lambert model calculated by \df\ and ASA.]{\textbf{Comparison of optical generation rate as a function of wavelength and position using the Beer-Lambert model calculated by \df\ and ASA.}}
	\label{fig:DF_vs_ASA_beer_lambert}
\end{figure}

\FloatBarrier
\subsubsection{Equilibrium states}

\begin{figure}[h!]
\centering
	\includegraphics[width=\textwidth]{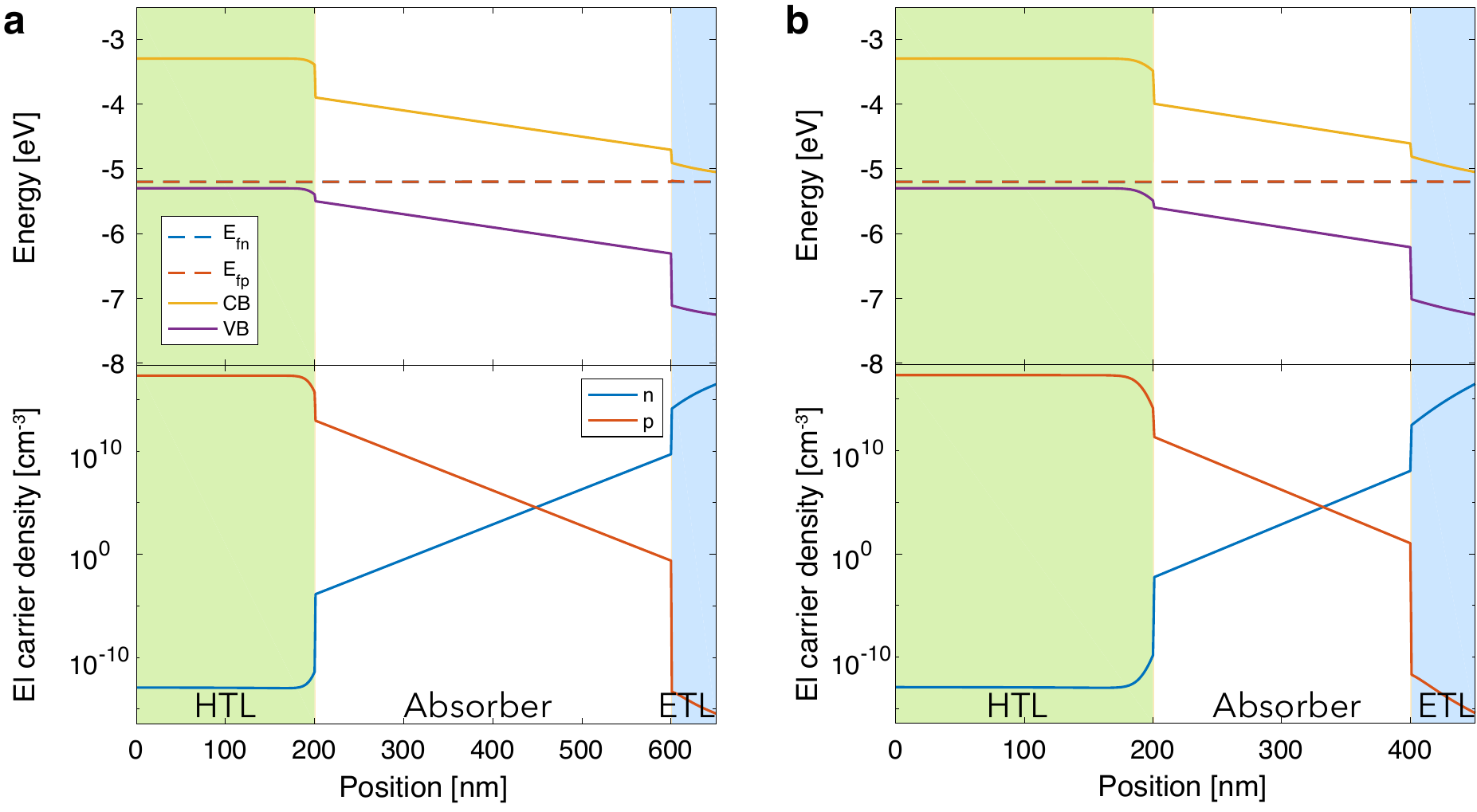}
	\caption[Equilibrium energy level diagrams and carrier densities for Parameter Sets 1a and 1b.]{\textbf{Equilibrium energy level diagrams and carrier densities for Parameter Sets 1a and 1b.} Equilibrium states for (\textbf{a}) Parameter Set 1a and (\textbf{b}) Parameter Set 1b.}
	\label{fig:DF_vs_ASA_PS1a_PS1b_ELnpx_equilibrium}
\end{figure}

\begin{figure}[h!]
\centering
	\includegraphics[width=\textwidth]{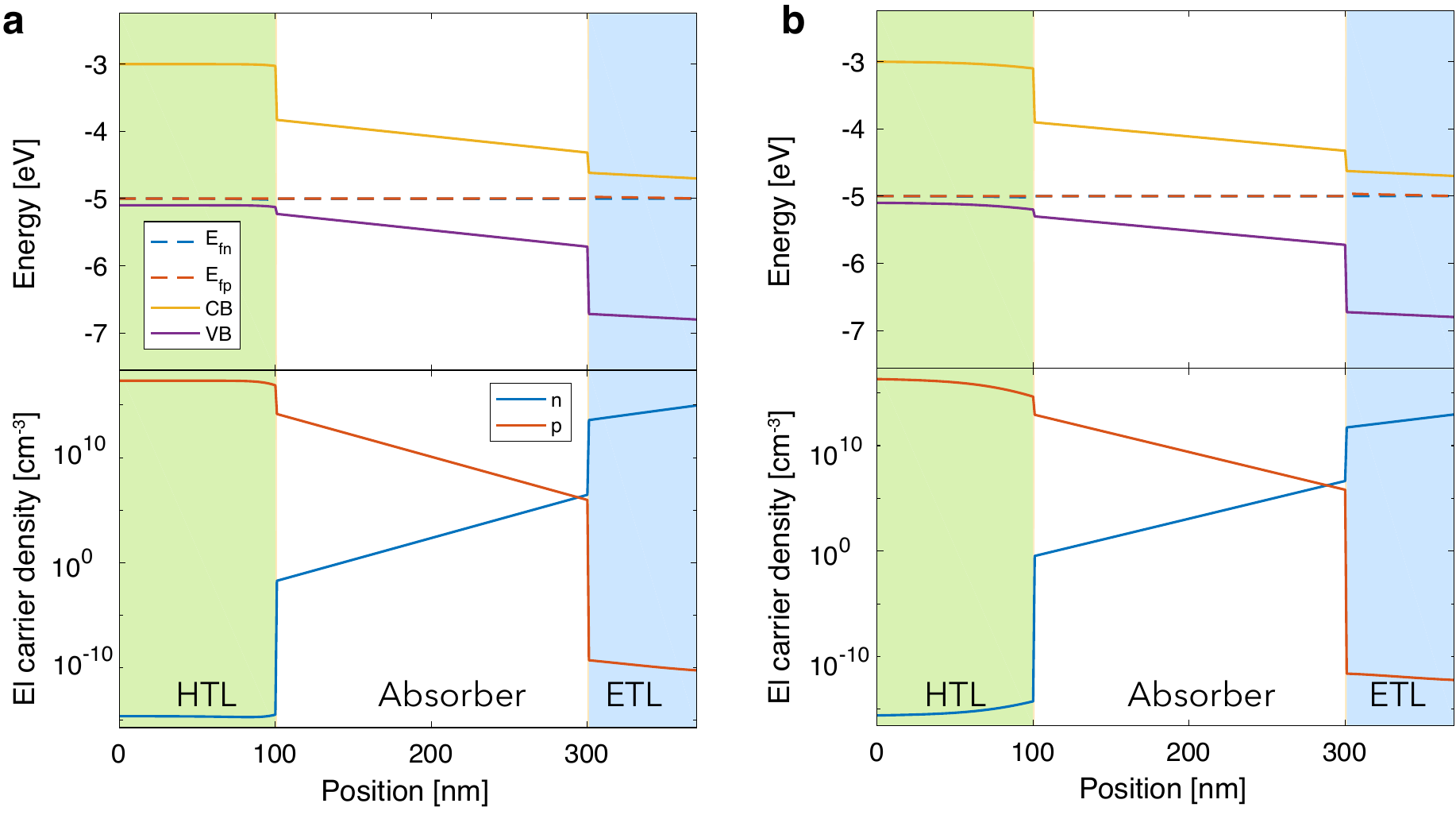}
	\caption[Equilibrium energy level diagrams and carrier densities for Parameter Sets 2a and 2b.]{\textbf{Equilibrium energy level diagrams and carrier densities for Parameter Sets 2a and 2b.} Equilibrium states for (\textbf{a}) Parameter Set 2a and (\textbf{b}) Parameter Set 2b.}
	\label{fig:DF_vs_ASA_PS1a_PS1b_ELnpx_equilibrium2}
\end{figure}

\FloatBarrier
\newpage
\subsubsection{Additional data}
\label{sssec:ASA_additional_data}

\begin{figure}[h!]
\centering
	\includegraphics[width=0.6\textwidth]{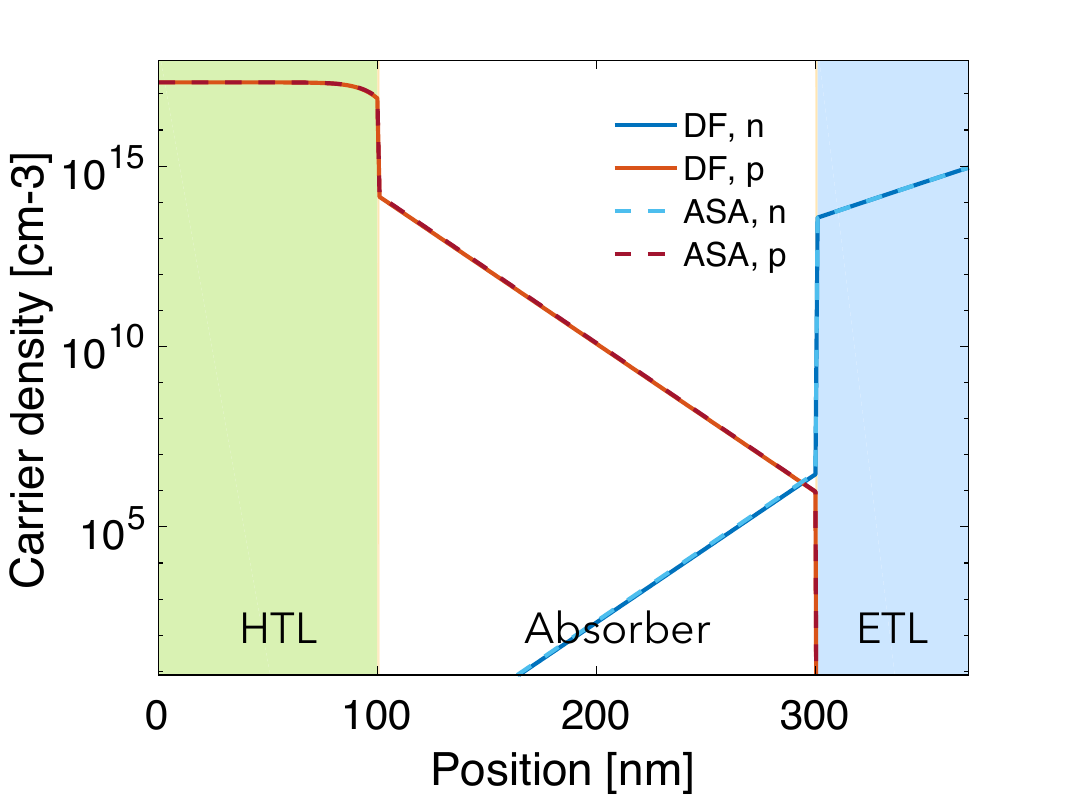}
	\caption[Carrier densities at equilibrium for a simulated device comparing \df\ with ASA using parameter set 2]{\textbf{Carrier densities at equilibrium for a simulated device comparing \df\ with ASA using parameter set 2.} Results from \df\ are indicated by solid curves whereas results from ASA are indicated by dashed curves.}
	\label{fig:npx_DF_vs_ASA_ps2_0V}
\end{figure}

\begin{figure}[h!]
\centering
	\includegraphics[width=0.6\textwidth]{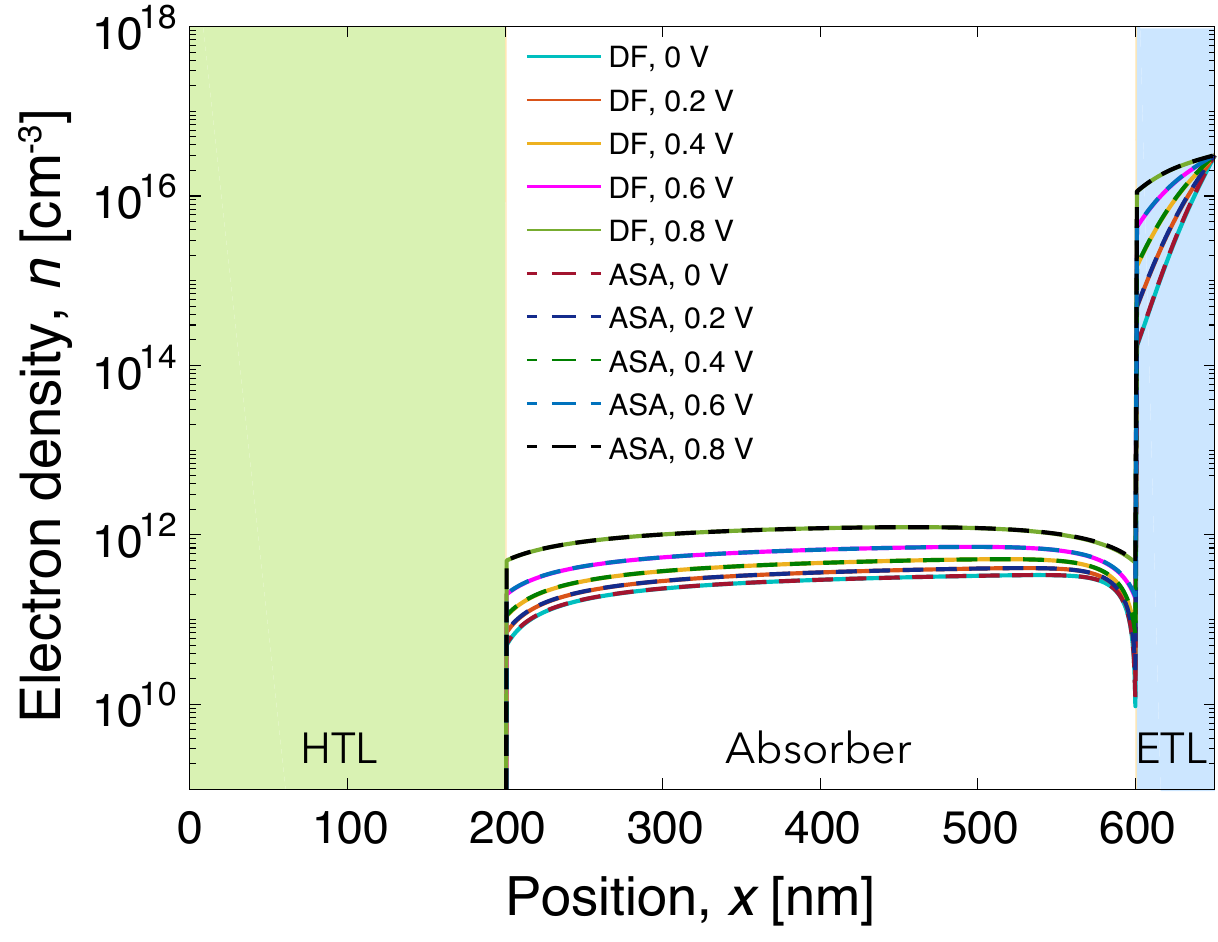}
	\caption[Comparison of calculated \df\ and ASA electron density profiles during \JV\ scan for a three-layer device with Parameter Set 1 under illumination.]{\textbf{Comparison of calculated \df\ and ASA electron density profiles during \JV\ scan for a three-layer device with Parameter Set 1 under illumination.}}
	\label{fig:DF_vs_ASA_nx_m1_1S}
\end{figure}

\begin{figure}[h!]
\centering
	\includegraphics[width=0.5\textwidth]{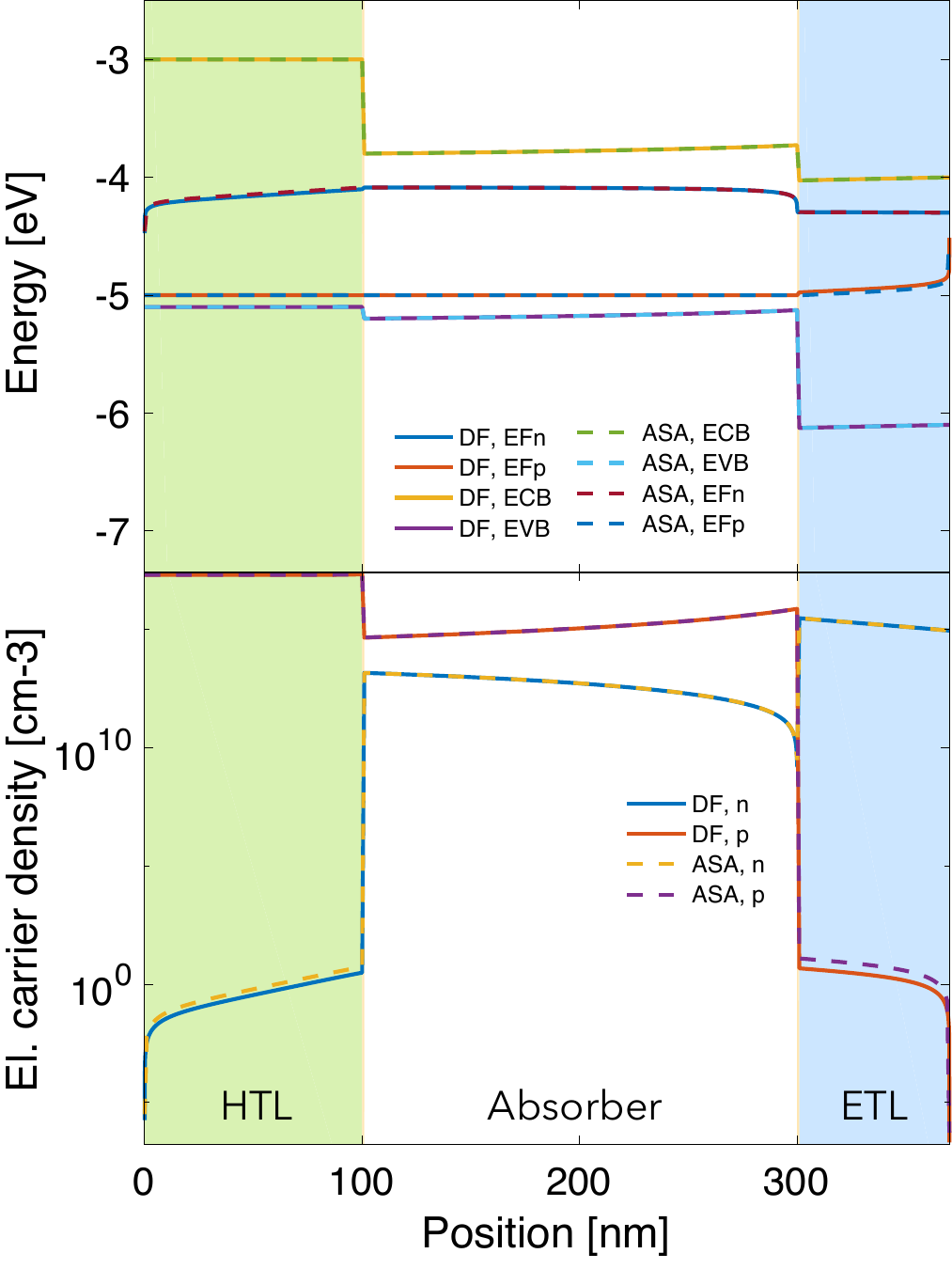}
	\caption[Energy level diagram and carrier densities for a device with properties defined by Parameter Set 2 under illumination at Vapp = 0.7 V.]{\textbf{Energy level diagram and carrier densities for a device with properties defined by Parameter Set 2 under illumination at Vapp = 0.7 V.} Results from \df\ indicated by the solid lines, while those from ASA are indicated by dashed lines.}
	\label{fig:DF_vs_ASA_m2_0p7V1S}
\end{figure}

\begin{figure}[h!]
\centering
	\includegraphics[width=0.6\textwidth]{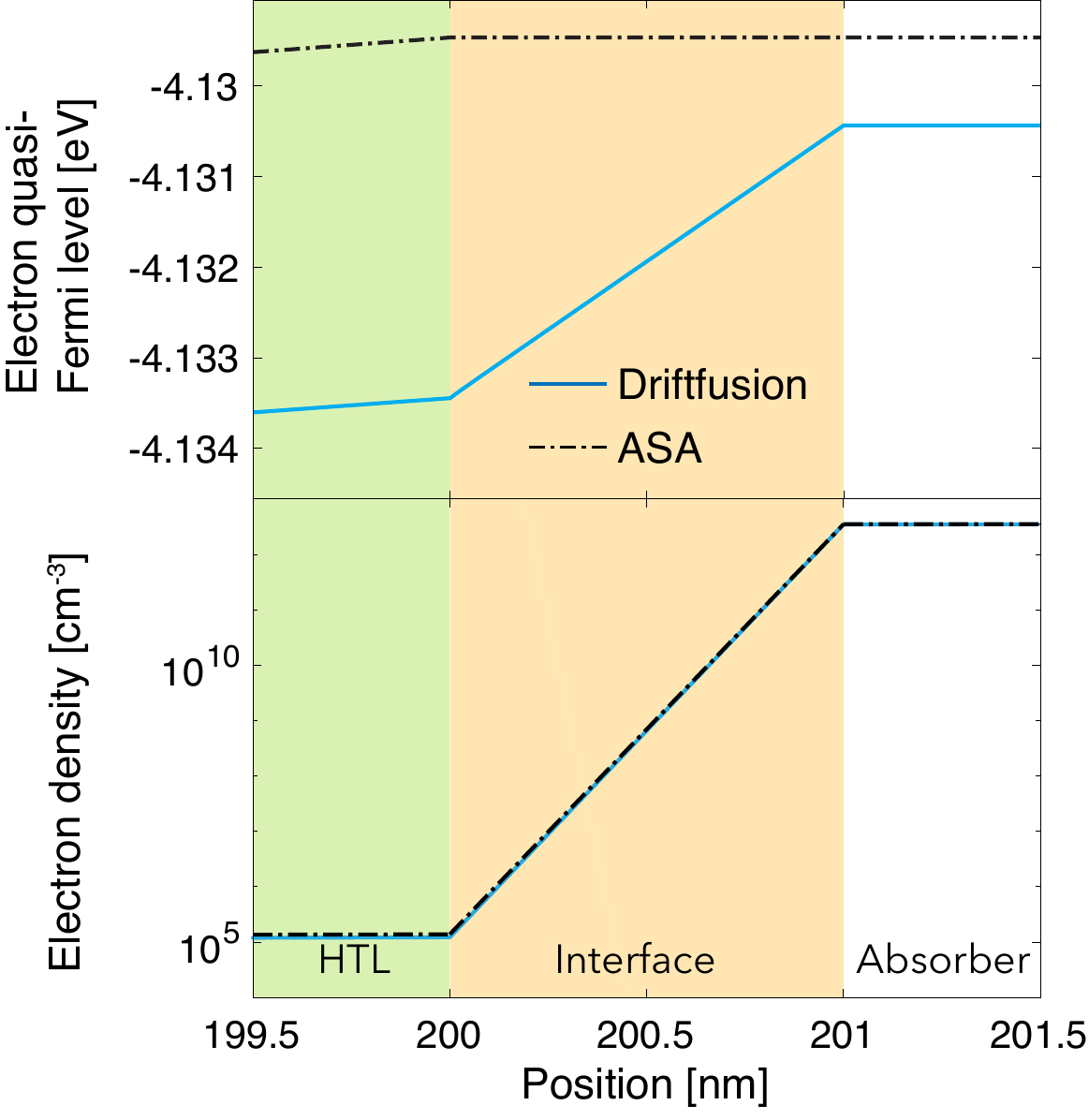}
	\caption[Electron quasi-Fermi level and carrier density under illumination at Vapp = 1 V comparison between \df\ and ASA.]{\textbf{Electron quasi-Fermi level and carrier density under illumination at Vapp = 1 V comparison between \df\ and ASA.} Results from \df\ are indicated by solid blue curves whereas results from ASA are indicated by dashed black curves.}
	\label{fig:ELnpx_DF_vs_ASA_zoom}
\end{figure}

\FloatBarrier
\clearpage
\subsection{Numerical solution for a three-layer device with mobile ions: comparison with IonMonger}
\label{sec:IM_comp_SI}
\begin{tabular}{l l}
Repository branch: & \texttt{Methods\_IonMonger\_comparison}\\
Parameters files:  & \texttt{./Input\_files/IonMonger\_default\_bulk.csv} \\
 					& \texttt{./Input\_files/IonMonger\_default\_IR.csv}\\
Scripts:			& \texttt{df\_methods\_IonMonger\_comparison\_bulk.m} \\	
					& \texttt{df\_methods\_IonMonger\_comparison\_IR.m}\\	
\end{tabular}

\begin{table}[h]
\small
\caption{\textbf{Key layer-specific parameters for simulation comparison with IonMonger.} $^{\ast}$Ions in these layer are immobile and are balanced by a static counter charge hence they do not contribute to mobile ionic space charge. $^{\dagger}$The limiting ion densities have been set to a very high value since IonMonger does not include a steric model similar to that described in Section \ref{ssec:transport_drift_diffusion}. ${\ddagger}$ Minority carriers are not simulated in the transport layers in IonMonger, hence recombination was effectively switched off in these regions. HTL and ETL denote hole and electron transport layers respectively.}
\label{tbl:Par_ion_monger_layers}
\begin{tabular}{p{3cm}lp{1.5cm}p{1.5cm}p{1.5cm}p{1.5cm}l}
\hline\noalign{\smallskip}
Variable 					& Symbol 			&								&	Layer		&									&	Unit  \\
							&					& HTL							& Absorber		& ETL								&	\\
\noalign{\smallskip}\hline\noalign{\smallskip}
Optical parameters material reference	&	-	&	SiO$_2$				&	MAPICl				&	SiO$_2$				&	- \\
Thickness 					& $d$ 				& $1 \times 10^{-5}$	& $4 \times 10^{-5}$  	& $2 \times 10^{-5}$ 	&	cm \\
Electron affinity			& $\Phi _{EA}$		& $-3.1$				& $-3.7$				& $-4.0$				&	eV	\\
Ionisation potential		& $\Phi _{IP}$		& $-5.1$				& $-5.4$				& $-6.0$				&	eV	\\
Equilibrium Fermi energy	& $E_\mathrm{F0}$	& $-5.0$				& $-4.55$				& $-4.1$				&	eV	\\
Acceptor density			& $N_\mathrm{A}$			& $1.03 \times 10^{18}$& $8.32 \times 10^{7}$	& $0$					&	\densunit	\\
Donor density				& $N_\mathrm{D}$			& $0$					& $0$					& $1.03 \times 10^{18}$&	\densunit	\\
SRH trap energy				& $E_\mathrm{trap}$		& $-4.1$				& $-4.55$				& $-5.0$				&	eV	\\
eDOS conduction band		& $N_\mathrm{CB}$			& $5 \times 10^{19}$ 	& $8.1 \times 10^{18}$	& $5 \times 10^{19}$ 	&	\densunit	\\
eDOS valence band			& $N_\mathrm{VB}$			& $5 \times 10^{19}$ 	& $5.8 \times 10^{18}$ & $5 \times 10^{19}$ 	&	\densunit	\\
Intrinsic cation density	& $N_\mathrm{cat}$			& $5 \times 10^{19\ast} $ 	& $5 \times 10^{19}$ 	& $5 \times 10^{19\ast}$ 	&	\densunit	\\
Intrinsic anion density	& $N_\mathrm{ani}$			& $5 \times 10^{19\ast}$ 	& $5 \times 10^{19\ast}$ 	& $5 \times 110^{19\ast}$ 	&	\densunit	\\
Limiting cation density	& $c_\mathrm{max}$			& $10^{100 \dagger}$		 	& $10^{100 \dagger}$			& $10^{100 \dagger}$			&	\densunit	\\
Limiting anion density		& $a_\mathrm{max}$			& $10^{100 \dagger}$	 		& $10^{100 \dagger}$			& $10^{100 \dagger}$			&	\densunit	\\
Electron mobility			& $\mu _e$			& $0.2$					& $20$					& $0.1$					&	\mobunit	\\
Hole mobility				& $\mu _h$			& $0.02$				& $2$					& $0.001$				&	\mobunit	\\
Cation mobility				& $\mu _c$			& $0^{\ddagger}$					& $3.12 \times 10^{-12}$	& $0^{\ddagger}$				&	\mobunit	\\
Anion mobility				& $\mu _a$			& $0^{\ddagger}$					& $0^{\ddagger}$					& $0^{\ddagger}$					&	\mobunit	\\
Relative dielectric constant	&	$\varepsilon _r$	&	$3$						& $24.1$						& $10$							&	-	\\
Direct recombination coefficient	&	$k_\mathrm{rad}$	&	$0^{\ddagger}$		& $1 \times 10^{-12}$			& $0^{\ddagger}$				&	cm$^3$s$^{-1}$	\\
SRH recombination electron lifetime	&	$\tau _n$	&	$10^{100 \ddagger}$	& $3 \times 10^{-10}$ 			& $10^{100 \ddagger}$ 			& s \\
SRH recombination hole lifetime		&	$\tau _p$	&	$10^{100 \ddagger}$	& $3 \times 10^{-8}$ 			& $10^{100 \ddagger}$ 			& s \\
\noalign{\smallskip}\hline
\end{tabular}
\end{table}

\begin{table}[h]
\small
\caption{\textbf{Interface-specific parameters for simulation comparisons with IonMonger.} (\textbf{a}) and (\textbf{b}) denote the bulk and interfacial recombination dominated devices respectively.* The relative dielectric constant is set artificially high in this parameter set to reduce the electric field strength within the interfacial regions: this modification facilitates a more direct comparison with the abrupt interface model.}
\label{tbl:Par_ion_monger_interface}
\begin{tabular}{lllll}
\hline\noalign{\smallskip}
Variable									&	Symbol				& 	HTL-Absorber 											& ETL-Absorber 			& 	Unit \\
Thickness 									& 	$d$ 				& 	$10^{-7}$												&	$10^{-7}$			&	cm	\\
Electron surface recombination velocity	&	$s_n$				& 	(\textbf{a})	$10^{-20}$	(\textbf{b})	$10$		&	(\textbf{a})	$10^{-20}$		(\textbf{b})	$10^7$				&	cm s$^{-1}$	 \\
Hole surface recombination velocity		&	$s_p$				& 	(\textbf{a})	$10^{-20}$	(\textbf{b})	$10^7$		&	(\textbf{a})	$10^{-20}$		(\textbf{b}) 	$10^3$				&	cm s$^{-1}$	 \\
Relative dielectric constant				&	$\varepsilon _r$	&	$10^3*$													& $10^3*$					&	-	\\
\noalign{\smallskip}\hline
\end{tabular}
\end{table}

\begin{table}[h]
\small
\caption{\textbf{Key device-wide parameters for simulation comparisons with IonMonger.}}
\label{tbl:Par_ion_monger_device_wide}
\begin{tabular}{p{6cm}lp{1.5cm}l}
\hline\noalign{\smallskip}
Variable 					& Symbol 			& 	Unit  \\
\noalign{\smallskip}\hline\noalign{\smallskip}
Left-hand boundary Fermi energy						&	$\Phi _{l}$	&	$-4.1$		&	eV	\\
Right-hand boundary Fermi energy						&	$\Phi _{r}$	&	$-5.0$		&	eV	\\
Left-hand boundary electron extraction coefficient	&	$s_{n,l}$	&	$10^{7}$	&	cm s$^{-1}$	\\
Right-hand boundary electron extraction coefficient	&	$s_{n,r}$	&	$0$			&	cm s$^{-1}$	\\	
Left-hand boundary hole extraction coefficient		&	$s_{p,l}$	&	$0$			&	cm s$^{-1}$	\\	
Right-hand boundary hole extraction coefficient		&	$s_{p,r}$	&	$10^{7}$	&	cm s$^{-1}$	\\
\noalign{\smallskip}\hline
\end{tabular}
\end{table}

\FloatBarrier
\subsubsection{Equilibrium state}
\label{sssec:IM_additional_data}

\begin{figure}[h!]
\centering
	\includegraphics[width=0.6\textwidth]{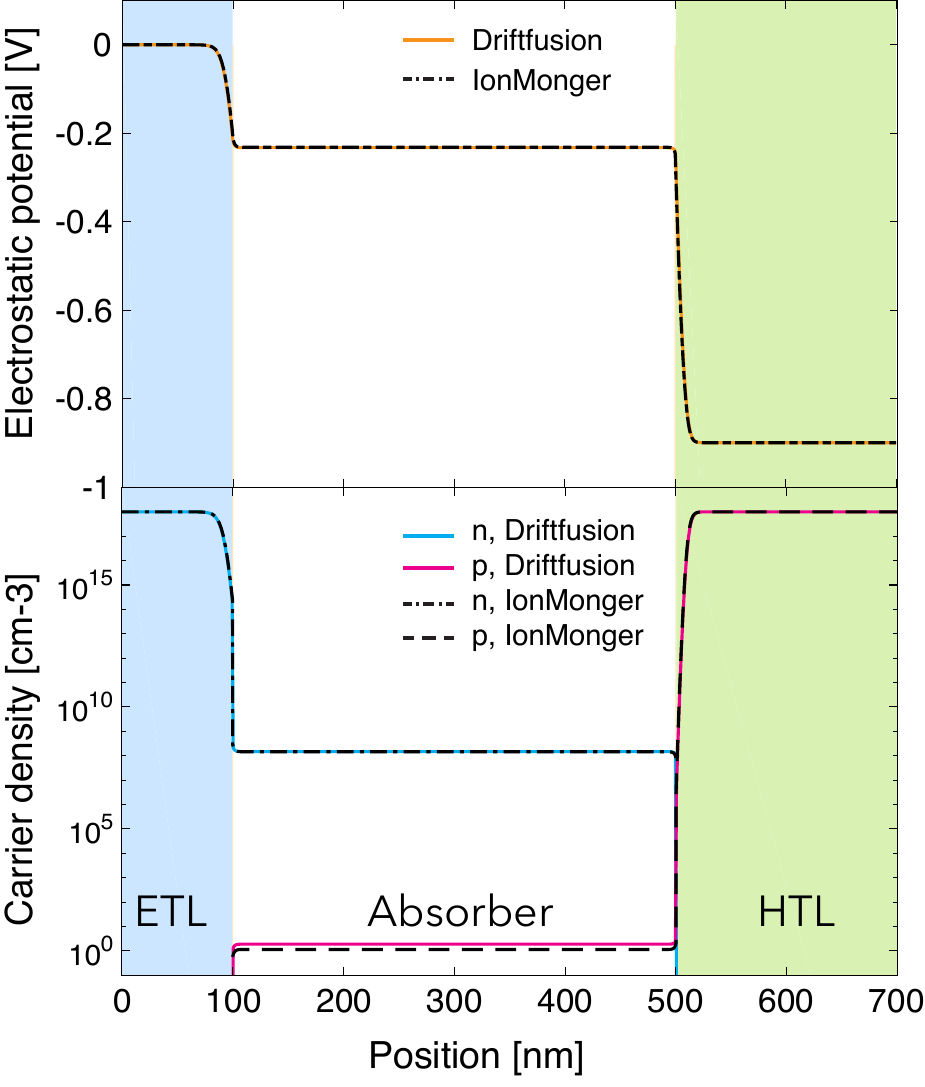}
	\caption[Comparison of the equilibrium state electrostatic potential and electronic carrier densities of the bulk carrier recombination device calculated by \df\ and \im . ]{\textbf{Comparison of the equilibrium state electrostatic potential and electronic carrier densities of the bulk carrier recombination device calculated by \df\ and \im .} Results from \df\ (DF) are indicated by solid coloured lines, whereas those from \im\ (IM) are indicated by dashed black lines. Results for the device dominated by interfacial recombination (not shown) were very similar.}
	\label{fig:df_vs_im_equilibrium}
\end{figure}

\FloatBarrier
\subsubsection{Additional data}

\begin{figure}[h!]
\centering
	\includegraphics[width=\textwidth]{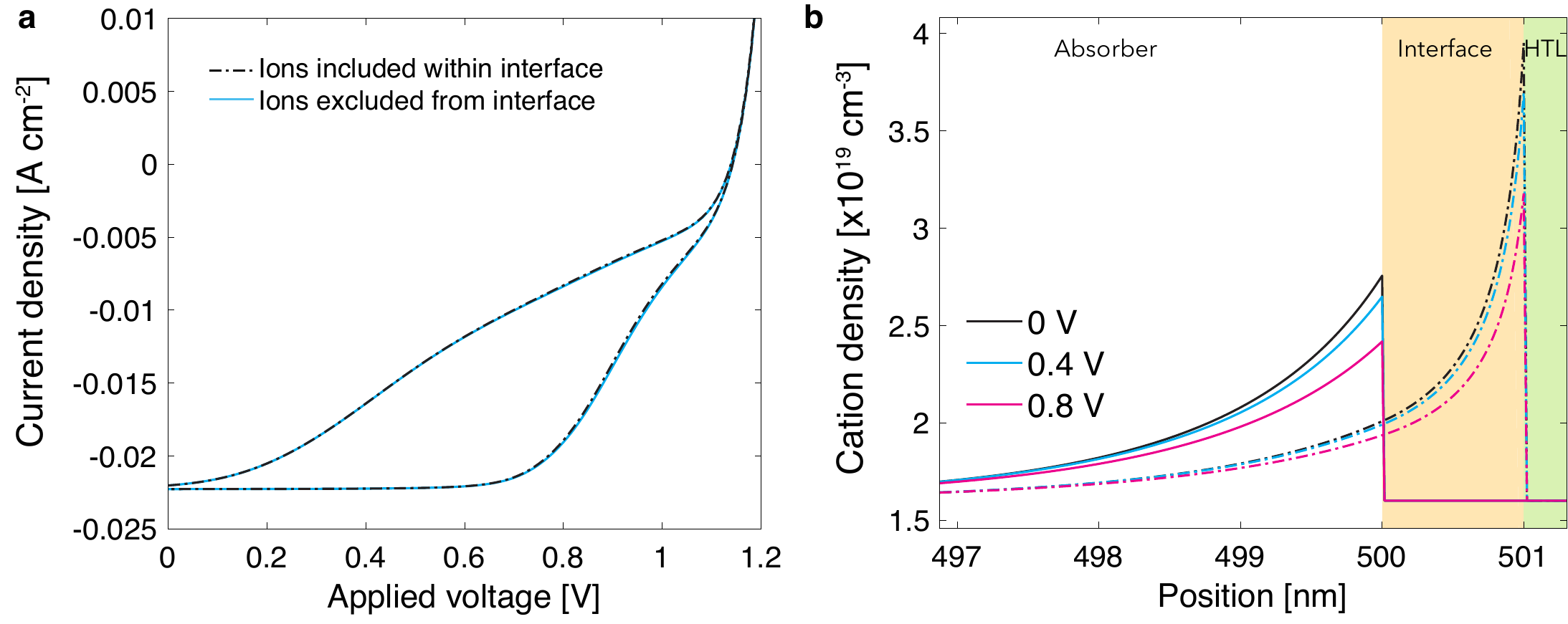}
	\caption[Comparison of current-voltage characteristics and ionic carrier density profiles for a three-layer device calculated with \df .]{\textbf{Comparison of current voltage characteristics and ionic carrier density profiles for a three-layer device calculated with \df .} (\textbf{a}) Current-voltage scans for the three-layer device described by the parameters given in Tables \ref{tbl:Par_ion_monger_layers} and \ref{tbl:Par_ion_monger_device_wide} at a scan rate of $1$ Vs$^{-1}$. (\textbf{b}) Corresponding ionic carrier densities for the two conditions at increasing applied voltage. Solid lines indicate calculations where ions were excluded, whilst dashed lines indicate cases where ions were included in the interface regions (yellow background). The green background indicates the hole transport layer.}
	\label{fig:Ion_inclusion}
\end{figure}

\begin{figure}[h!]
\centering
	\includegraphics[width=0.6\textwidth]{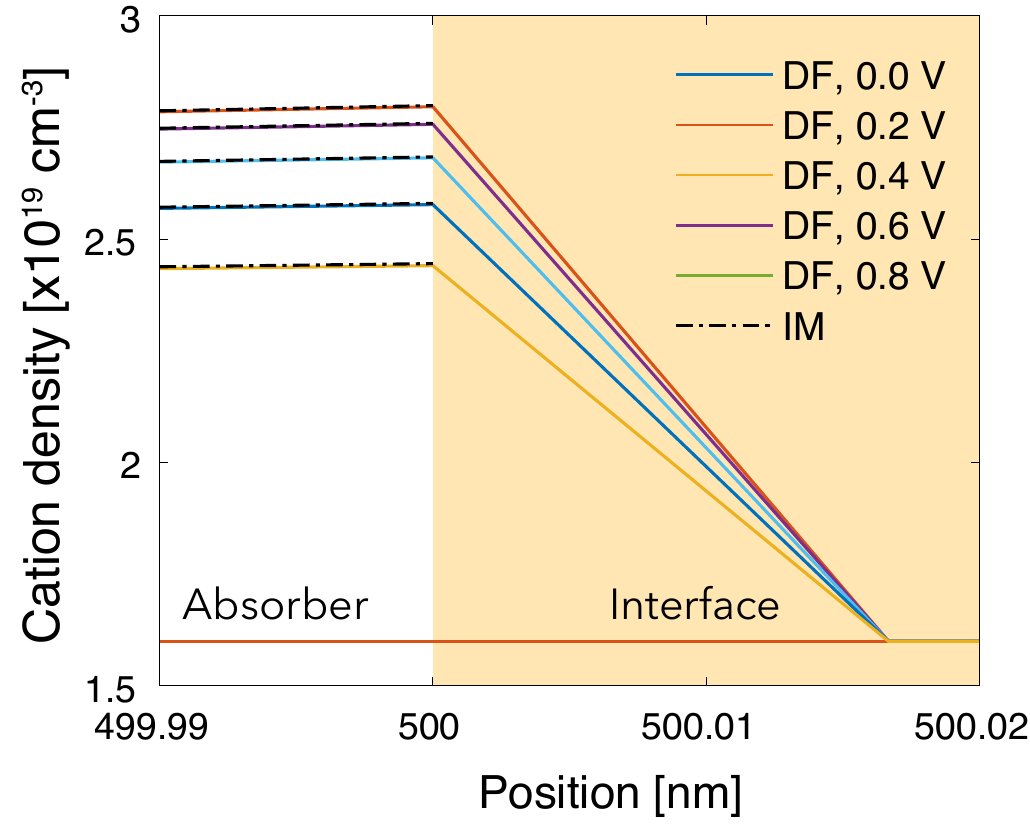}
	\caption[Zoom of ionic carrier densities at the absorber-HTL interface for a three-layer device calculated with \df and \im . ]{\textbf{Zoom of ionic carrier density profiles for a three-layer device calculated with \df .} Results from \df\ (DF) are indicated by solid coloured lines, whereas those from \im\ (IM)are indicated by dashed black lines.}
	\label{fig:Ion_interfaces}
\end{figure}

\begin{figure}[h]
\centering
	\includegraphics[width=\textwidth]{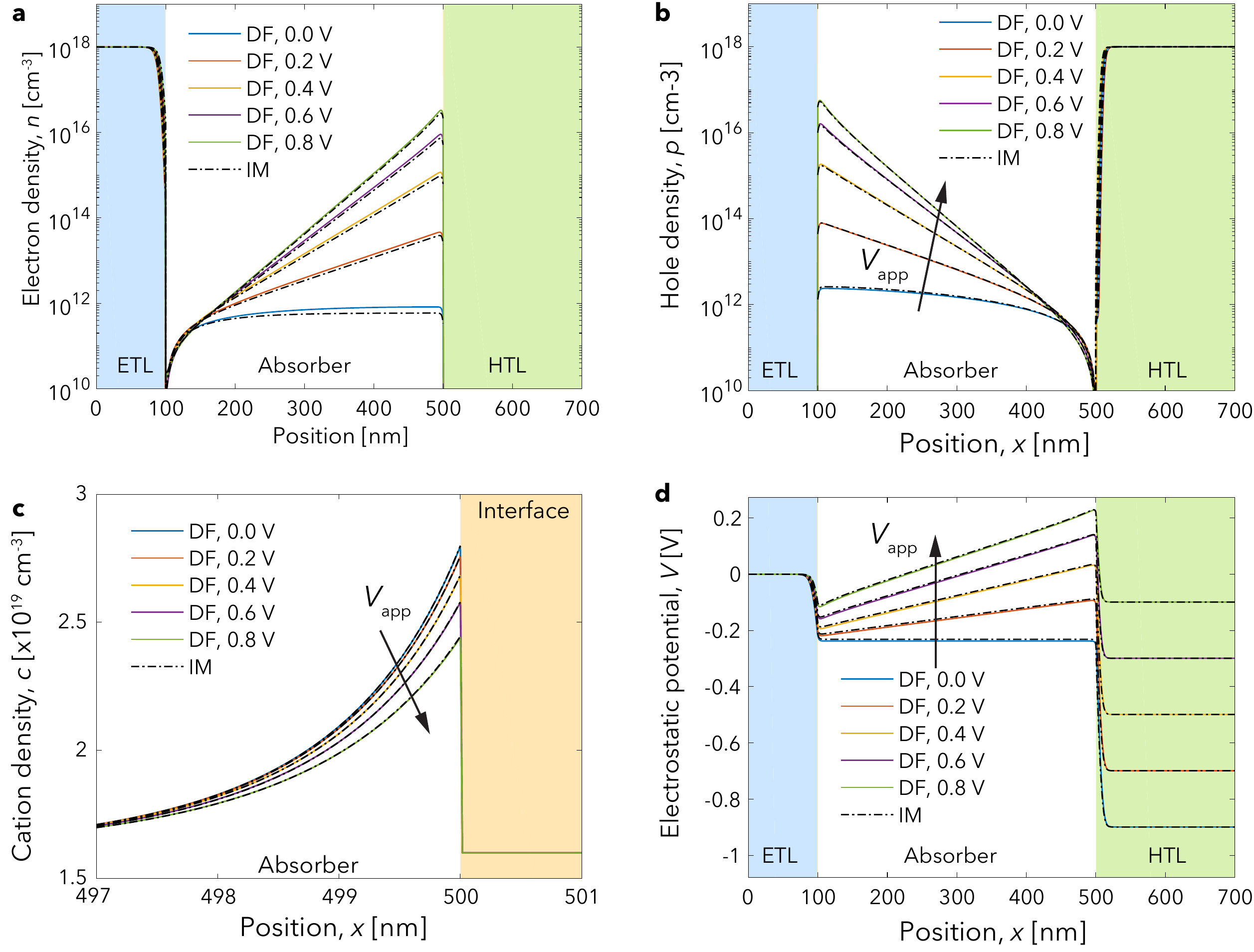}
	\caption[Comparison of results calculated using \df\ and \im\ for a three-layer solar cell dominated by bulk recombination including mobile ionic carriers in the absorber layer.]{\textbf{Comparison of results calculated using \df\ and \im\ for a three-layer solar cell dominated by bulk recombination including mobile ionic carriers in the absorber layer.} (\textbf{a}) Electron density (\textbf{b}) Hole density, (\textbf{c}) Cation density, and (\textbf{d}) electrostatic potential profiles at increasing voltage during a $1$ V s$^{-1}$ forward scan (c.f. Figure \ref{fig:DF_vs_IM}). Results from \df\ (DF) are indicated by solid coloured lines, whereas those from \im\ (IM) are indicated by dashed black lines. The complete parameter sets for the simulations are given in Tables \ref{tbl:Par_ion_monger_layers}, \ref{tbl:Par_ion_monger_interface} and \ref{tbl:Par_ion_monger_device_wide}.}
	\label{fig:DF_vs_IM_solx}
\end{figure}

\begin{figure}[h]
\centering
	\includegraphics[width=\textwidth]{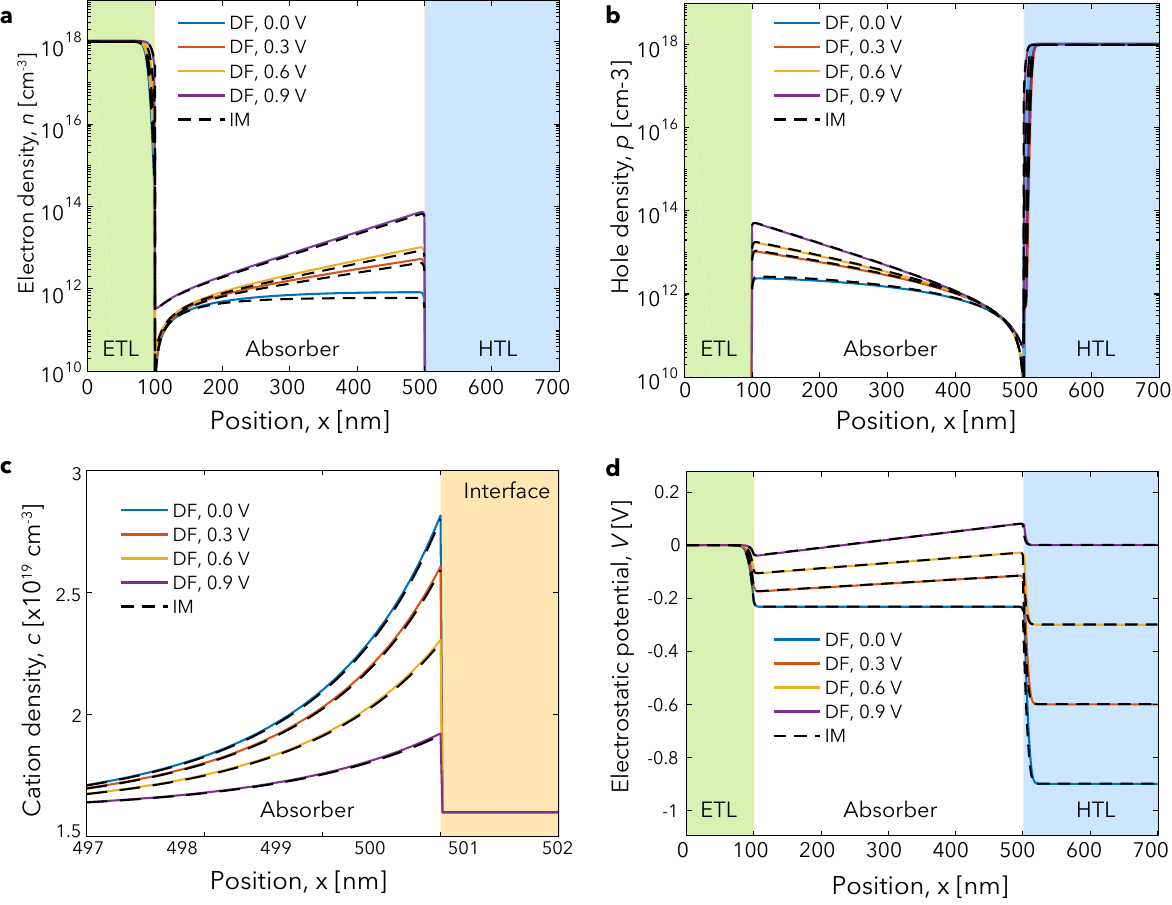}
	\caption[Comparison of results calculated using \df\ and \im\ for a three-layer solar cell dominated by interfacial recombination including mobile ionic carriers in the absorber layer.]{\textbf{Comparison of results calculated using \df\ and \im\ for a three-layer solar cell dominated by interfacial recombination including mobile ionic carriers in the absorber layer.} (\textbf{a}) Electron density (\textbf{b}) Hole density, (\textbf{c}) Cation density, and (\textbf{d}) electrostatic potential profiles at increasing voltage during a $100$ mV s$^{-1}$ forward scan (c.f. Figure \ref{fig:DF_vs_IM}). Results from \df\ (DF) are indicated by solid coloured lines, whereas those from \im\ (IM) are indicated by dashed black lines. The complete parameter sets for the simulations are given in Tables \ref{tbl:Par_ion_monger_layers}, \ref{tbl:Par_ion_monger_interface} and \ref{tbl:Par_ion_monger_device_wide}.}
	\label{fig:DF_vs_IM_solx_IR}
\end{figure}

\begin{figure}[h!]
\centering
	\includegraphics[width=\textwidth]{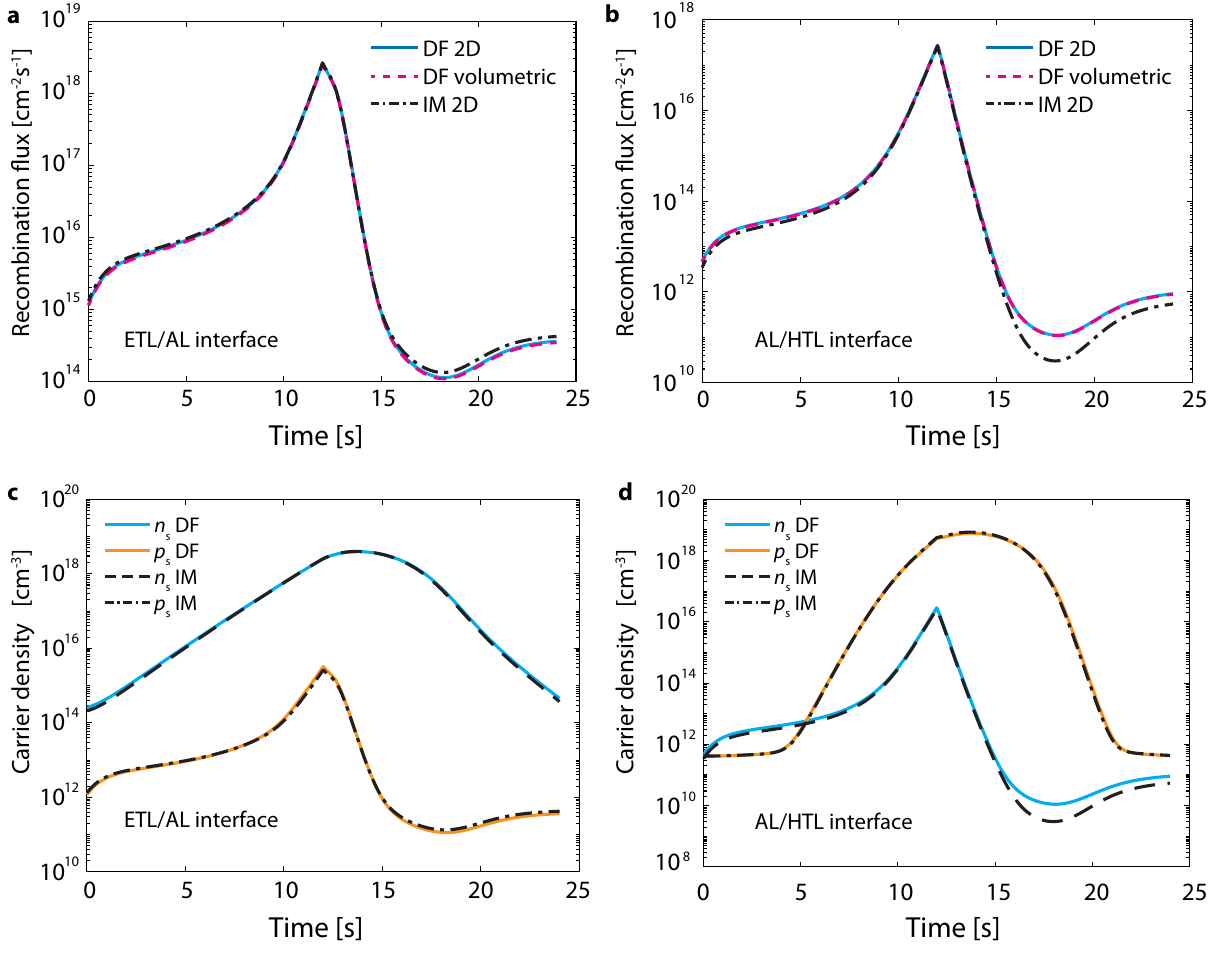}
	\caption[Comparison of interfacial recombination fluxes and interface surface carrier densities as a function of time during a 100 mV s$^{-1}$ \textit{J-V} scan calculated using \df\ and \im\ for a three-layer solar  cell dominated by interfacial recombination and including mobile ionic carriers in the absorber layer.]{\textbf{Comparison of interfacial recombination fluxes and interface surface carrier densities as a function of time during a 100 mV \textit{J-V} scan calculated using \df\ and \im\ for a three-layer solar  cell dominated by interfacial recombination and including mobile ionic carriers in the absorber layer.} Recombination fluxes for (\textbf{a}) the electron transport layer - active layer (ETL/AL) interface and (\textbf{b}) the active layer - hole transport layer interface (AL/HTL). (\textbf{c}) and (\textbf{d}): Corresponding interfacial surface carrier densities $n_\mathrm{s}$ and $p_\mathrm{s}$ (see Figure \ref{fig:Interface_schematic} for illustration) for the ETL/AL and AL/HTL interfaces respectively. The two dimensional recombination flux for \df\ (DF 2D, solid coloured curves) have been back-calculated from the solution using $n_\mathrm{s}$ and $p_\mathrm{s}$ and the SRH surface recombination expression in Equation \ref{eq:SRH_2D}. The actual recombination fluxes calculated in \df , using the volumetric scheme described in Section \ref{ssec:recombination_interfaces} are labelled as DF volumetric (dashed coloured curves). The recombination fluxes calculated using IonMonger\cite{Courtier2019a} are also back-calculated using the expression in Ref \cite{Courtier2019}, where additional factors are included with the $n_i^2$, which have been neglected here. The volumetric interfacial recombination scheme in \df\ shows a high level of self-consistency with the 2-D model. Some variation is observed between \df\ and \im , which can be attributed to the difference in electron density at the AL/HTL interface owing to small differences in the ionic carrier distributions shown in Figure \ref{fig:DF_vs_IM_solx_IR}.}
	\label{fig:DF_vs_IM_rec_flux_compare}
\end{figure}

\begin{figure}[h]
\centering
	\includegraphics[width=0.7\textwidth]{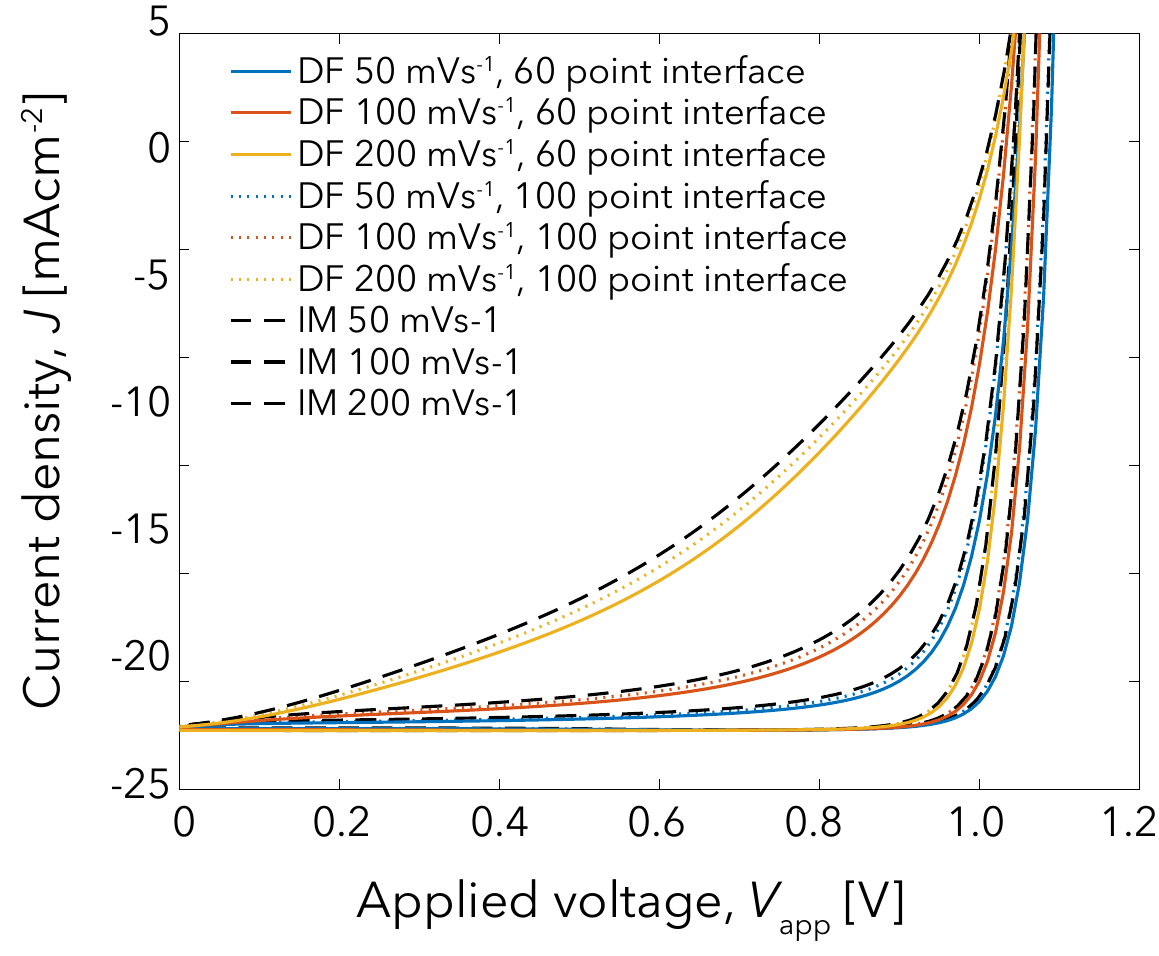}
	\caption[Effect of change in band energy and mesh spacing on accuracy of the volumetric surface recombination model currents.]{\textbf{Effect of change in band energy and mesh spacing on accuracy of the volumetric surface recombination model currents.} Current density-voltage scans for devices with the parameter set given in Table \ref{tbl:Par_ion_monger_layers} with energetic barriers to minority carriers of $0.8$ eV as opposed to $0.4$ eV for both transport layers. The larger values of $\alpha$ and $\beta$ (Main Text, Equations \ref{eq:alpha_interface} and \ref{eq:beta_interface}) for the $0.8$ eV barriers (solid curves) lead to a larger divergence from the abrupt interface model (black dashed curves) than the $0.4$ eV barriers shown in Figure \ref{fig:DF_vs_IM} of the Main Text. The accuracy of the volumetric recombination scheme can be improved by doubling the interface thickness to $2$ nm (thus reducing $\alpha$ and $\beta$) and increasing the number of points in the interfaces from $60$ to $100$ (Coloured, dotted curves).$  $}
	\label{fig:DF_vs_IM_interface_points}
\end{figure}

\bibliographystyle{spphys}       
\bibliography{library}   

%
%